\documentclass[12pt]{article}
\usepackage{caption}
\usepackage{authblk}
\usepackage{geometry}
\usepackage{amsmath}
\usepackage{amssymb}
\usepackage{amsthm}
\usepackage{mathrsfs}
\usepackage{customcommands}
\usepackage{bm}
\usepackage{Nettastyle}
\usepackage{dsfont}
\usepackage{comment}
\usepackage[utf8]{inputenc}
\usepackage{calc}
\usepackage{accents}
\usepackage{amsfonts}

\usepackage{braket}
\usepackage[normalem]{ulem}
\usepackage{color}
 \usepackage{booktabs}
\usepackage{ulem}
\usepackage{mathtools}
\usepackage{tensor} 
\usepackage{tikz}
\usetikzlibrary{arrows, positioning, quotes, intersections, arrows.meta}
\usetikzlibrary{decorations.pathmorphing}
\usepackage{subcaption}
\usepackage{graphicx}  

\usepackage{thmtools, thm-restate}
\declaretheorem{theorem}

\newtheorem{corollary}{Corollary}

\newcommand{\rhobar}{{\bar{\rho}}}
\newcommand{\calN}{\mathcal{N}}
\newcommand{\calV}{\mathcal{V}}
\newcommand{\calM}{\mathcal{M}}
\newcommand{\calG}{\mathcal{G}}
\newcommand{\calE}{\mathcal{E}}
\newcommand{\calR}{\mathcal{R}}
\newcommand{\bbZ}{\mathbb{Z}}
\newcommand{\calC}{\mathcal{C}}
\newcommand{\calO}{\mathcal{O}}
\newcommand{\calK}{\mathcal{K}}

\newcommand{\eq}[1]{\begin{align}#1\end{align}}
\newcommand{\ignore}[1]{}

\title{Cosmic Censorship in the Landscape of SUSY-Broken AdS Vacua}

\author{Netta Engelhardt}
\author{and Richard Nally}

\affiliation{Center for Theoretical Physics -- a Leinweber Institute\\ Massachusetts Institute of Technology \\Cambridge, MA 02139, USA}

\emailAdd{engeln@mit.edu}
\emailAdd{rnally@mit.edu}

\preprint{MIT-CTP/6154}

\abstract{We investigate  signatures of weak cosmic censorship in non-supersymmetric vacua of both IIA and IIB string theory. In particular, we examine the Penrose Inequality, which can be proven holographically using Ryu-Takayanagi, and can be independently proven in SUSY AdS vacua for spherically symmetric initial data. We prove that any EFT of scalars coupled to gravity in AdS, including in particular those of string compactifications, satisfies both the Penrose Inequality and a positive mass theorem if a function that we call a ``weak fake superpotential" exists. We demonstrate that this condition is satisfied globally for the DGKT skew-whiffed vacua in IIA. Since there is no known nonperturbative instability in this DGKT vacuum, this provides evidence that the \textit{IIA SUSY-broken AdS vacua may be absolutely stable}. We also find the weak fake superpotential exists  locally in field space in some SUSY-broken AdS vacua in a KKLT construction in IIB. Thus, within the framework considered here, some false SUSY-broken KKLT AdS vacua are stable against vacuum decay. Conversely, we find that some KKLT vacua do not have a weak fake superpotential, and correspondingly violate the Penrose Inequality and fail to have a positive mass theorem; the DGKT vacua only violate the Penrose Inequality in the presence of designer gravity boundary conditions.  In all cases studied, this is due to false vacuum decay. We discuss the implications of these violations for holography and black hole thermodynamics in these SUSY-broken AdS vacua. }

\begin{document}

\maketitle

\section{Introduction} \label{sec:intro}

The censorship conjectures of classical General Relativity~\cite{Pen69, Pen79, GerHor79, ChoGer69, Chr91, Chr99, FriSch93, GalSch99} aim to prevent semiclassical observers under varying assumptions from accessing physics that fails to admit a semiclassical description. The well-studied weak cosmic censorship conjecture (WCCC) asserts the absence of naked singularities that form from generic nonsingular initial data; see~\cite{GerHor79} for a precise formulation. The conjecture is notoriously difficult to test because by definition a naked singularity is one that does not lie behind a horizon; as horizons are defined teleologically from future infinity, $i^{+}$, it is challenging to ascertain whether a given singularity is naked or not with finite time evolution. See~\cite{HerHor04b, HubLiu04} for some work investigating WCCC in string theory and holography.

Fortunately,  there are methods of detecting the existence of certain trapped surfaces  -- which by the Penrose Singularity theorem~\cite{Pen65} lead to singularities -- outside of horizons that do not require a confrontation with the teleology problem. The Penrose Inequality (PI)~\cite{Pen73} is a test that is devised precisely to be implemented on a single initial data slice. In an asymptotically AdS spacetime with AD mass $M$, the inequality asserts~\cite{ItkOz11, EngHor19}:
\begin{equation}
    \frac{{\rm Area}[{\rm apparent \ horizon}]}{{\rm Area}[{\rm largest \ stationary \ BH \ w/ \ mass \ }M]}\leq 1.\label{eq:PIintro}
\end{equation}
This inequality can be derived from the joint assumptions of black hole equilibration and weak cosmic censorship whenever the ``apparent horizon'' here is taken to mean a minimar surface~\cite{EngWal17b, EngWal18}. The PI was also derived directly in~\cite{EngHor19} from the covariant holographic entanglement entropy prescription~\cite{RyuTak06, HubRan07} without the assumption of cosmic censorship. There are various proofs of the PI in limited non-holographic contexts~\cite{HuiIlm01, Bra99, Mar09}, but moreover, there are now established violations in AdS spacetimes satisfying the null energy condition~\cite{Fol22,Fol25}. 

Let us review the setup of the violations found in \cite{Fol22,Fol25}, as they form the basis of our investigation. \cite{Fol22} found a numerical dichotomy: in BF-allowed scalar truncations of eleven- and ten-dimensional supergravities known to admit a UV-complete embedding into string theory (as in e.g. \cite{CveDuf99}), no spherically symmetric initial data sets were found to violate the PI, but analogous initial data sets for NEC-satisfying nonsupersymmetric BF-allowed tachyonic potentials without a known stringy embedding, such as ordinary quartic scalars with suitably tuned coefficients,  had large robust PI violations.

The existence of violations in non-supersymmetric EFTs combined with a proof of the PI using holography raises an immediate question. Is the PI, and by extension some loose version of WCCC, enforced by supersymmetry or by a UV complete description, i.e. string theory -- even when supersymmetry is broken? We remind the reader that WCCC is tied to the validity of the statistical interpretation of black hole thermodynamics~\footnote{This is explained in greater detail in~\cite{EngFol24a}: if naked singularities rather than black holes are typical states, then the fact that black holes behave as if their microstates are indeed typical is a remarkable coincidence that requires an unlikely conspiracy.}. Furthermore, to the best of our knowledge there is no good candidate for observed violations of WCCC in our universe, and our low energy universe does not have unbroken supersymmetry. 

We thus ask the following sharp reformulation of the above question: do EFTs of valid SUSY-broken string vacua violate the PI? That is, we aim to ask whether the findings of~\cite{Fol22,Fol25}, that violations do not occur in supergravity but do occur in the absence of supersymmetry,  are a consequence of those non-SUSY theories being in the swampland, or simply a consequence of being non-supersymmetric. 

Surprisingly, we find that violations of the Penrose Inequality are directly related to instability due to  false vacuum decay in effective field theory, even though Coleman-de Luccia (CdL) bubbles~\cite{ColDel80} do not typically involve trapped surfaces\footnote{Arguably this should not be completely surprising:~\cite{Bra99} already noted the relationship between the positive mass theorem and the Penrose Inequality in the Riemannian context. However, the Riemannian context is dramatically different from the more general Lorentzian case, and we find it surprising that there is a connection that remains robust.}.  More precisely, we prove that the existence of a \textit{weak fake superpotential}, a function obeying:
\begin{equation}
V \geq 2|\nabla P|^2 - 3P^2,
\end{equation}
forbids violations of the (spherically symmetric) PI.  Furthermore, the existence of this function results in a \textit{positive mass theorem} in the corresponding effective field theory, even beyond spherical symmetry. Armed with this result, we investigate the validity of both the AdS PI and Positive Mass in explicit SUSY-broken AdS vacua of string theory. 

In the skew-whiffed SUSY-broken AdS vacua of DGKT~\cite{DewGir05,NarTri10} in type IIA, we explicitly construct a global, analytic weak fake superpotential. This immediately implies that there can be no Coleman-de Luccia (CdL) bubbles facilitating the decay of these vacua with standard boundary conditions. In addition to false-vacuum decay via CdL-like instantons, nonsupersymmetric string vacua are believed to be unstable to ten-dimensional effects mediated by branes, such as brane-mediated transitions between flux vacua with different values of the flux quanta. Intriguingly, there exist choices of flux in the model we study for which the most obvious decay route is not believed to exist. More precisely, \cite{MarQui22} argued that, at least to the order of approximation to which they worked, the only flux-changing transitions that can occur are constructed from D8 branes, and moreover that there exist choices of flux for which the tadpole bound prevents those decays. Thus, since this vacuum is not known to admit any brane-mediated decay routes and we rule out the existence of CdL-type decays in the false vacuum EFT, our results provide tantalizing evidence that \textit{this tachyonic vacuum might be absolutely stable}, at least with standard boundary conditions. In contrast, with nonstandard ``designer gravity'' boundary conditions~\cite{HerHor04b}, we do find violations of a weak version of the PI, but cannot test the strongest version of the PI in full generality, and find that no  Positive Mass result exists.

We also find violations of the AdS PI \textit{with standard boundary conditions} in  explicit nonsupersymmetric AdS KKLT \cite{KKLT} vacua in IIB string theory, constructed in ~\cite{McaMor24,McaNal26WIP}. As a warmup, we first consider two analytic toy models for KKLT, the original toy model introduced in \cite{KKLT} and a refined model introduced in \cite{McaMor24}. Within these toy models, we numerically search over spherically symmetric initial data sets, optimizing for violations of~Eq.~\ref{eq:PIintro}.

In IIB, we find large violations of the PI in a codimension-0 region of parameter space in the toy model of \cite{McaMor24}; since string compactifications themselves do not admit parameter spaces, we replace the scan over parameter space in the toy model with a scan over a landscape of 66 distinct nonsupersymmetric AdS KKLT vacua, and find that some, but not all, of the KKLT vacua admit PI-violating initial data.  Unlike in DGKT, where we only find violations with designer gravity boundary conditions, the violations we find in KKLT vacua exist with standard boundary conditions. Nevertheless,  violations of the Penrose Inequality are not \textit{immediate} upon breaking SUSY: most of the vacua we study admit no violations that we can find. 

This point is worth belaboring: in those SUSY-broken IIB vacua without violations, we find a positive mass EFT result within the field space that we can explore.   We say ``EFT positive mass theorem'' to emphasize the fact that the theorem applies to the given flux compactification in its regime of validity, but does not include nonperturbative effects such as antibrane metastability in KKLT \cite{KacPea01}. This, together with our finding that the DGKT vacua do not admit CdL decays and have no known brane-mediated decay routes,  strongly suggests that certain SUSY broken AdS vacua are more stable than previously expected.

What are the implications for weak cosmic censorship? The numerical evidence, presented in Sec.~\ref{sec:theorem} and in more detail in Sec. \ref{sec:kklt}, suggests that violations stem from a new bubble (of something or of nothing): violations do not appear to result from a naked singularity that can be observed by an infinitely long lived asymptotic observer (i.e. a naked singularity as defined in~\cite{Pen79}, which does not include a big crunch).  While this certainly constitutes a real violation of weak cosmic censorship, it is milder than the scenario of a negative mass Schwarzschild-like singularity forming from collapse, which is closer to Penrose's original concern~\cite{Pen79}. 

The proofs of the PI fail in these violating vacua for several reasons: first,  trapped surfaces exist without horizons. This subtle point is discussed in Sec.~\ref{sec:fail}, where we remind the reader that an event horizon is defined in terms of $i^{+}$, which does not exist in a CdL-type geometry.  Second, there is no area theorem for $\partial J^{-}[\mathscr{I}]$ if $\mathscr{I}$ is incomplete.   Third, in the case of the holographic proof, the dominant contribution to the thermal ensemble in some putative theory encompassing both true and false vacua is likely a black hole over the true vacuum rather than one over the false vacuum.  This last point suggests that violations of the PI need not immediately provide evidence against a Wheeler-DeWitt (WDW) patch version of the covariant holographic entanglement entropy prescription in these less stable vacua; we restrict to the WDW patch of a single boundary time slice in vacua that are known to have a nonvanishing probability to decay because an argument in \cite{HorOrg07,Har10} implies that such vacua have AdS-scale lifetimes.

Thus, in summary, we found one base result that guarantees both the AdS Penrose Inequality and a Positive Mass Theorem (with standard boundary conditions): the existence of a weak fake superpotential.  We find SUSY-broken AdS vacua that satisfy this requirement and thus have no Coleman-de Luccia type decay, do not violate the PI, and admit a Positive Mass result. Moreover, while PI-violating vacua do exist, they may be somewhat better behaved than one might have feared.

We now proceed to a review of the KKLT prescription for constructing AdS vacua in type IIB string theory, followed by the actual models and constructions used in this article.  Compactifying ten-dimensional type IIB string theory on an orientifold of a Calabi-Yau threefold (CY3) yields a four-dimensional effective action for three types of scalar fields: the axiodilaton $\tau$, which encodes the string coupling, the complex structure moduli $z^I$, which parameterize deformations of the shape of the underlying CY3, and the Kahler moduli $T_i$, which describe its size. Roughly speaking, KKLT gave a protocol for balancing the dynamics of these moduli against each other to obtain supersymmetric AdS vacua and then further proposed breaking supersymmetry by adding an anti-D3 brane to obtain nonsupersymmetric vacua. While the goal of KKLT was to construct de Sitter (dS) vacua, by judiciously selecting fluxes one can tune the scale of supersymmetry breaking to construct nonsupersymmetric AdS vacua as well, although it is more difficult to do so because the relevant fine tuning problem is harder in the AdS context.

Ref.~\cite{McaMor24}, as a byproduct of their construction of explicit KKLT dS vacua, also produced three nonsupersymmetric AdS vacua, of which one is tachyonic. Ongoing follow-up work \cite{McaNal26WIP} has yielded hundreds of additional vacua. As effective theories, these models are completely explicit: the effective potential for the hundreds of moduli fields, including in principle an infinite series of nonperturbative corrections, is known in closed form. That makes these models excellent candidates for directly applying the framework of \cite{Fol22,Fol25}, which involves only repeated numerical evaluations of the potential, to the study of PI violations in these EFTs. 

In this work, we approach the testing of the AdS PI, and correspondingly weak cosmic censorship, in the following contexts:

\begin{itemize}
    \item \textbf{The KKLT Toy Model:} In \cite{KKLT}, the existence of nonsupersymmetric vacua like those discovered in \cite{McaMor24} was predicted based on the study of the dynamics of a single complex modulus $\rho = u + i\sigma$, encoded in the holomorphic superpotential \eq{W(\rho) = W_0 + e^{2\pi i \rho},\label{eq:introKKLT}} where $W_0$ is a constant summarizing the effects of complex structure moduli stabilization, which KKLT do not model explicitly. From the superpotential $W(\rho)$, one can construct a supersymmetric potential $V_F$; KKLT model the effects of a supersymmetry-breaking anti-D3 brane by working not with $V_F$ but instead with the full potential \eq{V(\rho,\rhobar) = V_F + \frac{Z}{(\operatorname{Im}{\rho})^2}.\label{eq:intro_d3bar}} This model has a two-dimensional family of nonsupersymmetric solutions, parameterized by $Z$ and $W_0$; the relative size of these two parameters determines whether the vacuum one ends up with has a positive or negative cosmological constant. This model was rejected by~\cite{McaMor24} on account of its simplicity: it cannot admit AdS potentials with additional critical points before uplift, but generic supersymmetric AdS vacua of the KKLT type do have additional critical points. The potential is too trivial to be an accurate representation of the potential inherited by an actual Calabi-Yau flux compactification obtained via the KKLT protocol. For roughly the same reason, this potential is likewise too simple for Penrose Inequality  violations.

    \item \textbf{A Refined KKLT Toy Model:}  In \cite{McaMor24} it was suggested that the additional critical points that were missing from the original KKLT toy model can be accommodated by working with a generalization of the superpotential in~Eq. \ref{eq:introKKLT}, \eq{W(\rho) = W_0 + e^{2\pi i \rho} - A e^{4\pi i \rho}.\label{eq:intro_2term}} The full potentials of the compactifications constructed in \cite{McaMor24} are characterized by competition between many instantons, and Eq. \ref{eq:intro_2term} is an attempt to capture that competition in a single-field model. From this superpotential, one is free to construct a nonsupersymmetric potential as in Eq. \ref{eq:intro_d3bar}, and one now finds a three-parameter family of nonsupersymmetric vacua, parameterized by $W_0$, $Z$, and $A$. This refined toy model has a very rich phase diagram; varying these parameters yields a variety of solutions, including PI violation both with and without tachyons. To the best of our knowledge, this is the first theory known to admit a PI violation without a tachyon at the vacuum; we find this to be a novel and informative feature of this model.  
    
    \item \textbf{Candidate Anti de Sitter Vacua:} In \cite{McaMor24}, candidate de Sitter vacua of KKLT type were constructed for the first time. As a byproduct, three nonsupersymmetric AdS vacua of this type were also constructed, one of which admits a BF-allowed tachyon. In ongoing work, many more vacua of this type have been constructed \cite{McaNal26WIP}. We work with a landscape of 66 nonsupersymmetric AdS vacua obtained from an early state of the ongoing scan in \cite{McaNal26WIP}; these are fully explicit flux compactifications on orientifolds of CY3s yielding nonsupersymmetric AdS vacua with all moduli stabilized, with known potentials. We have scanned for Penrose violations in each of these geometries, and found violations in ten of these 66 models. We will demonstrate below that any vacuum that is the global minimum of its potential must necessarily be PI-safe, as was essentially proven in \cite{Fol22}; surprisingly, we find that some vacua to which this doesn't apply, and which therefore might be expected to be able to decay to other, lower-energy vacua, are nonetheless PI-safe, at least in a region of field space.    
    
    \item \textbf{Nonsupersymmetric DGKT Vacua:} The most explicit known compactifications in type IIA string theory were constructed by DGKT \cite{DewGir05}; while DGKT focused on supersymmetric compactifications, their framework equally well admits nonsupersymmetric solutions, called ``skew-whiffed" vacua because certain flux quanta are the negatives of the quanta in supersymmetric vacua \cite{NarTri10}. The skew-whiffed model described in \cite{DewGir05,NarTri10} is, at the level of its effective theory, a totally explicit nonsupersymmetric vacuum of string theory, and is known to admit a single tachyon of mass $m^2L^2 = -2$.  Nevertheless, we find that with standard boundary conditions it admits no PI-violating solutions, and indeed has a global weak fake superpotential everywhere. 
\end{itemize}

The paper is structured as follows. In Sec.~\ref{sec:PIReview}, we review the statement and arguments for the Penrose Inequality in AdS. In Sec.~\ref{sec:theorem} we prove the Penrose Inequality and a positive mass theorem from the existence of the ``weak fake superpotential'', describe our protocol for diagnosing violations of the PI, and examine two different types of violations. In Sec.~\ref{sec:kklt}, we study the PI safety of two toy models for KKLT. Sec.~\ref{sec:real} describes the PI and Positive Mass results for the class of KKLT AdS vacua that we study, while Sec.~\ref{sec:dgkt} does the same for DGKT.  In Sec.~\ref{sec:fail} we discuss the assumptions of the proofs of the PI that fail in the vacua with violations, which are all due to CdL bubbles. We conclude with a discussion of implications in Sec.~\ref{sec:conclusion}.

\section{Review of the Penrose Inequality}\label{sec:PIReview}
The goal of the Penrose Inequality (PI) is to test whether a given initial data set results, upon time evolution, in a violation of weak cosmic censorship. The construction of such a test \textit{prima facie} appears counterintuitive given the teleological nature of event horizons; we begin with a brief review of the original motivation for the PI. 

The argument takes as input two assumptions: first, that weak cosmic censorship holds, and in particular that trapped surfaces lie behind event horizons, and second, that a dynamical black hole in classical General Relativity equilibrates to a stationary black hole. In the case of the AdS PI, the logic is then as follows:

\begin{enumerate}
    \item Consider the boundary $\sigma$ of the trapped region $T$ on some time slice. This will frequently be a ``minimar'' surface -- i.e. marginally trapped and satisfying certain stability properties~\cite{EngWal18}.
    \item By the first assumption, the interior of the trapped region $T$ lies behind an event horizon, so $\sigma$ also lies behind or at worst on an event horizon, which we shall denote as ${\cal H}^{+}$.
    \item It is possible to prove directly from the definition of a minimar surface that  $\sigma$ has less area than any surface circumscribing it on some slice $\Sigma$. Since $\sigma$ lies behind (or on) ${\cal H}^{+}$, ${\cal H}^{+}\cap \Sigma$ circumscribes $\sigma$. Thus:
    \begin{equation}
        {\rm Area}[\sigma]\leq {\rm Area}[{\cal H}^{+}\cap \Sigma].
    \end{equation}
    \item Again by the first assumption of weak cosmic censorship, the event horizon ${\cal H}^+$ satisfies the Hawking area theorem, so its cross sectional area grows. Thus:
    \begin{equation}
        {\rm Area}[\sigma]\leq {\rm Area}[{\cal H}^{+}\cap \Sigma]\leq {\rm Area}[{\cal H}^{+} {\rm \ asymptotically}].
    \end{equation}
    \item By assumption of equilibration of black holes, the asymptotic value of the area is identical to the area of a stationary black hole. In AdS$_{4}$,~\cite{XiaYan22} showed that of the stationary AdS$_{4}$ black holes, Schwarzschild has the largest area. This yields the AdS Penrose Inequality: 
     \begin{align}
        {\rm Area}[\sigma]\leq {\rm Area}[{\cal H}^{+}\cap \Sigma]\leq {\rm Area}[{\cal H}^{+} {\rm \ asymptotically}]&= {\rm Area}[{\cal H}^{+} {\rm \ stationary}]\\ & \leq {\rm Area}[\rm Schwarzschild-AdS].\label{eq:penroseLeq}
        \end{align}
        More precisely, if the spacetime has an initial AD mass $M$, then the relevant Schwarzschild-AdS black hole is the one with mass $M$.
\end{enumerate}

Thus violations of the Penrose Inequality indicate a failure of either weak cosmic censorship or of black hole thermalization (or both). In particular, the expectation is that failures of the PI result in trapped surfaces outside of horizons. As we will see below, this is not in fact borne out in the string vacua that we have studied: the failures of the Penrose Inequality in such vacua are a consequence of the failure of equilibration to a final state black hole. This is in fact also true in the EFT studies of~\cite{Fol22,Fol25}.

It is possible to prove the PI in various restricted contexts without assuming weak cosmic censorship: e.g.~\cite{Bra99} in the Riemannian setting and~\cite{Fol22} under the assumption of spherical symmetry and the dominant energy condition. In Sec.~\ref{sec:theorem} below, we prove that the Penrose Inequality is satisfied if $V(\phi)$ is bounded from below by a fake superpotential, Eq.~\ref{eq:weakDef}. 

There is one proof that is worth our while to describe in detail: the proof in the holographic setting. The argument of~\cite{EngHor19} shows that (1) the covariant holographic entanglement entropy prescription (HRT)~\cite{HubRan07}, together with (2) the standard AdS/CFT duality between high energy thermal states at large-$N$ and black holes, immediately imply the Penrose Inequality. 

The argument is as follows: if $\sigma$ is a minimar surface on some initial data slice, then its area computes the so-called simple entropy:

\begin{equation}
    \frac{{\rm Area}[\sigma]}{4G}=S^{\rm simple}[\rho_{\rm bdy}]=\max\limits_{\rho\in {\cal H}}S_{\rm vN}[\rho]=S_{\rm vN}[\rho_{\rm simple}],
\end{equation}
where ${\cal H}$ is the set of all density matrices that agree with $\rho$ on the expectation values of simple operators in the boundary dual. This formula follows directly from the characteristic initial value problem in General Relativity and the HRT prescription. See also ~\cite{DutFau19} for a path integral justification of this formula that follows in a similar vein to Lewkowycz-Maldacena~\cite{LewMal13}. 

It then immediately follows that, at large-$N$:

\begin{equation}
    \frac{{\rm Area}[\sigma]}{4G}=S_{\rm vN}[\rho_{\rm simple}]\leq S_{\rm micro}[M]=\frac{{\rm Area [SAdS \ w/ \ mass \ M]}}{4G}
\end{equation}
where the inequality follows from the fact that all states in ${\cal H}$ agree on the energy $M$, and the last equality follows from the duality between the microcanonical ensemble and the static black hole~\cite{Mar18}. 

Thus a failure of the PI would call into question the validity of either HRT or the equivalence between the microcanonical ensemble and black holes in any proposed holographic dual to excited states of non-supersymmetric AdS vacua.

\section{A Penrose Safety Theorem and Its Consequences}\label{sec:theorem}

Given an initial data slice containing a minimar surface $\sigma$, the AdS PI in Eq. \ref{eq:penroseLeq} can be phrased most easily as a constraint on the ratio $\calR$ between the area of $\sigma$ and the area of the Schwarzschild black hole with the same mass: \eq{\calR \equiv \frac{\operatorname{Area}[\sigma]}{\operatorname{Area}[\text{Schwarzschild-AdS}]}\leq 1.\label{eq:calR}} In principle, to diagnose PI violations in a given gravitational EFT, one can simply scan numerically over explicit choices of initial data and look for ones with $\calR>1$, but without some way of specifically targeting violations, this is impractical, especially in high-dimensional field spaces like those of the KKLT vacua we will examine below. 

Instead, it is both more convenient and more instructive to prove abstractly that violations cannot occur, even in principle, on some region of field space; we will spend the remainder of this section developing tools that allow us to guarantee nonviolation of the PI. It is known that supersymmetry gives precisely such a guarantee; the goal of this section is to determine if there is a condition weaker than SUSY that -- without evolving the initial data -- guarantees the absence of violations. It turns out that there is: the notion of a ``(weak) fake superpotential'', which we now define. 

Supersymmetric potentials in theories with at least $\calN=1$ supersymmetry, called \textit{F-term potentials}, can be written in terms of two auxiliary functions, the holomorphic superpotential $W$ and the real K\"ahler potential $K$, as  \eq{V_F = e^K \left(g^{a\bar{b}} D_a W \overline{D}_{\bar{b}}\overline{W} - 3 |W|^2\right)\label{eq:VF},} where the indices $a,\bar{b}$ run over all fields and their complex conjugates, respectively, the metric on field space is defined by \eq{g_{a\bar{b}} = \partial_a \bar{\partial}_{\bar{b}}K\label{eq:gab}} and is positive-definite, and the covariant derivatives are defined by \eq{D_a W = \partial_a W + W\partial_aK.} Motivated by this structure, we define a weak fake superpotential as follows. Consider scalar fields $\phi$ exploring a potential $V(\phi)$, and assume we have an AdS vacuum at $\phi_0$ with AdS length $L^2 = -3/V(\phi_0)$. We say a differentiable, real, single-valued function $P$ is a \textit{fake superpotential} for $V$ on a region of field space $C \ni \phi_0$ if \begin{subequations}
    \eq{P(\phi_0) &= \frac{1}{L} \\ dP\Big|_{\phi=\phi_0} &= 0 }
    \label{eq:anchor}
\end{subequations}
and moreover if it satisfies
\eq{V(\phi) = 2\left|\nabla P(\phi)\right|^2 - 3|P|^2 \label{eq:fakeDef}}
for all $\phi \in C$. More generally, we say $P$ is a \textit{weak fake superpotential} if, in addition to Eq. \ref{eq:anchor}, we have \eq{V(\phi) \geq 2\left|\nabla P(\phi)\right|^2 - 3|P|^2 \label{eq:weakDef}} It will frequently be convenient to rewrite the condition for a weak fake superpotential as \eq{\calE \equiv V(\phi) + 3|P|^2 - 2|\nabla P|^2 \geq 0.\label{eq:calEdef}} Weak fake superpotentials will be the key technical tool used throughout the paper, so before we proceed we give two motivating examples.

Consider a four-dimensional $\calN=1$ supersymmetric theory with an F-term potential as in Eq. \ref{eq:VF}. Defining \eq{P = e^{K/2}|W|,} we find that Eq. \ref{eq:VF} becomes Eq. \ref{eq:fakeDef} exactly. Moreover, if we have a vacuum  where $D_aW=0~\forall a$, called a \textit{supersymmetric vacuum}, then Eq. \ref{eq:anchor} is satisfied as well. Thus, a supersymmetric vacuum in any theory with a genuine superpotential necessarily admits a fake superpotential; the point of a fake superpotential is that the analytic structure of the F-term potential, rather than e.g. the existence of fermions, is the aspect of supersymmetric theories that is relevant for positive mass theorems and PI safety, and that (weak) fake superpotentials can allow even nonsupersymmetric theories to have the same analytic structure.

Next, assume that $\phi_0$ is itself the \textit{global} minimum of $V$, and let $P = 1/L$. Then \eq{\calE \geq \frac{-3}{L^2} + \frac{3}{L^2} - 0 = 0,} and we have a weak fake superpotential everywhere in field space. We will see below that the existence of a fake superpotential implies PI-safety, so any vacuum that is the global minimum of its potential must satisfy the PI for all choices of initial data, subject to mild assumptions on boundary conditions that we will discuss later. Note that this is just the dominant energy condition (DEC); it was already shown that, for spherically symmetric initial data, vacua satisfying the DEC are PI-safe in \cite{Fol22}, and the above argument just rephrases that result in the language of (weak) fake superpotentials.

Note that the existence of a weak fake superpotential does not imply the existence of a fake superpotential. To see that, consider a single compact scalar $\theta$ with potential \eq{V = -3 + 4(1-\cos\theta),\label{eq:NoFakeExample}} and consider the vacuum $\theta=0$. $P = 1$ furnishes a weak fake superpotential, since \eq{\calE = 4(1-\cos\theta)\geq0,} but no fake superpotential exists. The proof of this last statement is analytic: a differentiable, nonconstant function $P$ on $S^1$ has both a minimum and a maximum. From Eq. \ref{eq:fakeDef} we see that critical points of a fake superpotential $P$ are necessarily critical points of $V$, and moreover at such a critical point we have that $V = -3|P|^2\leq0$. Thus, any potential for a compact scalar with a fake superpotential must either have two critical points with nonpositive energy or come from a constant fake superpotential, and therefore must itself be constant. The choice of $V$ in Eq. \ref{eq:NoFakeExample} has two critical points, the vacuum at $\theta = 0$ and one at $\theta = \pi$ with energy $V(\pi) = -3+8=5$. Thus, it cannot have a fake superpotential, even though it admits a weak fake superpotential. Note that this potential obeys the DEC, and so is PI-safe even in the absence of a fake superpotential; this disproves a conjecture in \cite{Fol22,Fol25} that the existence of a fake superpotential is necessary for PI-safety. 

Let field space be equipped with a metric $G$; in the models we will consider here, $G$ will be obtained from a Kahler potential as in Eq. \ref{eq:gab}. As usual consider a vacuum at $\phi_0$, and let $\rho(\phi)$ denote the square of the metric distance between $\phi$ and $\phi_0$ in field space. Further assume that the potential strictly obeys the BF bound, so that the lightest canonically normalized mode has $m^2L^2>-9/4$. Then \eq{LP = 1 + \frac{3}{8}\rho\label{eq:Puniversal}} always furnishes a weak fake superpotential in a small ball around the origin; this is because \eq{L^2\calE = L^2\left(V-V_0\right) + \frac{9}{8}\rho + \frac{27}{64}\rho^2,} but near the origin we can Taylor expand $V-V_0\sim \frac{1}{2}m^2\rho \geq-9\rho/8L^2$ so that \eq{L^2\calE \geq - \frac{9}{8}\rho + \frac{9}{8}\rho + \frac{27}{64}\rho^2 \geq0.} Thus, any physically allowed AdS vacuum\footnote{Technically the proof as written does not cover BF-saturating vacua with $m^2L^2=-9/4$ exactly, but the proof can be extended to that case.} admits a \textit{local} weak fake superpotential in some sufficiently small neighborhood about $\phi_0$.

Nevertheless, this local weak fake superpotential can fail to extend to a global weak fake superpotential, i.e. one where Eq. \ref{eq:weakDef} holds everywhere in field space. Initial data slices whose field profiles only explore regions of field space where a weak fake superpotential exists enjoy significant protections; in particular, they obey positive mass theorems and are provably always PI-safe. Thus, to find PI violations, the natural first step is to first find weak fake superpotentials with as large a domain of validity as possible, and then only explore field excursions outside that range; this is the strategy we will employ in the remainder of the paper. Before we do so, however, we will explore some more properties of weak fake superpotentials, including proving the above mentioned positive mass and PI-safety theorems. 

In the first few subsections, we will work exclusively with \textit{standard boundary conditions}, where fields $\phi$ asymptotically obey \eq{\phi \sim r^{-\Delta},\label{eq:standard}} where \eq{\Delta = \frac{3}{2}+\sqrt{\frac{9}{4}+m^2L^2};} demanding reality of $\Delta$ yields the usual BF bound \cite{BreFre82} \eq{m^2L^2\geq-\frac{9}{4}.} We begin in Sec. \ref{sec:asmund} by setting conventions and definitions. Next, using standard boundary conditions, we will prove a positive mass theorem for any theory with a weak fake superpotential in Sec. \ref{sec:pmt}, and a PI safety theorem for the same theories in Sec. \ref{sec:thm}, before in Sec. \ref{sec:profiles} explaining when exactly the PI can be violated. Next, in Sec. \ref{sec:designer} we introduce a larger class of boundary conditions that only exist for a subset of theories, the \textit{designer gravity} boundary conditions, where \eq{\phi(r) \sim \frac{\alpha}{r^{\Delta_-}} + \frac{\beta}{r^{\Delta_+}}} with \eq{\Delta_\pm = \frac{3}{2}\pm\sqrt{\frac{9}{4}+m^2L^2},\label{eq:deltapm}} and explain to what extent our positive mass and PI-safety theorems can be extended to this more general context.  In Sec. \ref{sec:cosh} we demonstrate how all of these ideas work out in an explicit example from \cite{CveDuf99}. Finally, note that throughout this section and the remainder of the paper, we assume spherical symmetry.

\subsection{Initial Data and the PI Ratio}
\label{sec:asmund}
We begin by laying out a framework for diagnosing spherical PI violations in general multifield potentials, and along the way  establish some basic definitions that we  use throughout the paper. For one-field models, such a framework was presented in \cite{Fol22,Fol25}; we will employ a slight generalization of that approach to theories with multiple fields. 

Consider as usual AdS gravity coupled to scalars\footnote{We will only leave the index on the fields $\phi^I$ explicit when the particular indices are relevant, i.e. when they are being contracted.} $\phi^I$ with a field space metric $G_{IJ}$ and a potential $V$. As reviewed in Sec. \ref{sec:PIReview}, the PI concerns initial data sets; we will restrict to spherically symmetric initial data throughout, and here will restrict to time-symmetric initial data for simplicity before relaxing this condition in Sec. \ref{sec:thm}. To specify the initial data, we need to choose a field profile $\phi(r)$ for the radial evolution of our fields and a minimar surface $\sigma$ on some time slice $\Sigma$; because of spherical symmetry, $\sigma$ takes the form of a sphere of some radius, which we call $r_0$. 

Define \eq{U(\phi) = V(\phi)-V_0,} where $V_0 = -3/L^2$ is the cosmological constant, and the field-space kinetic term \eq{\calK(r) = G_{IJ}[\phi(r)] \frac{d\phi^I}{dr}\frac{d\phi^J}{dr},\label{eq:calKdef}} and write the induced 3-metric on $\Sigma$ as \begin{subequations}
    \eq{ds_\Sigma^2  &= \frac{dr^2}{F(r)} + r^2d\Omega_2^2 \\ F(r) &= 1 + \frac{r^2}{L^2} - \frac{\mu(r)}{r}.}
    \label{eq:SigmaMetric}
\end{subequations}
The Hamiltonian constraint for this configuration reads \eq{\mu'(r) = r^2\left\{U\left[\phi(r)\right] + \frac{1}{2}\calK(r)F(r)\right\},} or, equivalently, \eq{\mu'(r) + \frac{r}{2}\calK(r)\mu(r) = r^2\left[U(r) + \frac{1}{2}\left(1+\frac{r^2}{L^2}\right)\calK(r)\right]\label{eq:mu'}.} 

Following \cite{Fol22}, we define \eq{H(r) = \frac{1}{2}\int_{r_0}^r d\rho~ \rho \calK(\rho),\label{eq:H}} so that \eq{\mu(r) = e^{-H(r)} \left\{\mu\left(r_0\right) + \int_{r_0}^r d\rho~ e^{H(\rho)}\rho^2\left\{U[\phi(\rho)] + \frac{1}{2}\calK(\rho)\left(1+\frac{\rho^2}{L^2}\right)\right\}\right\}\label{eq:mu}} solves Eq. \ref{eq:mu'} subject to the boundary condition \eq{\mu\left(r_0\right) = r_0 + \frac{r_0^3}{L^2}} which ensures that $\sigma$ is minimal. Valid field profiles have $\mu(r)$ such that $F(r)>0$ for all $r>r_0$ so that $\sigma$ is the outermost minimal sphere. 

Given the metric on $\Sigma$ in Eq. \ref{eq:SigmaMetric} with $\mu(r)$ as in Eq. \ref{eq:mu}, for standard boundary conditions the mass parameter of $\Sigma$ is simply given by \eq{M = \lim_{r\to\infty}\mu(r).} Define $r_{SAdS}$ to be the solution of \eq{r_{SAdS} + \frac{r_{SAdS}^3}{L^2} = M,\label{eq:rSAdS}} i.e. the radius of the Schwarzschild-AdS black hole of mass $M$. The PI ratio $\calR$ defined in Eq. \ref{eq:calR} is then given by \eq{\calR = \frac{\operatorname{Area}[\sigma]}{\operatorname{Area}_{SAdS}[M]} = \left(\frac{r_0}{r_{SAdS}}\right)^2.} Thus, given an explicit field profile $\phi(r)$ on $\Sigma$, to compute $\calR$ and diagnose PI violation, we need only compute the single one-dimensional numerical integral in Eq. \ref{eq:mu}. If $\calR>1,$ the initial data violate the PI; otherwise, they are PI-safe.

\subsection{Weak Fake Superpotentials Imply Positive Mass}
\label{sec:pmt}
The Witten-Nester proof of the positive mass theorem \cite{Wit81,Nes81}, generalized to AdS supergravity in \cite{GibHul83}, proceeds in a supersymmetric theory by explicitly manipulating the fermions. It was pointed out in \cite{Bou84,Tow84} that actual supersymmetry is unnecessary for this argument, and that a global fake superpotential as in Eq. \ref{eq:fakeDef} is sufficient; see e.g. \cite{FreNun04} for a more modern perspective. Finally, it was observed no later than (and possibly before) \cite{EldJoy14} that the same argument applies equally well to weak fake superpotentials; we now briefly review this well-known argument.

The point of the Boucher-Townsend proof of PMTs from fake superpotentials is that all of the relevant structure obtained from genuine supersymmetry can also be obtained from (weak) fake superpotentials. Consider a scalar potential $V(\phi)$, and work in units where $L=1$ so that the constraint for a weak superpotential is that there exists a function $P(\phi)$ such that \eq{\calE = V - 2G^{IJ}\partial_IP\partial_JP + 3P^2 \geq0,\label{eq:pmtCalE}} in addition to the usual constraints in Eq. \ref{eq:anchor}. In terms of $P$, we define a \textit{fake supercovariant derivative} $\hat{\nabla}$ that acts on spinors $\epsilon$ as \eq{\hat{\nabla}_\mu\eps = \left(\nabla_\mu \eps + \frac{1}{2}P\gamma_\mu\right)\eps,} and pick an epsilon satisfying \eq{\gamma^i\hat{\nabla}_i\eps = 0.\label{eq:killingSpinor}} Next, we define a ``fake" variation \cite{AmsMar06} \eq{\delta\lambda_{fake}^I = \frac{1}{\sqrt{2}}\left(\gamma^\mu\partial_\mu \phi^I - 2G^{IJ}\partial_JP\right)\eps,} along with a positive semidefinite norm \eq{\left|\delta\lambda_{fake}\right|^2 = G_{IJ}\left(\delta\lambda_{fake}^I\right)^\dagger\delta\lambda_{fake}^J \geq 0.} This is exactly the structure of a genuinely supersymmetric theory,   for an appropriately chosen $\hat{\nabla}$ and  dilatini $\lambda$.

With these ingredients in hand, \cite{Bou84,Tow84,EldJoy14} show that the so-called Nester energy can be rewritten as  \eq{E_N = \int_\Sigma d\Sigma\ \left(2\left|\hat{\nabla}_i\eps\right|^2 + \left|\delta\lambda_{fake}\right|^2 + \calE|\eps|^2\right).\label{eq:E_N}} Since $\calE\geq0,$ all three of these terms are positive-semidefinite, and we once again find \eq{E_N\geq0.} With standard boundary conditions as in Eq. \ref{eq:standard}, the Nester energy $E_N$ is simply the mass $M$ of $\Sigma$, so we recover the positive mass theorem \eq{M\geq0} for theories with a globally defined weak fake superpotential. For designer boundary conditions more care will need to be taken; we return to PMTs for designer gravity in Sec. \ref{sec:designer}.  Finally, we note that a global weak fake superpotential is not necessary to have a local version of a PMT: if $P$ satisfies Eq. \ref{eq:weakDef} on some region $R$ of field space, then any initial data whose field profile $\phi(r)$ is contained entirely within $R$ will obey the PMT, regardless of how $P$ behaves outside of $R$. While we will only consider spherically symmetric initial data in this work, the proof of the positive mass theorem does not require spherical symmetry; theories with a weak fake superpotential obey a positive mass theorem even for non-spherically symmetric initial data.

\subsection{Weak Fake Superpotentials Imply PI Safety}
\label{sec:thm}
We now prove the theorem that will be our main technical tool throughout the paper: with standard boundary conditions, theories with a weak fake superpotential do not admit PI-violating initial data. We begin by stating the theorem precisely:

\begin{theorem}[Weak Fake Superpotentials Imply PI Safety]
    Consider four-dimensional AdS gravity coupled to scalars $\phi^I$ exploring a potential $V(\phi)$ on a field space with positive-definite metric $G_{IJ}$, with a vacuum at $\phi_0$ whose masses sit strictly above the BF bound, and consider spherically symmetric initial data $\phi(r)$ on a time slice $\Sigma$ containing a future marginally trapped surface $\sigma$ whose exterior is outward untrapped, which we take to be a sphere of radius $r_0$ and extending to the conformal boundary. Assume that $V(\phi)$ is such that a weak fake superpotential, i.e. a function $P$ satisfying Eq. \ref{eq:anchor} and \ref{eq:weakDef}, exists on an open neighborhood $B$ of the field space image of $\phi(r)$ and moreover is $C^1$ on $B$ and $C^2$ at $\phi_0$. Then, additionally assuming that all scalars $\phi^I$ obey the standard boundary conditions in \ref{eq:standard}, we have that \eq{\calR\leq1,} i.e. the spherically symmetric AdS Penrose inequality is satisfied for all choices of initial data with field profiles contained in $B$.  
\end{theorem}

\begin{proof}
Define $\calK$ as in Eq. \ref{eq:calKdef}, define $F$ as in \ref{eq:SigmaMetric} for time-symmetric initial data or as $F=(\nabla r)^2$ more generally, and for convenience define \eq{\mu_0 = \mu(r_0) = r_0 + \frac{r_0^3}{L^2}} and \eq{U = V-V_0.} We assume that we have a weak fake superpotential so that we have \eq{U = \frac{3}{L^2} + 2|\nabla P|^2 - 3 P^2 + \calE,\label{eq:up}} where $\calE$ is defined in Eq. \ref{eq:calEdef}. In Eq. \ref{eq:mu'} we gave the Lorentzian Hamiltonian constraint for spherically symmetric and time-symmetric initial data; weakening the restriction of time symmetry generalizes this constraint to \eq{\mu' = r^2\left(U + \frac{1}{2}F\calK + \Xi\right),\label{eq:mu'xi}} where $\Xi\geq0$; $\Xi=0$ for time symmetric data. Define $n^\mu$ to be the future-directed unit timelike vector normal to the time slice $\Sigma$, and $s^\mu$ to be the outward-pointing spacelike tangent to $\Sigma$ that is orthogonal to $\sigma$; we have \eq{n^\mu n_\mu = -1, s^\mu s_\mu = 1, n^\mu s_\mu = 0.} For later use, we define \eq{w = \frac{n^\mu\partial_\mu r}{s^\mu \partial_ \mu r}.} We also define \eq{\pi^I &= n^\mu \partial_\mu\phi^I \\ x^I &= s^\mu\partial_\mu \phi^I.} One can show that \eq{\Xi = \frac{1}{2}G_{IJ} \left(\pi^I - wx^I\right)\left(\pi^J - wx^J\right) \geq 0.\label{eq:Xi}} In terms of \begin{subequations}
\eq{
    \delta &= \mu-\mu_0 \\
    F_0 &= 1+\frac{r^2}{L^2} - \frac{\mu_0}{r} \\
    H' &= \frac{1}{2}r\calK\label{eq:H'},
}
\end{subequations}
Eq. \ref{eq:mu'xi} becomes \eq{\delta' + H'\delta = r^2\left(U + \frac{1}{2}F_0\calK + \Xi\right)} which implies that \eq{\left(e^H\delta\right)' = e^Hr^2\left(U + \frac{1}{2}F_0\calK + \Xi\right).\label{eq:ehd'}}

So far we have simply reshuffled the constraint equation around, essentially for algebraic convenience; now we will actually prove the theorem. An equivalent formulation of the PI is \eq{M = \lim_{r\to\infty}\mu(r) \geq \mu_0.\label{eq:muineq}} So, to actually get at the PI, we need a way to compare the asymptotic behavior of $\mu(r)$ to its behavior at $\sigma$. One way of doing that, roughly speaking, is to construct a function $\calC(r)$ that interpolates between $\mu_0$ and $M$; if such a function has positive-semidefinite derivative, then we must satisfy Eq. \ref{eq:muineq}, and we have proved the PI. Such a function is given by \eq{\calC(r) = e^{H(r)}\left\{\mu(r) - \mu_0 + a(r)\left[P(r)^2 - \frac{1}{L^2}\right]\right\},} where \eq{a(r) = r^3-r_0^3} and we have defined $P(r) = P[\phi(r)]$ for brevity.

To prove the PI, we need to show that $\calC'\geq0$ everywhere. Differentiating, plugging in Eq. \ref{eq:ehd'} and \ref{eq:up}, and then using the identities \begin{subequations}
    \eq{rF_0 &= r - r_0 + \frac{a}{L^2} \\ \frac{dP}{dr} &= G_{IJ}\partial^IP\frac{d\phi^J}{dr}}
\end{subequations}
allows us to write \eq{e^{-H}\calC' = 2r^2|\nabla P|^2 + 2aPG_{IJ}\partial^IP\frac{d\phi^J}{dr} + H'\left(r-r_0 + aP^2\right) + r^2\left(\calE + \Xi\right).} Finally, we complete the square to find the main result, \eq{e^{-H}\calC' = 2r^2\left|\nabla P + \frac{aP}{2r^2}\phi'\right|^2 + H' \left(r-r_0 + \frac{ar_0^3}{r^3}P^2\right) + r^2\left(\calE + \Xi\right).\label{eq:calC'}} The point is that, for $r\geq r_0$, each of these is nonnegative; the first term is nonnegative because $G_{IJ}$ is positive-semidefinite, $\calE\geq0$ by hypothesis, $\Xi\geq0$ by Eq. \ref{eq:Xi}, $H' \geq 0$ by Eq. \ref{eq:H'}, and $a\geq0$ by definition. Thus, $\calC(r)$ is a nondecreasing function between $r_0$ and $r=\infty$, and we have \eq{\calC(r_0) \leq \lim_{r\to\infty}\calC(r).} 

What remains is to compare $\calC(r_0)$ to $\lim_{r\to\infty}\calC(r)$. By construction, we have \eq{\calC(r_0) = 0.} By Eq. \ref{eq:anchor} and the assumption that $P$ is $C^2$ at $\phi_0$, we have that \eq{P^2 - \frac{1}{L^2} \sim \calO[(\phi-\phi_0)^2]} as $r\to\infty$. Moreover, we assumed that the vacuum $\phi_0$ is strictly above the BF bound, so we have $\Delta>3/2$, and we work with standard boundary conditions, so that \eq{a\left(P^2-\frac{1}{L^2}\right) \sim \calO\left(r^{3-2\Delta}\right) \to 0} as $r\to\infty$. This allows one to show that $\lim_{r\to\infty}H$ is finite and nonnegative; call its value $H_\infty$. We therefore have that \eq{\lim_{r\to\infty}\calC(r) = e^{H_\infty}\left(M-\mu_0\right).} We therefore have, at long last, our desired result: \eq{0 = \calC(r_0) \leq \lim_{r\to\infty}\calC(r) = e^{H_\infty}(M-\mu_0).} Since $H_\infty\geq0$, this implies \eq{M\geq \mu_0,} or equivalently \eq{\calR\leq 1,} and we have proven the PI. 
\end{proof}

In Sec \ref{sec:PIReview}, we took our initial data to contain a spherical minimar surface $\sigma$; as written, the proof above does not apply to generic minimar $\sigma$. This is essentially because the key assumption of the proof is that $\sigma$ has an untrapped exterior, i.e. that we have $F(r)>0$ for all $r>r_0$, but not every minimar $\sigma$ necessarily satisfies this criterion. Nevertheless, the theorem can be extended to the minimar context, essentially because by definition $\sigma$ has smaller area than any surface enclosing it:

\begin{corollary}
Let all of the above hypotheses hold, except now consider the case where $\sigma$ is an arbitrary spherically symmetric minimar surface (instead of assuming that it has an untrapped exterior) such that the time slice on which $\sigma$ has smaller area than any other surface circumscribing it is $\Sigma$ itself. Then the Penrose inequality $\calR\leq1$ holds.
\end{corollary}

\begin{proof}
 If $\sigma$ has an untrapped exterior on $\Sigma$, then the preceding theorem applies, and we once again have $\calR\leq1$. Thus, we need only consider the case where $F$ has additional zeroes outside of $\sigma$ on $\Sigma$. Let $\sigma'$ be the outermost surface on $\Sigma$ where $F$ vanishes. Because the initial data are spherically symmetric, the \textit{outermost} zero must occur everywhere on a  spherically symmetric surface as well, which must therefore circumscribe $\sigma$. It therefore has an untrapped exterior on $\Sigma$, and the above theorem implies the PI for $\sigma'$, i.e. $\operatorname{Area}[\sigma'] \leq \operatorname{Area}[\text{Schwarzschild-AdS}].$ By spherical symmetry, $\sigma'$ encircles $\sigma$. The minimarity of $\sigma$ then implies that the area of $\sigma$ is smaller than that of $\sigma'$, and thus we have that \eq{\operatorname{Area}[\sigma] \leq \operatorname{Area}[\sigma'] \leq \operatorname{Area}[\text{Schwarzschild-AdS}],} establishing the PI for $\sigma$.  
\end{proof}

This is one of our main results and is a key tool that we will use in all calculations for the remainder of the paper; it allows us to rule out PI violations on entire regions of field space without actually constructing initial data. Importantly, like the PMT it is local in field space: if $P$ furnishes a weak fake superpotential on some region $R$, then initial data with $\phi(r)$ lying entirely within $R$ cannot possibly violate the PI. Thus, one very effective way to rule out PI violations is to start with a weak fake superpotential defined near the vacuum, such as the universal one in Eq. \ref{eq:Puniversal}, and try numerically extending it as far as possible; this is a key tool we use throughout, especially in the large-scale numerical searches in Sec. \ref{sec:real}. Of course, the construction of a global weak fake superpotential instantly guarantees that no PI violations can occur in that theory, at least with standard boundary conditions.

\subsection{Diagnosing PI Violations}
\label{sec:profiles}

So far we have given a sufficient condition for avoiding PI violations. We have not yet described how to find them in situations where the theorem in the previous subsection does not apply. To do so, we must explicitly specify the field profiles $\phi(r)$ on initial data slices. For single field models, \cite{Fol22,Fol25} used the power-law field profile 
\eq{\phi(r) = \left(\frac{\left|c_1\right|}{r}\right)^\Delta \left[\operatorname{sign}(c_1) + \operatorname{sign}(c_2)\left(\frac{\left|c_2\right|}{r}\right)^2 + \operatorname{sign}(c_4)\left(\frac{\left|c_4\right|}{r}\right)^4\right] \label{eq:asmundRationalPhiProfile}}
and the logarithmic field profile
\eq{\phi(r) = \left\{\begin{array}{cc}
    0 & r\geq R_0~~  \\
    v \log\frac{r}{R_0}~~ & r_0 \leq r \leq R_0 
\end{array}\right.,
\label{eq:asmundLog} }
where $c_i,v,\text{ and }R_0$ are constants; we will make use of multi-field generalizations of these throughout.  

However, we will frequently find it more useful to work with a more exotic type of field configuration. To have PI violations at all, there must be a region with $V(\phi)<V(\phi_0)$ somewhere in field space; if $\phi_0$ is the global minimum of the potential then we have already proven it to be PI-safe and there is no need to scan over field configurations in the first place. We can therefore look for Coleman-de Luccia (CdL)-like solutions interpolating between the false vacuum at $\phi_0$ and these lower-energy regions. CdL instantons \cite{ColDel80} themselves have exactly zero energy, as measured from the false vacuum, and do not have a trapped surface, so the PI cannot sensibly be applied to them. However, given any CdL instanton, one can always build PI-violating initial data by deforming the data to include a trapped surface of sufficiently small $r_0$.

To see this, let $\mu_{CdL}$, $\phi_{CdL}(r)$ be the initial data giving rise to a CdL instanton, and write the Euclidean metric for a CdL instanton abstractly as \eq{ds^2_E = d\zeta^2 + r^2(\zeta)d\Omega_3^2,} where \eq{\left(\frac{dr}{d\zeta}\right)^2 = F(r).} The Euclidean Hamiltonian constraint \eq{\left(\frac{dr}{d\zeta}\right)^2 = 1 + \frac{r^2}{3}\left[\frac{1}{2}G_{IJ}\frac{d\phi^I}{d\zeta}\frac{d\phi^J}{d\zeta} -V(\phi)\right]} implies  that \eq{\mu_{CdL}(r) &=  \frac{r^3}{3} \left[V(\phi)-V_0 - \frac{1}{2}\left(\frac{d\phi}{d\zeta}\right)^2\right] \\ &= \frac{r^3}{3} \left[V(\phi)-V_0 - \frac{1}{2}F(r)G_{IJ}\frac{d\phi^I}{dr}\frac{d\phi^J}{dr}\right].} A straightforward but tedious calculation involving field space Christoffel symbols shows that this form of $\mu_{CdL}$ satisfies the Lorentzian Hamiltonian constraint in Eq. \ref{eq:mu'}, and from it we see that $\lim_{r\to\infty}\mu_{CdL}=0$, so that the CdL instanton is a zero-energy solution, as expected. 

From this we can construct PI-violating initial data. \footnote{Readers who are less interested in the technical construction itself may want to skip ahead to the discussion following Eq. \ref{eq:382}.} Pick some radius $r_0$ for the trapped surface, and consider a field profile $\phi_{new}(r)$ defined to equal $\phi_{CdL}(r)$ for $r>r_0$, and define a corresponding mass function $\mu_{new}(r)$ to be given by the unique solution to Eq. \ref{eq:mu} with the usual boundary condition \eq{\mu_{new}(r_0) = r_0 + \frac{r_0^3}{L^2};\label{eq:mu0}} we specify initial data on the remainder of $\Sigma$ by reflecting across $\sigma$. Note that, although our theorem applies to more general choices, we work with time-symmetric initial data throughout. Because of Eq. \ref{eq:mu0}, even though both $\mu_{CdL}$ and $\mu_{new}$ obey Eq. \ref{eq:mu'}, we do not have $\mu_{CdL}(r) = \mu_{new}(r)$, and in particular \eq{\lim_{r\to\infty}\mu_{new} \neq \lim_{r\to\infty}\mu_{CdL} = 0.} Instead, we have that \eq{\lim_{r\to\infty}\left(\mu_{new}-\mu_{CdL}\right) = \left[\mu_{new}\left(r_0\right)-\mu_{CdL}\left(r_0\right)\right]\Gamma(r_0),} where \eq{\Gamma(r) = \exp\left[-\frac{1}{2}\int_r^\infty d\rho~ \rho G_{IJ}\frac{d\phi^I}{d\rho}\frac{d\phi^J}{d\rho}\right].\label{eq:gammaDef}} From the definitions of $\mu_{new}$ and $F_{CdL}$, we have that \eq{\mu_{new}(r_0) - \mu_{CdL}(r_0) = r_0 + \frac{r_0^3}{L^2} - \mu_{CdL}(r_0) = r_0 F_{CdL}(r_0),} and, since $\lim_{r\to\infty}\mu_{CdL}=0$, we therefore have that \eq{\lim_{r\to\infty}\mu_{new}(r) = r_0\Gamma(r_0)F_{CdL}(r_0).}

As mentioned above, we find PI-violating initial data for sufficiently small $r_0$, so now let us take $r_0\to0$; in this limit we have $F_{CdL}\sim 1 + \calO(r_0^2)$, and therefore  \eq{\lim_{r\to\infty}\mu_{new}(r) \sim r_0\Gamma(0) + \calO\left(r_0^3\right).} Since the field space metric $G$ is positive definite, we have that $|\phi'|_G^2>0$ for at least some value of $r$, so $\Gamma(0)<1$; for normalizable modes, i.e. for scalars above the BF bound with $\phi\sim r^{-\Delta}$ at large $r$, the integral in $\Gamma(0)$ converges, so that $\Gamma(0)>0$. We thus have \eq{0<\Gamma(0)<1,} so that \eq{\lim_{r\to\infty}\mu_{new}(r) \sim \Gamma(0)r_0 < r_0.} Eq. \ref{eq:rSAdS} then implies that \eq{r_{SAdS} \sim \Gamma(0)r_0,} and hence the PI ratio is given by \eq{\calR = \left(\frac{r_0}{r_{SAdS}}\right)^2 \sim \frac{1}{\Gamma(0)^2} > 1.\label{eq:382}} We therefore have the promised PI violation for sufficiently small $r_0$.

\begin{figure}
    \centering
    \includegraphics[width=.85\linewidth]{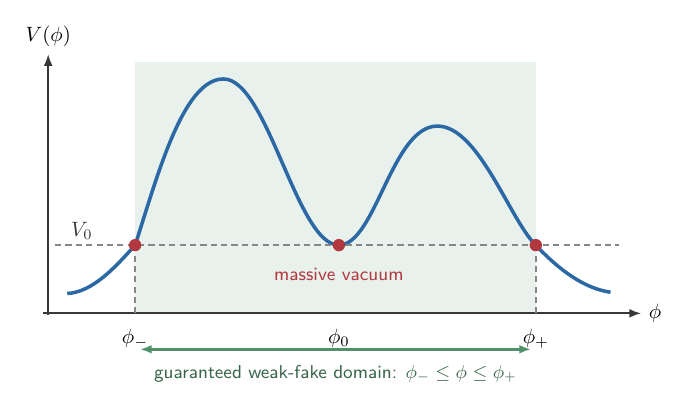}
    \caption{Schematic drawing of the barrier surrounding a false vacuum that is a strict local minimum. Here, the barrier extends from $\phi_0$ to the two points $\phi_\pm$; in higher dimensions, these two points are in general replaced by a curve $\phi_*$. Within the barrier, $P=1/L$ furnishes a weak fake superpotential, exactly as with true vacua, and PI violations are forbidden.}
    \label{fig:barrier}
\end{figure}

 Initial data whose field profiles live inside the barrier of some vacuum can never violate the PI. Consider for simplicity a vacuum $\phi_0$ that is a strict local minimum of its potential; a version of this argument applies more generally, but we restrict to this case for pedagogical clarity. Further assume that $\phi_0$ is a false vacuum, so that the DEC theorem of \cite{Fol22} does not automatically imply its PI-safety. Because $\phi_0$ is nontachyonic, $V$ rises along all directions in field space as you move away from $\phi_0$, but because $\phi_0$ is a false vacuum $V$ must at some point turn back around and descend below $V_0$. Thus, as is familiar, $\phi_0$ is surrounded by a positive-energy barrier (or, more precisely, a barrier of positive \textit{relative energy} $V-V_0$); this is shown schematically in Figure \ref{fig:barrier}. Importantly, between the vacuum $\phi_0$ and the locus $\phi_*$ where $V(\phi_*) = V_0$ on the far side of the barrier, a weak fake superpotential \textit{must} exist, and PI violations are forbidden. As with global minima, within the barrier $P = 1/L$ always furnishes a weak fake superpotential, regardless of the details of $V$ itself. Thus, no initial data whose field excursions lie solely within the barrier of any given vacuum can ever give rise to PI violations.  

\begin{figure}
    \centering
    \begin{tikzpicture}[
  x=2.2cm,y=2.2cm,
  line cap=round,line join=round,
  boundary/.style={draw=black,
                   line width=.65pt},
  centre/.style={draw=black,line width=.8pt},
  wall/.style={draw=black,line width=.85pt},
  null/.style={draw=black,line width=.5pt,dash pattern=on 3pt off 2.6pt},
  initial/.style={draw=black,line width=.65pt},
  singularity/.style={draw=black,line width=.95pt,decorate,
    decoration={zigzag,amplitude=.8pt,segment length=4.6pt}},
  every node/.style={text=black,inner sep=0pt},
]
  \pgfmathsetmacro{\edge}{pi/2}
  \pgfmathsetmacro{\crunchheight}{rad(acos(-.6))}

  \draw[centre] (0,-\crunchheight)--(0,\crunchheight);
  \draw[boundary] (\edge,-\edge)--(\edge,\edge);

  \draw[null] (0,0)--(\edge,\edge);
  \draw[null] (0,0)--(\edge,-\edge);

  \draw[wall,samples=141,domain=-pi/2:pi/2,variable=\t]
    plot ({rad(acos(.7*cos(deg(\t))))},{\t});

  \draw[initial] (0,0)--(\edge,0);
  \node[anchor=south,font=\large] at (1.2,.075) {$\Sigma$};

  \draw[singularity,samples=101,domain=0:pi/2,variable=\x]
    plot ({\x},{ rad(acos(-.6*cos(deg(\x))))});
  \draw[singularity,samples=101,domain=0:pi/2,variable=\x]
    plot ({\x},{-rad(acos(-.6*cos(deg(\x))))});

  \node[anchor=west,font=\large] at (\edge+.1,0.5) {$\mathscr I$};
\end{tikzpicture}
    \caption{Penrose diagram of the big crunch geometries we encounter throughout this paper. Numerical evidence shows that the PI-violating initial data we encounter here evolve to a big crunch like the one shown here, rather than to a naked singularity.}
    \label{fig:CrunchPenroseDiagram}
\end{figure}
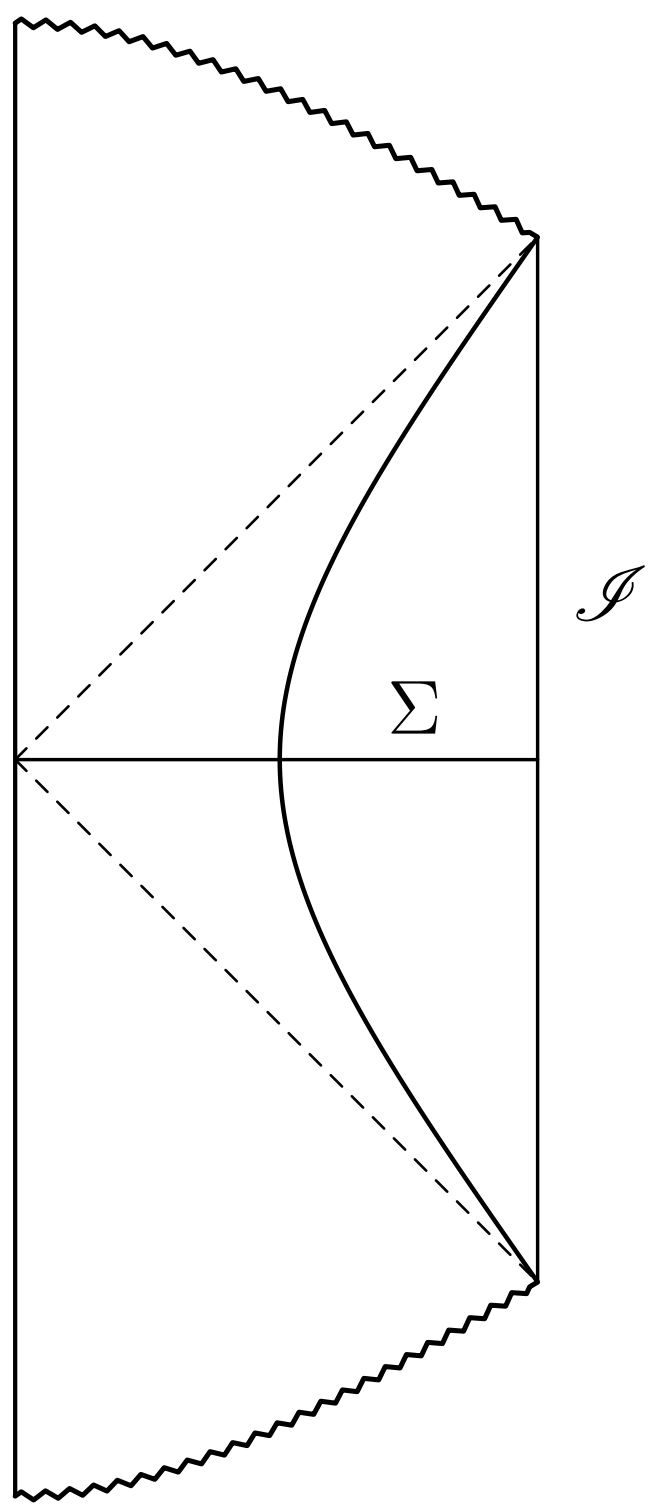

This argument, along with the fact that PI-violating initial data can be built from any CdL instanton, suggests that these violations might be due to vacuum decay. We have extensive numerical evidence supporting this intuition. Given PI-violating initial data, one can numerically evolve it forward to determine the ultimate fate of its universe. In all cases of PI violation encountered in this paper where we were able to do so, the numerical evolution strongly suggests that the universe evolves into a big crunch, rather than a naked singularity. In particular, we find that the boundary collapses after finite time, as in the big crunch Penrose diagram in Figure \ref{fig:CrunchPenroseDiagram}. While we do not have a theorem that naked singularity formation is impossible, in light of the above argument about PI-safety inside each vacuum's barrier we find this extremely suggestive that no observer in a false vacuum can detect a PI violation before their universe collapses. 

\subsection{Attack of the Tachyons: Fake Superpotentials in Designer Gravity}
\label{sec:designer}
So far we have worked with standard boundary conditions, also sometimes called fast falloff boundary conditions or more simply Dirichlet conditions, where $\phi(r)\sim r^{-\Delta}$ for $r\to\infty$. These are standard because they define normalizable modes for fields of any BF-allowed mass. However, some BF-allowed tachyons admit a second normalizable mode. Define $\Delta_\pm$ as in Eq. \ref{eq:deltapm}, and consider the slow falloff mode $\phi\sim r^{-\Delta_-}$ for $r\to\infty$. This defines a normalizable mode iff \eq{-\frac{9}{4}<m^2L^2 < -\frac{5}{4}.\label{eq:window}} Inside this window the most general allowed boundary conditions are  therefore 
\eq{\phi(r) = \frac{\alpha}{r^{\Delta_-}} + \frac{\beta}{r^{\Delta_+}} + \cdots. }
There are three canonical choices for $\alpha,\beta$. First, if $\alpha=0$, we recover standard boundary conditions. Next, setting $\beta=0$ so that only the slow mode is turned on yields \textit{alternative}, or Neumann, boundary conditions. Finally, the so-called \textit{designer boundary conditions}, introduced in \cite{HerHor04b} and studied in e.g. \cite{Fol25,AmsMar06,AmsHer07,HenMar06} occur when
\eq{\beta = w'(\alpha)\label{eq:designerw}}
for some $w$, which we always take to satisfy $w(0)=0$; these are so named because they allow one to ``design" the effective potential of the holographic boundary theory. 

In this paper, we will encounter only two theories in the designer gravity mass window, and both have $m^2L^2=-2$, and so have $\Delta_+ = 2, \Delta_-=1$. While all of the formulas in the remainder of the section can be derived for any $m^2$ in the designer window, the general formulas are not especially instructive, so we specialize to $m^2L^2=-2$ for the remainder of the section, so that designer boundary conditions become \eq{\phi(r) = \frac{\alpha}{r} + \frac{\beta}{r^2} + \cdots. \label{eq:designer}} A particularly nice family of designer boundary conditions are those which preserve the AdS symmetries of the standard boundary conditions; for $m^2L^2=-2$, these occur when \cite{FauHor10}
\eq{w(\alpha) = \frac{f}{3}\alpha^3 \implies \beta = f\alpha^2, \label{eq:designerf}}
and can be interpreted holographically in terms of a multitrace deformation of the boundary theory \cite{Wit01b,KlebWit99}. 

We saw above that the existence of a weak fake superpotential in a theory with standard boundary conditions implies a PMT and PI safety. For designer gravity, the story is more complicated; PMTs and PI safety must be proven separately for each choice of $f$ in Eq. \ref{eq:designerf}. As usual, we write \eq{F(r) = 1 + \frac{r^2}{L^2} - \frac{\mu(r)}{r}.} With standard boundary conditions, the physical mass $M$ of the system is simply related to $\mu$ by \eq{M = \lim_{r\to\infty}\mu(r).} In designer gravity, this relationship no longer holds, and we must distinguish between the physical mass $M$ and the coefficient $M_0$ in the asymptotic metric. In the presence of the slow mode, we have \eq{\mu(r) = -\frac{\alpha^2}{2L^2}r + M_0 + \calO(r^{-1})} at large $r$; \cite{AmsMar06} showed that the metric coefficient $M_0$ is related to the physical mass $M$ by \eq{M = M_0 + \frac{1}{L^2}\left[\alpha\beta + w(\alpha)\right] \label{eq:designerE}.} For $\alpha = 0$ we recover $M=M_0,$ as expected. 

We now proceed to the proofs of the PMT and PI safety. Consider a weak fake superpotential admitting a series expansion of the form
\eq{P_s = \frac{1}{L}\left(1+\frac{\phi^2}{4} - \frac{s}{6}|\phi|^3 + \cdots\right), \label{eq:Ps}}
 and write $E_{N,s}$ for the Nester energy in Eq. \ref{eq:E_N} obtained from this weak fake superpotential. With standard boundary conditions, we have \eq{M= E_{N,s},} which since $E_{N,s}\geq0$ implies the usual positive mass theorem. In designer gravity, by contrast, we have \eq{M= E_{N,s} + \frac{1}{L^2}\left[w(\alpha) + \frac{s}{3}|\alpha|^3\right] \geq \frac{1}{L^2}\left[w(\alpha) + \frac{s}{3}|\alpha|^3\right]\label{eq:designerEsgeq}.}  
Similarly, the same argument that gives the PI safety theorem for standard boundary conditions, \eq{M \geq r_0 + \frac{r_0^3}{L^2},} gives in designer gravity \eq{M \geq r_0 + \frac{r_0^3}{L^2} + \frac{1}{L^2}\left[w(\alpha) + \frac{s}{3}|\alpha|^3\right] \label{eq:designerMs}.} 

One additional subtlety arises in the PI for designer gravity: the proof of the PI from cosmic censorship asserts that the area of $\sigma$ should be less than that of the black hole with the largest area at fixed AD mass; the proof in \cite{XiaYan22} that this is the Schwarzschild-AdS black hole does not apply in designer gravity. Thus, with more general boundary conditions, there might in principle exist hairy black holes with larger area than that of SAdS. If such black holes exist, then  there would be some state with $\operatorname{Area}[\text{SAdS}]\leq\operatorname{Area}[\text{Hairy}]$, and the strongest form of the PI would read \eq{\operatorname{Area}[\sigma]\leq\operatorname{Area}[\text{Hairy}]\label{eq:hairy};} initial data with $\calR>1$, i.e. $\operatorname{Area}[\sigma]>\operatorname{Area}[\text{SAdS}]$, may well satisfy the weaker inequality in Eq. \ref{eq:hairy}. Without knowing the area of these highly entropic hairy black holes, or indeed knowing whether or not they exist in the first place, the best we can do is to work with the usual Schwarzschild-AdS PI (SAdS-PI) as a proxy for Eq. \ref{eq:hairy}, subject to the caveat that initial data that violate the SAdS-PI might nonetheless satisfy Eq. \ref{eq:hairy}. We expect that, in the event that these hairy black holes exist, the holographic proof of the PI will reflect the highest-entropy black hole in the ensemble, and thus yield Eq. \ref{eq:hairy} as well.

Thus, we have a PMT and SAdS-PI safety if \cite{FauHor10} \eq{w(\alpha) + \frac{s}{3}|\alpha|^3\geq0 ~ \forall \alpha.} For the AdS-compatible boundary conditions in Eq. \ref{eq:designerf}, this occurs only for \eq{|f| \leq s. \label{eq:fs}} Note that different choices of weak fake superpotential yield different values of $s$, and so imply PMTs and PI safety for different values of $f$. In general, a choice of $f$ being excluded by one choice of weak fake superpotential does not prohibit it from being allowed by a different choice, so the existence of one weak fake  superpotential yielding $s<|f|$ does not mean that choice of $f$ necessarily gives rise to PI violating initial data.

\subsection{A Worked Example}
\label{sec:cosh}
We conclude this section by demonstrating all of the above ideas in a single example. Consider the potential \eq{L^2V(\phi) = -2 - \cosh{\sqrt{2}\phi},\label{eq:cosh}} derived from dimensional reduction of M-theory in \cite{CveDuf99} and studied in e.g. \cite{HerHor04,HerHor05}. This potential has a single vacuum at $\phi=0$ with mass $m^2L^2=-2$, and is unbounded from below. It is straightforward to verify that \eq{P = \frac{1}{L}\cosh\frac{\phi}{\sqrt{2}}\label{eq:coshP}} furnishes a global fake superpotential. We therefore know that, with standard boundary conditions, all choices of initial data are  PI-safe. This was seen to be true numerically for this example in~\cite{Fol22}, and we now understand that those results follow directly from the theorem in Sec.~\ref{sec:thm}.

\begin{figure}
    \centering
    \includegraphics[width=0.6\linewidth]{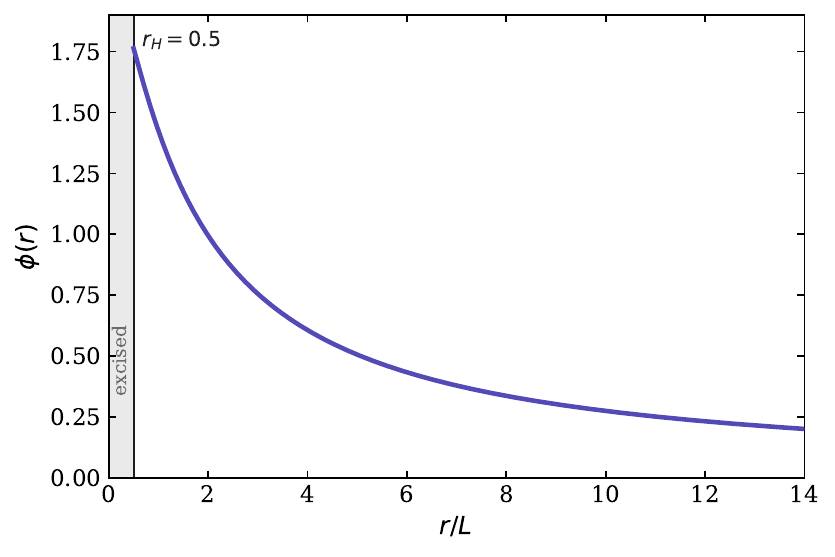}
    \caption{A Penrose-violating field profile for the hyperbolic cosine potential in Eq. \ref{eq:cosh}. This field profile obeys designer gravity boundary conditions, rather than those of standard quantization, so the theory is allowed to admit PI-violating initial data even though it admits a fake superpotential.}
    \label{fig:coshProfile}
\end{figure}

A CdL-like instanton in this theory was constructed in \cite{HerHor04}; in light of the discussion in Sec. \ref{sec:profiles}, we therefore expect that it should admit PI-violating initial data. Indeed, one can numerically construct radial profiles that violate the PI. We have plotted one such field configuration with $r_0 = 0.5L$ in Figure \ref{fig:coshProfile}; this configuration has $\calR = 1.14$, and so violates the SAdS-PI.

One might wonder how this is compatible with the ``weak fake superpotential implies PI safety" theorem earlier in the section. The resolution to this seeming problem lies in the boundary conditions: weak fake superpotentials only guarantee PI safety for field configurations with standard boundary conditions, and theories allowing more general ones can admit PI-violating initial data even in the presence of a fake superpotential.\footnote{Recall that, with designer gravity conditions, $\calR\geq1$ does not necessarily imply violation of the strongest-possible PI, because Schwarzschild-AdS has only been proven to be the largest black hole at fixed mass with standard boundary conditions.} This theory has $m^2L^2=-2$, so it is precisely in the designer gravity window. Indeed, asymptotically the field profile in Figure \ref{fig:coshProfile} approaches \eq{\phi(r) \sim \frac{2.969}{r} - \frac{2.204}{r^2} + \calO(r^{-3}),} so we have \eq{\beta = -\frac{1}{4}\alpha^2,} exactly as in Eq. \ref{eq:designerf} with $f=-1/4$.

\section{Toy Models for KKLT}
\label{sec:kklt}
In this section, we apply the above discussion to search for PI violations in two toy models for KKLT compactifications. We begin with an introductory overview of KKLT, leaving aside details of its implementation in the genuine vacua of \cite{McaMor24,McaNal26WIP}; we will provide further details on the mechanics of KKLT in Sec.~\ref{sec:real}. Readers familiar with KKLT may wish to skip ahead to the novel PI results beginning around Eq. \ref{eq:toy}.

A Calabi-Yau threefold (CY3) $X$ is a compact, Ricci-flat Kahler manifold; see e.g. \cite{Hub92,HorKat03} for reviews. CY3s play a special role in superstring theory, which is usually formulated in ten dimensions: product metrics of the form $M_{10} = \mathbb{R}^{1,3}\times X$ allow one to \textit{compactify} string theory to four dimensions while preserving supersymmetry in the resulting four-dimensional EFT \cite{CandHor85}.   Compactification of type IIB string theory on a CY3 $X$ yields a four-dimensional effective theory of $\mathcal{N}=2$ supergravity, whose matter content depends on the topology of $X$; in particular, in addition to the axiodilaton $\tau$, the matter spectrum includes scalars called complex structure and Kahler moduli whose numbers are determined by $X$. At the $\calN=2$ level, these moduli experience no potential whatsoever, and so the theory has no isolated vacua. 

To induce a potential for the moduli fields, one performs a quotient of the underlying CY3, called an \textit{orientifold}, and introduces fluxes for various ten-dimensional gauge fields. These choices have the effect of breaking $\calN=2$ to $\calN=1$ and inducing a potential for the moduli fields. As reviewed in Eq. \ref{eq:VF}, potentials in $\calN=1$ theories are defined in terms of a superpotential $W$ and a K\"ahler potential $K$. We will give the full details of $W$ and $K$ for KKLT vacua below, in Sec. \ref{sec:real}; for now it suffices to note that the superpotential for the Kahler moduli $T_i$ takes the schematic form \eq{W\supset \sum e^{-2\pi T_i}.\label{eq:Wnp}} A privileged class of solutions to theories with F-term potentials are those which preserve supersymmetry; these solutions occur exactly at points where \eq{D_aW = 0} for all $a$, and by Eq. \ref{eq:VF} have negative-semidefinite potentials.

The goal of KKLT is to build, from these supersymmetric solutions of a supersymmetric theory, nonsupersymmetric solutions of a nonsupersymmetric theory. They proposed breaking supersymmetry by sequestering a supersymmetry-breaking anti-D3 brane at the end of a Klebanov-Strassler throat \cite{KKLT,KleStr00,GidKac01,KacPea01}. This breaks the $\calN=1$ supersymmetry that survives the orientifold to a genuine $\calN=0$ theory and deforms the potential away from just the F-term potential to \eq{V = V_F + V_{up},\label{eq:V}} where $V_{up}$ is determined by the physical volume $\calV$ of the CY3 $X$ as \eq{V_{up} = \frac{Z}{\calV^{4/3}},} where the uplift parameter $Z$ is positive.

To find a vacuum of the nonsupersymmetric theory, one must find a critical point of $V$, with either purely positive eigenvalues (for dS vacua) or all eigenvalues obeying the BF bound (for AdS vacua), but random matrix arguments show that such critical points are exponentially rare \cite{MarMca11} in these potentials. Nevertheless, for sufficiently small $Z$, one can find such a critical point by starting with a genuine supersymmetric minimum of the F-term potential, called an \textit{AdS precursor} in \cite{McaMor24}, and using that as the starting point to numerically find roots in the full potential. This strategy, called \textit{uplifting} the vacuum, was essentially proposed in \cite{KKLT}, and is now the standard procedure for finding KKLT vacua. Importantly, the vacuum energy of the resulting vacuum is determined by the relative size of $V_{up}$ and the vacuum energy of the AdS precursor. At small $Z$, the energy of the uplifted vacuum remains negative, and we obtain a nonsupersymmetric AdS vacuum; as $Z$ grows, eventually the positive energy of the uplift overwhelms the negative energy of the precursor, and we land on a dS vacuum. Constructing dS vacua was the primary goal of KKLT, but the nonsupersymmetric vacua obtained at small $Z$ are perfectly legitimate, and in fact some AdS examples were constructed as a byproduct of the explicit KKLT dS vacua found in \cite{McaMor24,McaNal26WIP}. 

While by now it is standard to compute all of the terms in Eq. \ref{eq:V} totally explicitly, at least for certain types of CY3, when KKLT was written in 2003 actually computing these quantities was far out of reach. KKLT thus constructed a toy model by first imagining that the entire complex structure sector, including the axiodilaton, is fixed supersymmetrically at some heavy scale, so that its dynamics can be essentially ignored; we therefore model it only with a constant contribution $W_0$ to the superpotential. Next, they imagined a CY3 with a single complexified Kahler modulus $\rho = u+i\sigma$, so that there remains only one dynamical field in the model. 

We are thus led to consider the \textit{KKLT toy model}, where the dynamics of $\rho$ are governed by a holomorphic superpotential \eq{W = W_0 + e^{2\pi i \rho},\label{eq:toy}} where $W_0$ is the vev of the complex structure superpotential at its supersymmetric minimum,  and a real Kahler potential \eq{K(\rho,\rhobar) = -3\ln\big[-i\left(\rho-\rhobar\right)\big].} As described above, we begin by constructing supersymmetric vacua of this model, whose F-term potential is given as in Eq. \ref{eq:VF}. Supersymmetric minima occur at purely imaginary $\rho_* = i\sigma_*$, with $\sigma_*$ positive and fixed in terms of $W_0$ by \eq{W_0 = -e^{-2\pi \sigma_*} \left(1+\frac{4}{3}\pi \sigma_*\right).} At the supersymmetric minimum, we have \eq{V_0 = -\frac{2}{3}\pi^2\frac{e^{-4\pi \sigma_*}}{\sigma_*}.}

Next we proceed to uplift. For a single-modulus model, the CY volume is given by \eq{\calV = \left(\operatorname{Im}{\rho}\right)^{3/2},} so the antibrane potential scales as \eq{\calV^{-4/3} = \left[\left(\operatorname{Im}{\rho}\right)^{3/2}\right]^{-4/3} = \left(\operatorname{Im}{\rho}\right)^{-2}.} We therefore consider the nonsupersymmetric potential \footnote{In their model for the antibrane potential, KKLT \cite{KKLT} neglected the volume dependence of the warp factor, and so worked with  the uplift potential $V_{\bar{D3}} \sim 1/(\operatorname{Im}\rho)^3$. In real KKLT compactifications, the effects of warping are significant, so, following KKLMMT \cite{KacKal03b}, we work instead with the warped uplift potential $V_{\overline{D3}} \sim 1/(\operatorname{Im}\rho)^2$. While the precise formulas and numerical results quoted in the following two sections differ between the two potentials, the qualitative conclusions do not: the original KKLT toy model does not admit PI violations with either choice of uplift potential, and the two-instanton model admits both tachyonic and non-tachyonic violations with either choice.} \eq{V = V_F + \frac{Z}{\sigma^2}.\label{eq:toyV}} We once again have vacua at purely imaginary $\rho_* = i\sigma_*$, with
\eq{W_0 = -e^{-2\pi\sigma_*}\left(1+\frac{4}{3}\pi \sigma_*\right) - Z\frac{e^{2\pi\sigma_*}}{\pi\left(1+\pi\sigma_*\right)}.} 
The vev of the potential at these critical points is given by 
\eq{V_0 = -\frac{2}{3}\pi^2\frac{e^{-4\pi \sigma_*}}{\sigma_*} + \frac{\pi Z}{\sigma_*\left(1+\pi\sigma_*\right)}.}
We see that we obtain a nonsupersymmetric AdS vacuum, i.e. one with $V_0<0$, for 
\eq{0< Z< \frac{2\pi}{3}e^{-4\pi\sigma_*}\left(1+\pi\sigma_*\right).\label{eq:KKLTzbound}} 

We therefore have a two-parameter family of nonsupersymmetric AdS vacua, parameterized by $\sigma_*$ (or equivalently $W_0$) and $Z$. It is straightforward to show that any nonsupersymmetric AdS vacuum at fixed $\sigma_*,Z$ is the global minimum of its potential; we therefore conclude that the KKLT toy model is PI-safe without needing to perform any numerical scans.

When explicit supersymmetric AdS vacua were studied in \cite{McaMor24}, it was noted that the potentials of the genuine compactifications along the overall volume direction always admit a second critical point, in the form of a barrier to decompactification. While nonsupersymmetric AdS vacua in the KKLT toy model do have these additional critical points, it is a straightforward exercise to show that, before breaking supersymmetry, potentials in the supersymmetric KKLT toy model only ever admit a single critical point. Thus, the toy model fails to capture the full complexity of real vacua, and the absence of PI violations in the toy model is not sufficient to conclude the same for the potential of a genuine CY3 compactification.

The gap between the toy model and genuine CY3s is essentially due to the sum in Eq. \ref{eq:Wnp}; competition between the hundreds of instantons in the nonperturbative superpotential can yield potentials with very complicated topological features \cite{KleNalWIP}, all of which are invisible to the single-instanton superpotential in Eq. \ref{eq:toy}. To model these effects in a single-field toy model, it was suggested in \cite{McaMor24} that we instead consider a more general model, which we will call the \textit{two-instanton KKLT toy model}, with superpotential given by \eq{W(\rho) = W_0 + e^{2\pi i \rho} - A e^{4\pi i \rho},\label{eq:2term}} where the amplitude $A$ is positive; competition between these two instantons, which contribute to the superpotential with opposite signs, can mimic the features of the full superpotential of a CY3.

Let us begin by analyzing the vacuum structure of this model. Before breaking supersymmetry, we once again find a family of supersymmetric AdS vacua at purely imaginary $\rho_* = i\sigma_*$ with \eq{W_0 = -\frac{1}{3}e^{-4\pi\sigma_*}\left[e^{2\pi\sigma_*}\left(4\pi\sigma_*+3\right) - A \left(8\pi\sigma_*+3\right)\right].\label{eq:2termSusyMinimum}} At the minimum, we have \eq{V_0 = -\frac{2\pi^2}{3} \frac{e^{-8\pi\sigma_*}}{\sigma_*} \left(e^{2\pi\sigma_*} - 2A\right)^2.}

\begin{figure}
    \centering
    \includegraphics[width=.7\linewidth]{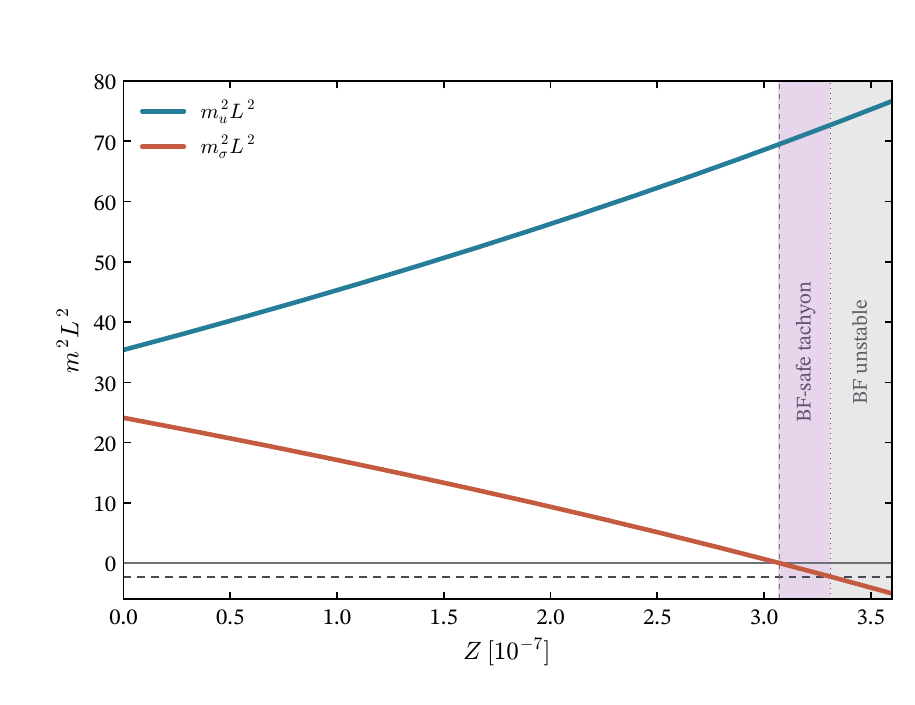}
    \caption{Masses of the canonically normalized modes of the nonsupersymmetric two-term toy model at $A=146$ and $\sigma_*=0.98$ as a function of the remaining parameter $Z$.}
    \label{fig:2termMasses}
\end{figure}

As usual, we now proceed to break supersymmetry, and work once again with Eq. \ref{eq:toyV}. We find nonsupersymmetric critical points at \eq{W_0 = -\frac{1}{3}e^{-4\pi\sigma_*} \big[-A \left(8\pi \sigma_* +3 \right) + e^{2\pi\sigma_*}\left(4\pi\sigma_* + 3\right)\big] - \frac{Ze^{4\pi\sigma_*}}{\pi\left[e^{2\pi\sigma_*}\left(1+\pi\sigma_*\right) -2A\left(1+2\pi\sigma_*\right)\right]}} 
with vacuum energy \eq{V_0 = - \frac{2\pi^2}{3}\frac{e^{-8\pi\sigma_*}}{\sigma_*}\left(e^{2\pi\sigma_*}-2A\right)^2 +\frac{\pi Z\left(e^{2\pi\sigma_*}-4A\right)}{\sigma_*\left[e^{2\pi\sigma_*}\left(1+\pi\sigma_*\right) - 2A\left(1+2\pi\sigma_*\right)\right]}.} One can show analytically that the spectrum of canonically normalized modes is allowed to contain tachyons, as seen in Figure \ref{fig:2termMasses}.

\begin{figure}[h!]
    \centering
    \begin{subfigure}{\linewidth}
        \centering
        \includegraphics[width=\linewidth]{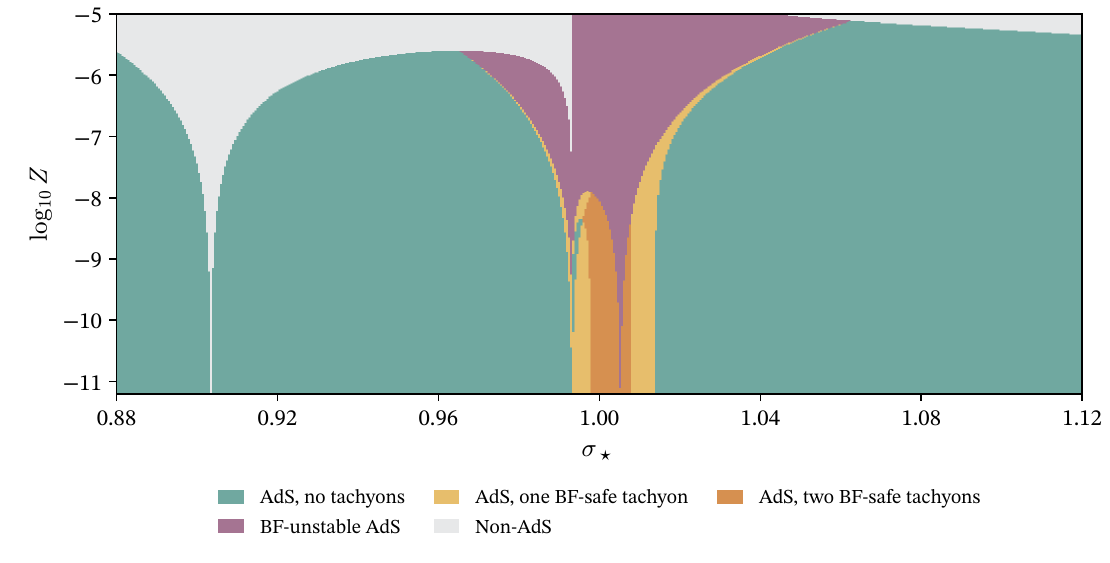}
        \caption{}
        \label{fig:2termPhase}
    \end{subfigure}

    \begin{subfigure}{\linewidth}
        \centering
        \includegraphics[width=\linewidth]{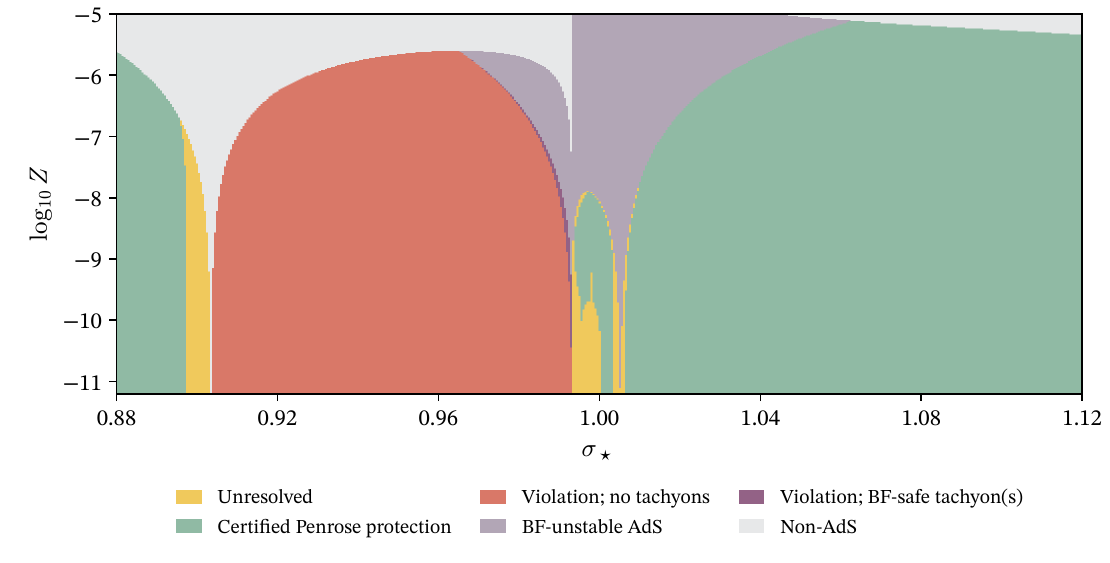}
        \caption{}
        \label{fig:2termPenrose}
    \end{subfigure}
    \caption{Phase diagrams of the two-term KKLT toy model at $A=146$ as a function of the parameters $Z$ and $\sigma_*$. The top diagram describes the vacuum energy and masses at each point in parameter space, and the bottom diagram summarizes the existence of PI violations, again at each point in parameter space. Here, ``unresolved'' points are those for which we can neither rule out nor establish the existence of PI-violating initial data.}
    \label{fig:2termPhaseDiagrams}
\end{figure}

We thus have a three-parameter family of nonsupersymmetric vacua, parameterized by $W_0,Z,A$. Varying these parameters yields a rich phase diagram of vacua, as shown in Figure \ref{fig:2termPhase}. For our purposes, it will suffice to restrict to the slice $A = 146$ and vary only $\sigma_*$ and $Z$. In particular, along this slice we encounter AdS vacua both with and without BF-allowed tachyons, as well as dS vacua which will not feature in this work.

We now proceed to study PI violations in this model. As shown in Figure \ref{fig:2termPenrose}, along the $A=146$ slice, we encounter codimension-zero regions of parameter space where we encounter each of the following behaviors:
\begin{itemize}
    \item PI-safe nonsupersymmetric AdS vacua with tachyons.
    \item PI-safe nonsupersymmetric AdS vacua without tachyons. 
    \item PI-violating nonsupersymmetric AdS vacua with tachyons.
    \item PI-violating nonsupersymmetric AdS vacua without tachyons.
\end{itemize}

We expect each of the regions shown in Figure \ref{fig:2termPenrose} to continue off of the $A=146$ slice, thus giving us full codimension-zero regions of parameter space where each of these behaviors occurs. Whereas in the genuine compactifications presented in the next section we will be unable to scan over the full, very high-dimensional moduli space, here we can frequently construct a weak fake superpotential everywhere in the moduli space, and therefore genuinely prove PI-safety and a PMT, at least with standard boundary conditions.

We will now analyze several points in parameter space that illustrate the wide variety of behaviors possible in this model. We begin with the slice $A=146$, $\sigma_* = 0.98$, leaving only the scale of supersymmetry breaking $Z$ as a free parameter; along this slice, we have a supersymmetric AdS precursor at $Z=0$ and nonsupersymmetric BF-allowed AdS vacua until  $Z = Z_{\text{BF}} \sim 3.31\times10^{-7}$, at which point the lightest canonically normalized mode dips below the BF bound, as shown in Figure \ref{fig:2termMasses}. At $Z=0$, we have a fake superpotential $P$ inherited from the genuine superpotential in Eq. \ref{eq:2term}, but PI-violating initial data can be found for any allowed point in parameter space with $Z>0$; this demonstrates that we cannot have a fake superpotential, or even a weak fake superpotential, after breaking supersymmetry.

For instance, consider the point $Z = 3.2\times10^{-7}$ along this slice; as shown in Figure \ref{fig:2termMasses}, this point is tachyonic, with canonical saxion mass \eq{m_\sigma^2L^2 = -1.19\ldots.} The initial data specified by Eq. \ref{eq:asmundRationalPhiProfile} with \eq{c_1 = 2.9, ~ ~ ~ c_2 = -3.5, ~ ~ ~ c_4 = 2.5,\label{eq:2termViolationParams}} with $r_0 = 2.5L$ as in \cite{Fol22}, describe an apparent horizon $\sigma$ with \eq{\frac{A[\sigma]}{A_{\text{SAdS}}[M]} = 4.613\ldots,} and so we have a well-controlled PI-violating solution. 

Similarly, consider the point $Z = 2\times10^{-7}$ along the same $A = 146$, $\sigma_*=0.98$ slice. Working once again with Eq. \ref{eq:asmundRationalPhiProfile} with parameters as in Eq. \ref{eq:2termViolationParams} and $r_0 = 2.5L$, we find a trapped surface $\sigma$ with \eq{\frac{A[\sigma]}{A_{\text{SAdS}}[M]} = 1.043\ldots,} and so we have a well-controlled PI-violating solution at this point in parameter space as well. However, unlike the vacuum at $Z = 3.2\times10^{-7}$ discussed above, the spectrum at this point is totally nontachyonic! The mass of the saxion is \eq{m_\sigma^2L^2 = 9.35,} and the mass of the axion is comfortably positive. To the best of our knowledge, this is the first known (AdS) PI violation without tachyonic modes.

\begin{figure}
    \centering
    \includegraphics[width=\linewidth]{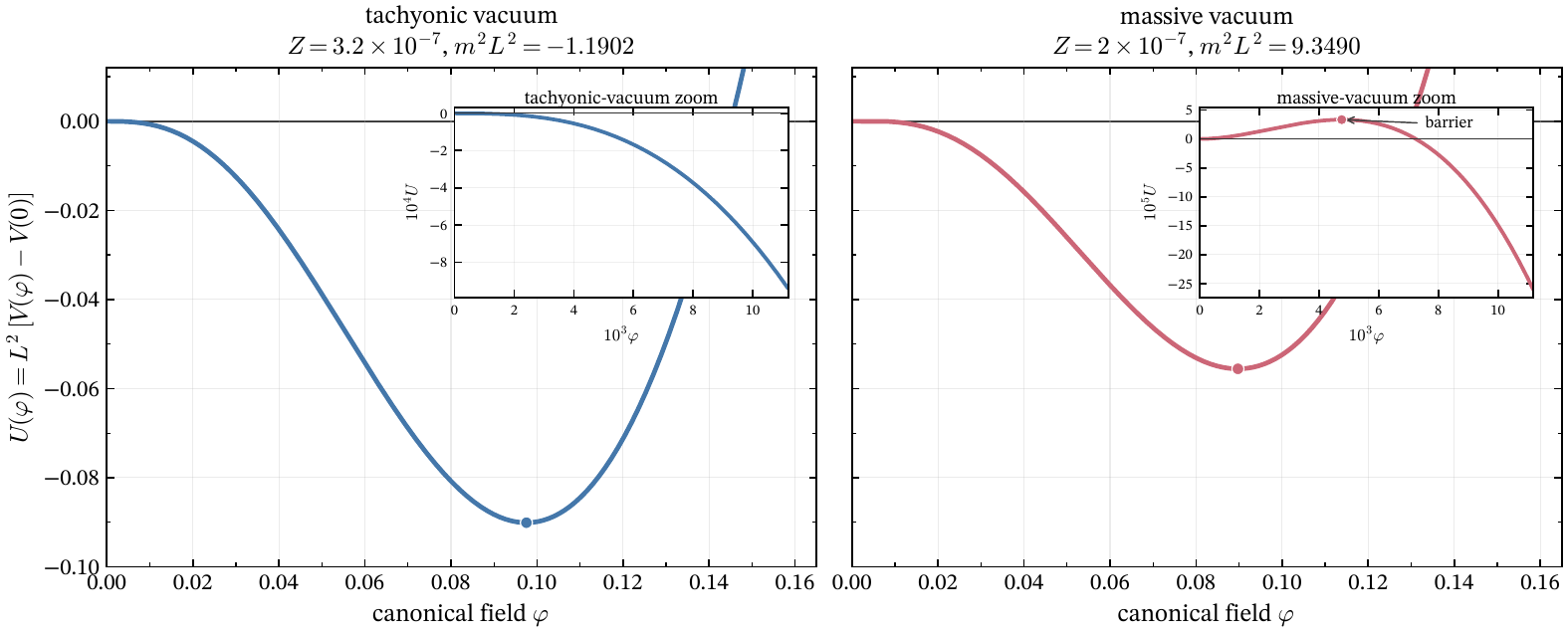}
    \caption{Potentials of the two-instanton toy models with $A=146$ and $\sigma_*=0.98$ at (left) a tachyonic point with $Z = 3.2\times10^{-7}$ and (right) a nontachyonic point with $Z=2\times10^{-7}$. Both of these points host Penrose-violating initial data, essentially because of the large negative well; the tiny barrier in the massive case does not prevent these solutions from accumulating enough negative energy to violate the Penrose inequality.}
    \label{fig:2termPotentials}
\end{figure}

In \cite{Fol22}, it was proved that (at least in the spherically symmetric case considered here and there) any theory satisfying the DEC is PI-safe, so to find PI violations we must study theories violating the DEC as well. While the presence of a BF-allowed tachyon in the spectrum, as happens in all theories considered in \cite{Fol22,Fol25}, guarantees DEC violation, it is not necessary for PI violation. We have plotted the potentials for both of these values of $Z$ in Figure \ref{fig:2termPotentials}; we see that both potentials have a large negative well, and so violate the DEC. The mechanism for PI violation even in the nontachyonic case is clear: PI-violating field configurations obtain large negative energy from the well, and the tiny amount of positive energy contributed by the positive-mass barrier is insufficient to overcome it. We note that all examples here satisfy the null energy condition. 

While one of the key features of the PI is that violations can be diagnosed purely on an initial data slice, without the need to numerically evolve them, to determine whether a naked singularity actually forms, doing the numerical evolution is necessary. In these PI-violating models with low-dimensional field spaces, it is comparatively easy to do so. We numerically studied the time evolution of the PI-violating initial data in Eq. \ref{eq:2termViolationParams} in both the tachyonic and massive models studied above, and found that their time evolution is consistent with evolving into a big crunch, rather than the formation of a naked singularity. Amongst other tests, we diagnose the existence of a big crunch singularity by studying the curvature invariant $L^4R_{abcd}R^{abcd}$, and match this to a CdL instanton by studying the profile $\phi(r)$ of the canonically normalized scalar mode $\phi = \sqrt{3/2}\log(\sigma/\sigma_*)$. In both theories, we find that there exists a CdL instanton with standard boundary conditions whose late time behavior closely matches that of the evolution of the initial data slice. In addition to these two profiles, we considered a landscape of 421 other choices of PI-violating initial data in these two theories; in no case did we find the formation of a naked singularity. These results are consistent with the interpretation that these PI violating trajectories correspond to false vacuum decay rather than the formation of a naked singularity.

\begin{figure}
  \centering
  \includegraphics[width=\linewidth]{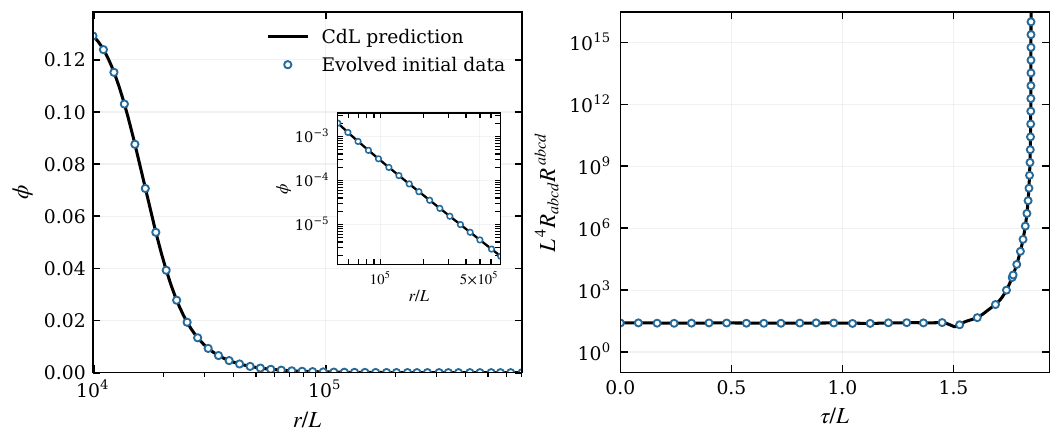}
  \par\vspace{3mm}
  \includegraphics[width=\linewidth]{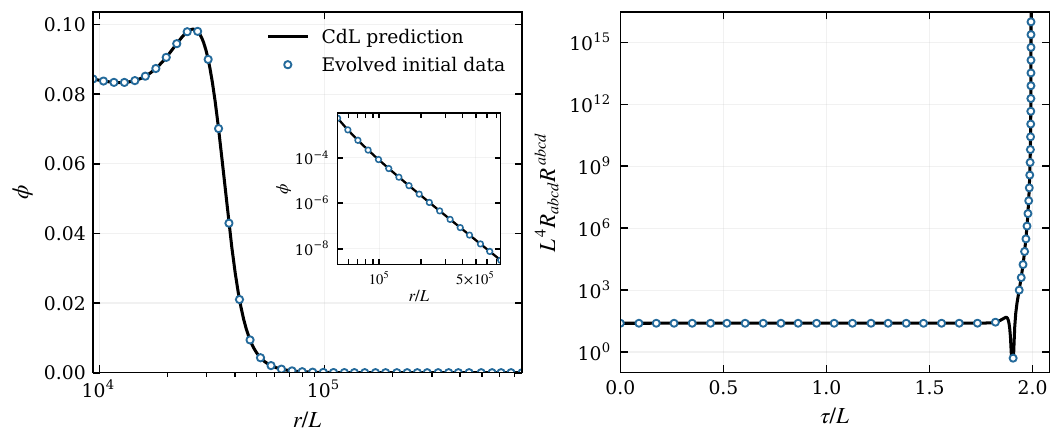}
  \caption{%
  The results of numerically evolving the PI-violating initial data sets given by Eq. \ref{eq:2termViolationParams} in the tachyonic (top) and massive (bottom) two-instanton models discussed in the main text. We find that, in each case, the theory admits a Euclidean CdL instanton with standard boundary conditions whose Lorentzian continuation has late-time behavior closely matching that of the evolved initial data, shown with open dots. In the left hand panel we show the late-time radial profile of $\phi$, zooming in on the large-radius behavior in the inset, and on the right we show the curvature invariant $L^4 R_{abcd}R^{abcd}$ along the trajectory of a freely falling observer as a function of the elapsed proper time $\tau$. In both cases, we find results consistent with evolving to a big crunch geometry, as expected from false vacuum decay.  
  }
  \label{fig:evolution}
\end{figure}

Next, we consider the nearby point $A=146$, $\sigma_* = 1.0565$, $Z = 5.6\times10^{-6}$, where the canonically normalized axionic mode is a BF-allowed tachyon of mass $m_u^2L^2 = -1.04$. While the analytic proof that ruled out the existence of the fake superpotential at $\sigma_*=0.98$ no longer applies here, both extensive numerical tests and an analytical argument indicate that there is no globally defined, everywhere differentiable fake superpotential. However, there is a global \textit{weak} fake superpotential, given by \eq{LP = 1 + \frac{9}{32\pi^2\sigma_*^2}\frac{\sigma_*}{\sigma} \left(1-\cos{2\pi u}\right),} and so this theory does not admit PI-violating initial data with standard boundary conditions; it is not in the designer gravity window, so no other boundary conditions need to be considered. In \cite{Fol25} it was conjectured that positive mass theorems and the absence of PI violations are equivalent. The existence of a fake superpotential implies the existence of a positive mass theorem, and as shown above also implies PI safety; this model demonstrates that the existence of a global fake superpotential is not necessary for PI safety.

As with the ordinary KKLT toy model, a key feature of this model is that one can find a vacuum for continuous families of $A,\sigma_*,Z$. This is in sharp contrast to real flux compactifications, where there are no continuous parameters. Instead, choices of fluxes yield vacua at isolated points in moduli space, so instead of a continuous family of models we obtain a discrete family of models for each choice of fluxes. In the remainder of the paper, we will study these isolated vacua; we find that all of the behaviors we have seen in this section are reflected in genuine compactifications.

\section{Explicit Nonsupersymmetric AdS KKLT Vacua}
\label{sec:real}
So far we have described toy models for KKLT; we turn now to their implementation in real CY3s. We will consider a landscape of 66 nonsupersymmetric AdS vacua constructed in \cite{McaMor24,McaNal26WIP}, and find PI violations in ten of them. Before describing those results, we will review the implementation of KKLT used in \cite{McaMor24,McaNal26WIP} and the models studied here; the reader familiar with KKLT can skip ahead to Sec. \ref{sec:vacua}.

\subsection{Review of KKLT}
The KKLT construction of nonsupersymmetric string vacua begins with Calabi-Yau compactification of type IIB string theory. The topology of a CY3 $X$ is characterized by a short list of data \cite{Wal66,ChaCon23,GenMac23}: the Hodge numbers $h^{1,1}$ and $h^{2,1}$, the symmetric rank three tensor $\kappa_{ijk}$ of intersection numbers, and the second Chern class $c_2$. Most important for our purposes will be the Hodge numbers; $h^{1,1}$ determines the number of homology classes of curves and divisors (embedded two- and four-dimensional submanifolds, respectively), and $h^{2,1}$ determines the number of three-cycles in the CY3. More importantly, the Hodge numbers also determine the number of moduli of the theory obtained by compactifying on $X$: $h^{1,1}$ counts the Kahler moduli, and $h^{2,1}$ counts the complex structure moduli. Including the axiodilaton, we therefore have $h^{1,1}+h^{2,1}+1$ complex moduli to stabilize.

Throughout this paper, as in \cite{McaMor24,McaNal26WIP,DemKim21}, we work with a class of CY3s known as toric hypersurface CY3s \cite{Bat93,KreSka00}. CY3s of this type are constructed from an appropriate (fine, regular, and star, or FRST for short) triangulation of one of the 473,800,776 four-dimensional reflexive polytopes, and are well-suited for computational study; in particular, the topological data specified above are easily computable for these geometries using the CYTools package \cite{DemRio22}. A given 4D reflexive polytope might admit many different triangulations of the appropriate type, and each of these triangulations may well give rise to a different CY3. In fact, given a reflexive polytope, it has recently been understood that a more general class of triangulations, the so-called ``vex triangulations" \cite{BerHub16}, also define smooth CY3s \cite{MacShe25}. Thus, while there are only 474 million polytopes, the number of CY3s built from them is much larger, although the total number is unknown \cite{ChaCon23,GenMac23,DemMca20,MacOre26}.  

\begin{figure}
    \centering
    \includegraphics[width=0.7\linewidth]{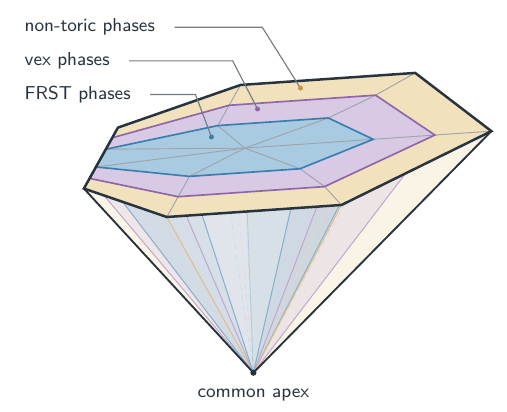}
    \caption{A sketch of the extended Kahler cone of a CY3 $X$, emphasizing the difference between internal boundaries between different CY3s and the global end of the moduli space. Note the containment structure: the cone of vex phases contains the cone of FRST phases, and both are contained in the full extended Kahler cone.}
    \label{fig:ekc}
\end{figure}

The CY3s built from a reflexive polytope, whether originating from an FRST or a vex triangulation, are connected geometrically in the Kahler moduli space of the compactification. The Kahler moduli space $\calM_K(X)$ of a given CY3 $X$ can be parameterized in one of two ways, either in terms of the volumes $t^i$ ($i = 1, \cdots, h^{1,1}(X)$) of curves or the volumes $T_i$ of divisors. The simplest way of defining $\calM_K(X)$ is as the set of $t$ such that all curves have positive volumes; this is clearly closed under taking positive multiples and addition, so $\calM_K(X)$ is a cone, usually called the Kahler cone of $X$.  As you approach the boundary of the Kahler cone, at least one curve approaches zero volume. Frequently, one can analytically continue the scalar potential and similar functions of the Kahler moduli to negative curve volumes to obtain sensible formulas for a new CY3 $X'$; these transitions, called ``flops," glue together the Kahler cones of different CY3s into an ``extended Kahler cone," consisting of the Kahler cones of all CY3s that can be reached by flops from $X$. Because the potential is smooth across a flop, the Kahler moduli space of compactifications on one CY3 $X$ is in fact the extended Kahler cone of $X$, rather than just the Kahler cone of $X$ itself. We have sketched a slice of an extended Kahler cone in Figure \ref{fig:ekc}. If $X$ is a toric hypersurface CY3, i.e. it is constructed from a triangulation of a 4D reflexive polytope, its extended Kahler cone automatically includes all other CY3s that can be built from other triangulations of the same polytope, whether FRST or vex, but in general other nontoric phases may be present in the extended Kahler cone as well. While by now computationally constructing and studying toric CY3s, whether obtained from FRST or vex triangulations, is straightforward \cite{DemRio22,DemKim23,MacShe25}, nontoric phases are much less amenable to systematic study, and so when we search for PI violations in these models we will only be able to look within the subcone of the extended Kahler cone corresponding to toric phases. 

In addition to the Kahler moduli space $\calM_K$, spanned by the $t^i$ or equivalently the $T_i$, there is a geometric complex structure moduli space $\calM_{CS}$, spanned by the complex structure moduli $z^a$ and axiodilaton $\tau$. KKLT \cite{KKLT} sketched how to construct a potential on the full joint complex and Kahler structure moduli space, and that sketch was fleshed out in \cite{McaMor24,McaNal26WIP}; we work in the EFT used there, working to tree level in $g_s$ but consistently incorporating all known $\alpha'$ corrections at that order. As discussed above, the nonsupersymmetric potential is defined by the superpotential $W$ and Kahler potential $K$ that specify the supersymmetric F-term potential, as well as a supersymmetry breaking term $V_{up}$; in terms of these components, the full potential is given as usual by \eq{V = e^K\left[\left(\partial_I\bar{\partial}_{\bar{J}}K\right)^{-1}D_IWD_{\bar{J}}\overline{W} - 3|W|^2\right] + V_{up}.} For full details of the various terms of the potential, we refer the reader to \cite{McaMor24}, but for completeness we will now sketch the various terms. 
\begin{itemize}
    \item \textbf{Superpotential:} The superpotential itself is the sum of two terms, one for the complex structure moduli and one for the Kahler moduli: \eq{W(z,\tau,T) = W_{flux}(z,\tau) + W_{np}(T).} The Gukov-Vafa-Witten \cite{GukVaf99} flux superpotential $W_{flux}$ is, as its name suggests, determined by the choice of integral flux vectors $f,h\in H^3(X,\mathbb{Z})$; once fluxes have been chosen, schematically $W_{flux}$ takes the form of a cubic polynomial in $z$ plus an infinite sum of shrinking exponentials in $z$, where the coefficients of both the polynomial and exponential pieces are determined by the fluxes as well as the topology of $X$ and its mirror. We work throughout with special flux choices, called conifold perturbatively flat vacua or coni-PFVs for short \cite{McaMor24,McaNal26WIP,DemKim21,DemKim19,DemKim20}, where the polynomial terms vanish so that the superpotential is entirely comprised of these exponentials, which are generated by worldsheet instantons in the mirror type IIA picture. These instanton-generated superpotentials are naturally exponentially small, as required by KKLT, and were shown in \cite{DemKim20} to be compatible with the existence of a KS throat, so are natural candidates for KKLT.

    The nonperturbative superpotential for the Kahler moduli is simpler in form: it is given by a sum over divisors in $X$ of contributions generated by nonperturbative objects such as Euclidean D3 branes and gaugino condensation on space-filling orientifold seven planes hosted on the divisor; this is because axionic shift symmetries prevent Kahler moduli from entering the superpotential perturbatively \cite{GidKac01} but not nonperturbatively \cite{Wit96a,Wit96b}. In the geometries considered in \cite{McaMor24,McaNal26WIP} that we will work with here, the nonperturbative superpotential is given explicitly by \eq{W_{np}(T) = \sum_{\text{divisors } D} A_D e^{-2\pi T_D/c_D},} where the coefficients $A_D$ are simply constant numbers, $T_D$ is the complexified Kahler modulus corresponding to the divisor $D$, and the dual Coxeter numbers $c_D$ are either one, if $D$ hosts an ED3 instanton, or 6, if it hosts an O7 plane; note that this is exactly the form predicted in \cite{KKLT}. Much of the complexity in dealing with the Kahler moduli comes from passing between the divisor volumes $T_i$ and the curve volumes $t^i$; at leading order in $\alpha'$, we have that \eq{\operatorname{Re}{T_i} = \frac{1}{2}\kappa_{ijk}t^jt^k,} but throughout we incorporate the $\alpha'$ corrections laid out in \S2.3 of \cite{McaMor24}, which modify this relation. 
    
    \item \textbf{Kahler Potential:} In the EFT of \cite{McaMor24,McaNal26WIP}, the Kahler potential $K$ is given by \eq{e^{-K/2} = 4\left[-i\int \Omega(z) \wedge \bar{\Omega}(z)\right]^{1/2}\calV g_s^{-3/2},} where the holomorphic three-form $\Omega(z)\in H^{3,0}(X)$ describes how the complex geometry of $X$ varies as a function of the complex structure moduli and $\calV$ is the $\alpha'$-corrected volume of $X$ as a function of the Kahler moduli. At leading order in $\alpha'$, we simply have that \eq{\calV = \frac{1}{6}\kappa_{ijk}t^it^jt^k,} but once again we incorporate the $\alpha'$ corrections laid out in \S2.3 of \cite{McaMor24}, which add additional terms to the volume formula. 

    The joint complex structure and Kahler moduli space itself is equipped with a Hermitian metric $\calG_{I\bar{J}}$, defined in terms of the Kahler potential as \eq{\calG_{I\bar{J}} = \partial_I \bar{\partial}_{\bar{J}} K.\label{eq:moduliSpaceMetric}} 
    
    \item \textbf{Uplift term:} So far we have specified the ingredients needed to assemble the supersymmetric F-term potential $V_F$; to break supersymmetry, we also need the antibrane potential $V_{up}$. For supersymmetry breaking in KKLT, it is important to engineer a Klebanov-Strassler (KS) throat \cite{KleStr00}. The role of the KS throat is to redshift the energy of a supersymmetry-breaking antibrane. KS throats come into existence near a certain type of boundary of the complex structure moduli space of $X$ called a conifold wall, and the  scale of redshifting is determined by the conifold modulus $z_{cf}$. In the strict $z_{cf}\to0$ limit, we sit on a wall of the moduli space, and the underlying CY3 $X$ becomes singular. The role of the coni-PFV flux choice is to stabilize $z_{cf}$ at an exponentially small but nonzero value, so that we retain both a strong redshift and a smooth CY3. In terms of $z_{cf}$ and $\calV$, we have that \cite{McaMor24} \eq{V_{up} = \eta \frac{1}{g_sM^2\tilde{\calV}^{2/3}}\frac{\left|z_{cf}\right|^{4/3}}{\calV^{4/3}},} where $\eta\sim2.6727$ is a numerical constant, $M$ is the flux through the cycle that shrinks to zero size as $z_{cf}\to0$, and $\tilde{\calV}$ is the volume of the mirror CY3 as a function of the complex structure moduli. For most purposes, the dependence of $V_{up}$ on the non-conifold complex structure moduli is very weak, and whether a supersymmetric precursor vacuum uplifts to dS or nonsupersymmetric AdS is determined almost entirely by $z_{cf}$.
\end{itemize}

As described above, the potential for the moduli is determined by $W$, $K$, and $V_{up}$. Given a CY3, then, to specify the full potential we require just topological data of $X$ and its mirror and a choice of flux vectors. By now the computation of this topological data is completely standard, at least for toric hypersurfaces \cite{DemRio22,DemKim23}, and powerful tools have been developed to scan over different choices of flux vectors \cite{DemKim21,McaMor24}. Given a choice of CY3 and flux, then, the potential is known completely explicitly as a function of the moduli, and to find vacua one only needs to be able to search numerically for critical points of the potential and evaluate the canonically normalized Hessian eigenvalues, which are defined with respect to the Kahler metric on moduli space given in Eq. \ref{eq:moduliSpaceMetric}; any critical point of the full potential where either (a) $V>0$ and all canonical mass-squareds are positive definite or (b) $V<0$ and all canonical mass-squareds satisfy the four-dimensional BF bound is a vacuum of the compactification.

\subsection{PI Violations in KKLT Vacua}
\label{sec:vacua}

\begin{figure}[t]
    \centering
    \includegraphics[width=.75\linewidth]{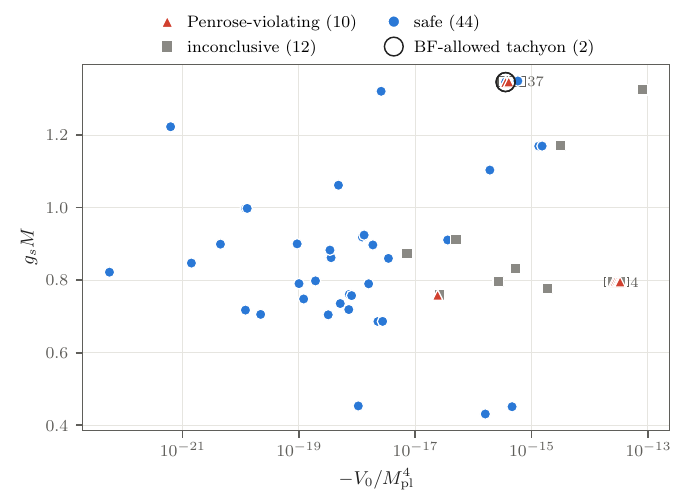}
    \caption{Properties of the 66 vacua considered here. For each vacuum, we specify the control parameter $g_sM$, the vacuum energy in Planck units, and  whether PI violations were found in the vacuum. More details on the clusters of vacua labeled 4 and 37 are given in Figures \ref{fig:decays-c4} and \ref{fig:decays-c37}, respectively.}
    \label{fig:vacuaScatter}
\end{figure}

By scanning over coni-PFVs and performing numerical searches for vacua in the associated potentials, a large landscape of nonsupersymmetric KKLT vacua, of both positive and negative vacuum energy, was found in \cite{McaMor24,McaNal26WIP}.  In the remainder of this section, we will describe searches for PI violation in a set of 66 nonsupersymmetric AdS vacua obtained from 41 flux vacua across 25 polytopes; these vacua were obtained early in the preparation of \cite{McaNal26WIP}\footnote{We thank the authors of \cite{McaNal26WIP} for allowing us to include these in-progress results in this work.}. We have summarized their properties in Table \ref{tab:vacua} and visually in Figure \ref{fig:vacuaScatter}. Note that only two of these nonsupersymmetric AdS KKLT vacua admit a tachyon; in neither case is the tachyon mass in the window allowing designer gravity boundary conditions, so we need only worry about standard boundary conditions.

\begin{table}[t]
\centering
\scriptsize
\setlength{\tabcolsep}{3pt}
\renewcommand{\arraystretch}{1.12}
\begin{tabular}[t]{@{}l c r r c l@{}}
\toprule
label & $h^{1,1}\!,h^{2,1}$ & $g_sM$ & $V_0/M_{\rm pl}^4$ & tach. & PI \\
\midrule
1 & $(96,6)$ & 1.172 & $-3.11\cdot10^{-15}$ & no & inc. \\
\addlinespace[2pt]2 & $(96,6)$ & 1.325 & $-8.10\cdot10^{-14}$ & no & inc. \\
\addlinespace[2pt]3 & $(96,6)$ & 0.778 & $-1.86\cdot10^{-15}$ & no & inc. \\
\addlinespace[2pt]4a & $(97,5)$ & 0.795 & $-2.43\cdot10^{-14}$ & no & inc.\\
4b &  & 0.795 & $-2.57\cdot10^{-14}$ & no & \textbf{viol.} \\
4c &  & 0.795 & $-2.65\cdot10^{-14}$ & no & safe \\
4d &  & 0.796 & $-2.70\cdot10^{-14}$ & no & \textbf{viol.} \\
4e &  & 0.796 & $-2.89\cdot10^{-14}$ & no & safe \\
4f &  & 0.796 & $-2.90\cdot10^{-14}$ & no & \textbf{viol.} \\
4g &  & 0.796 & $-3.07\cdot10^{-14}$ & no & inc. \\
4h &  & 0.796 & $-3.08\cdot10^{-14}$ & no & \textbf{viol.} \\
4i &  & 0.796 & $-3.18\cdot10^{-14}$ & no & \textbf{viol.} \\
4j &  & 0.796 & $-3.18\cdot10^{-14}$ & no & safe \\
4k &  & 0.796 & $-3.19\cdot10^{-14}$ & no & inc. \\
4l &  & 0.796 & $-3.32\cdot10^{-14}$ & no & \textbf{viol.} \\
4m &  & 0.796 & $-3.41\cdot10^{-14}$ & no & inc. \\
\addlinespace[2pt]5 & $(99,7)$ & 0.719 & $-7.24\cdot10^{-19}$ & no & safe \\
\addlinespace[2pt]6 & $(103,11)$ & 0.431 & $-1.60\cdot10^{-16}$ & no & safe \\
\addlinespace[2pt]7 & $(103,11)$ & 0.919 & $-1.24\cdot10^{-18}$ & no & safe \\
\addlinespace[2pt]8 & $(107,11)$ & 0.924 & $-1.33\cdot10^{-18}$ & no & safe \\
\addlinespace[2pt]9 & $(107,11)$ & 0.798 & $-1.94\cdot10^{-19}$ & no & safe \\
\addlinespace[2pt]10 & $(107,11)$ & 0.451 & $-4.63\cdot10^{-16}$ & no & safe \\
\addlinespace[2pt]11 & $(108,10)$ & 0.860 & $-3.47\cdot10^{-18}$ & no & safe \\
\addlinespace[2pt]12 & $(112,10)$ & 0.790 & $-1.59\cdot10^{-18}$ & no & safe \\
\addlinespace[2pt]13 & $(118,8)$ & 0.706 & $-2.22\cdot10^{-20}$ & no & safe \\
\addlinespace[2pt]14a & $(121,9)$ & 0.686 & $-2.29\cdot10^{-18}$ & no & safe \\
14b &  & 0.686 & $-2.76\cdot10^{-18}$ & no & safe \\
\addlinespace[2pt]15 & $(121,9)$ & 0.717 & $-1.21\cdot10^{-20}$ & no & safe \\
\addlinespace[2pt]16 & $(121,9)$ & 0.874 & $-7.23\cdot10^{-18}$ & no & inc.\\
\addlinespace[2pt]17 & $(121,9)$ & 1.321 & $-2.60\cdot10^{-18}$ & no & safe \\
\addlinespace[2pt]18 & $(121,9)$ & 0.748 & $-1.21\cdot10^{-19}$ & no & safe \\
\addlinespace[2pt]19a & $(121,9)$ & 0.911 & $-3.62\cdot10^{-17}$ & no & safe \\
19b &  & 0.913 & $-5.06\cdot10^{-17}$ & no & inc. \\
\bottomrule
\end{tabular}
\hfill
\begin{tabular}[t]{@{}l c r r c l@{}}
\toprule
label & $h^{1,1}\!,h^{2,1}$ & $g_sM$ & $V_0/M_{\rm pl}^4$ & tach. & PI \\
\midrule
20 & $(125,9)$ & 0.899 & $-4.51\cdot10^{-21}$ & no & safe \\
\addlinespace[2pt]21a & $(125,9)$ & 0.862 & $-3.48\cdot10^{-19}$ & no & safe \\
21b &  & 0.862 & $-3.59\cdot10^{-19}$ & no & safe \\
\addlinespace[2pt]22 & $(125,9)$ & 0.736 & $-5.16\cdot10^{-19}$ & no & safe \\
\addlinespace[2pt]23 & $(137,9)$ & 0.831 & $-5.33\cdot10^{-16}$ & no & inc.\\
\addlinespace[2pt]24 & $(137,9)$ & 0.822 & $-5.59\cdot10^{-23}$ & no & safe \\
\addlinespace[2pt]25 & $(137,9)$ & 0.796 & $-2.69\cdot10^{-16}$ & no & inc.\\
\addlinespace[2pt]26 & $(138,8)$ & 0.900 & $-9.38\cdot10^{-20}$ & no & safe \\
\addlinespace[2pt]27a & $(138,8)$ & 0.998 & $-1.21\cdot10^{-20}$ & no & safe \\
27b &  & 0.998 & $-1.22\cdot10^{-20}$ & no & safe \\
27c &  & 0.998 & $-1.30\cdot10^{-20}$ & no & safe \\
\addlinespace[2pt]28a & $(138,8)$ & 0.759 & $-2.44\cdot10^{-17}$ & no & \textbf{viol.} \\
28b &  & 0.759 & $-2.45\cdot10^{-17}$ & no & safe \\
28c &  & 0.759 & $-2.47\cdot10^{-17}$ & no & safe \\
28d &  & 0.759 & $-2.61\cdot10^{-17}$ & no & inc. \\
\addlinespace[2pt]29 & $(145,9)$ & 0.761 & $-7.38\cdot10^{-19}$ & no & safe \\
\addlinespace[2pt]30a & $(145,9)$ & 1.169 & $-1.33\cdot10^{-15}$ & no & safe \\
30b &  & 1.170 & $-1.52\cdot10^{-15}$ & no & safe \\
\addlinespace[2pt]31 & $(146,8)$ & 0.897 & $-1.88\cdot10^{-18}$ & no & safe \\
\addlinespace[2pt]32 & $(146,8)$ & 0.791 & $-1.01\cdot10^{-19}$ & no & safe \\
\addlinespace[2pt]33 & $(146,8)$ & 1.062 & $-4.81\cdot10^{-19}$ & no & safe \\
\addlinespace[2pt]34 & $(146,8)$ & 0.705 & $-3.21\cdot10^{-19}$ & no & safe \\
\addlinespace[2pt]35 & $(150,8)$ & 0.847 & $-1.43\cdot10^{-21}$ & no & safe \\
\addlinespace[2pt]36 & $(150,8)$ & 0.757 & $-8.10\cdot10^{-19}$ & no & safe \\
\addlinespace[2pt]37a & $(150,8)$ & 1.346 & $-3.58\cdot10^{-16}$ & yes & \textbf{viol.} \\
37b &  & 1.346 & $-3.58\cdot10^{-16}$ & yes & \textbf{viol.} \\
37c &  & 1.346 & $-3.59\cdot10^{-16}$ & no & safe \\
37d &  & 1.347 & $-4.05\cdot10^{-16}$ & no & \textbf{viol.} \\
37e &  & 1.349 & $-5.82\cdot10^{-16}$ & no & safe \\
\addlinespace[2pt]38 & $(150,8)$ & 1.103 & $-1.92\cdot10^{-16}$ & no & safe \\
\addlinespace[2pt]39 & $(155,11)$ & 0.453 & $-1.05\cdot10^{-18}$ & no & safe \\
\addlinespace[2pt]40 & $(159,7)$ & 0.883 & $-3.44\cdot10^{-19}$ & no & safe \\
\addlinespace[2pt]41 & $(347,11)$ & 1.223 & $-6.29\cdot10^{-22}$ & no & safe \\
\bottomrule
\end{tabular}
\caption{The 66 non-supersymmetric AdS vacua studied here, obtained from an early state of the search in \cite{McaNal26WIP}. Each of the 41 compactifications is assigned a number, and for compactifications with more than one vacuum, each of the vacua is then assigned a letter designation. The compactifications are ordered by increasing $Q = 2 + h^{1,1}+h^{2,1}$, and within a compactification the different vacua are ordered by decreasing cosmological constant $V_0$. We find violations only in three of the 41 compactifications. For each vacuum, we indicate its Hodge numbers $h^{1,1},h^{2,1}$, the parameter $g_sM$ which controls metastability of the KS throat, the cosmological constant $V_0$ in Planck units, whether the lightest canonically normalized mode is tachyonic, and whether PI violations were observed in that vacuum. Here, ``safe" indicates that no PI violation was found, not that we have proven that such violations cannot occur, ``viol." indicates that an explicit set of PI violating initial data was found, and ``inc" indicates that the scan for PI violations in that theory was inconclusive.}
\label{tab:vacua}
\end{table}

These polytopes have $h^{1,1}+h^{2,1}$ ranging from 102 to 358, and so the potential is a function of between 206 and 718 real fields. Actually doing a scan for PI violations in these rather complicated EFTs on very high-dimensional field spaces is quite computationally intensive. In particular, it is computationally infeasible to sample the full-dimensional field space, as was done in searches for PI violation in the toy models in Section \ref{sec:kklt} above. Instead, we restrict ourselves to the subspace of moduli space determined by the six lightest directions of the canonically normalized Hessian on the full complex structure and Kahler moduli space. This is because PI violations require regions of energy less than that of the vacuum that are also very close, in metric distance, to the original vacuum; it is implausible to us that such regions can occur along very heavy directions, and numerical evidence supports this intuition. 

Along the low-dimensional field space we consider, we begin by reconstructing the potential over as large a field range as possible. In complex structure moduli space, following \cite{McaMor24,McaNal26WIP} we work in a fixed phase of the mirror CY; conversely, within the extended Kahler cone we allow Kahler flops between toric phases, but in general do not allow flops to nontoric phases. We shoot rays out of the origin along each of the six lightest eigendirections, in both the positive and negative directions, and continue along the Born-Oppenheimer valley $\phi(s)$ defined by that direction until one of the following conditions is met:
\begin{enumerate}
    \item $U(\phi) > 2/L^2$. In a direction along which the potential grows to be significantly larger than the vacuum, it is implausible that a negative well sufficiently steep to allow for PI violations can also occur.
    \item $V(\phi)$ jumps by more than 1\% across a flop. The potential depends on an infinite set of topological data called the Gopakumar-Vafa (GV) invariants of (curves in) $X$. We always truncate this infinite set to a finite list. When we encounter a flop between a CY3 $X$ and a new phase $X'$, in the limit where all GVs are kept, the potential is known to be continuous. However, in the finitely truncated case, it need not be; we allow flops as long as our approximation to the potential does not jump significantly across the Kahler wall defining the flop.     
    \item A nontoric flop which we cannot reconstruct is encountered. As described above, it is easiest to work with CY3s obtained from (possibly vex) triangulations of a four-dimensional reflexive polytope; if we encounter a flop to another phase which cannot be described in this way, we frequently can no longer reconstruct the moduli space or the potential on it, and we end our trajectory. However, under certain conditions (namely, if the GV invariant of the vanishing curve is known), we can construct the topological data of the resulting CY3 from the known toric data; we allow for this limited class of nontoric flops, subject to the above constraint that the numerical approximation to the potential is sufficiently smooth across the flop.
    \item The moduli space ends. Sharp criteria for the boundaries of the moduli space of compactifications on a CY3 $X$ were worked out in \cite{wit96c}; if we encounter one of these boundaries, the physical moduli space ends, and we cannot continue.
    \item $\phi\geq0.4$ in canonically normalized metric distance. This is a numerical cutoff for computational reasons, rather than a principled physical cutoff; without such a cutoff, one ray could in principle continue through an infinite flop chain without bound.
\end{enumerate}

We thus reconstruct the potential along twelve paths out of any given vacuum. If $V>V_0$ everywhere along a trajectory, the constant weak fake superpotential $P = 1/L$ guarantees PI safety and a PMT along that trajectory, by the arguments of Sec. \ref{sec:theorem}, and we call that trajectory safe. Conversely, if $V<V_0$ somewhere along the trajectory, then we need to look more carefully. In particular, from the first $V<V_0$ point discovered along a given trajectory we launch a descent trajectory, and proceed until we find another critical point, leave the toric Kahler cone, encounter a true boundary of moduli space, or encounter a discontinuity in the potential that indicates we have badly reconstructed a flop. 

Once the descent trajectory terminates, we try to numerically evolve a fake superpotential along it. If the trajectory ends at another vacuum and the fake superpotential exists everywhere along the trajectory, we call it safe. Conversely, if the fake superpotential ceases to exist somewhere along the trajectory, we begin to scan for PI violating initial data whose field ranges include the regions where the theorem in Sec. \ref{sec:theorem} does not guarantee PI safety. In every case where the fake superpotential folds, we do find PI violating initial data. Finally, if the trajectory ends while the potential is still decreasing and the superpotential still exists, we call it inconclusive, because we cannot rule out PI violations existing further out in moduli space.\footnote{An interesting edge case occurs when a trajectory ends on an SU(2) wall, which is a genuine geometric boundary of moduli space; even if the potential is still descending when we hit the wall, there is nowhere else for the trajectory to go, and so we can rule these trajectories as safe even though they end at a boundary of moduli space rather than a minimum.}

If all trajectories starting at a given vacuum are safe, we call that vacuum itself PI-safe; note that this means only that we have been unable to find PI violating initial data, and not that we have rigorously proven that they do not exist. Such a proof would require sampling the entire full-dimensional moduli space of the compactification, including non-toric regions of the extended Kahler cone, and is computationally intractable. Conversely, if we find PI violating initial data along any of the rays, we say that vacuum is violating. Finally, if at least one of the trajectories starting at a vacuum is inconclusive and no rays yield PI-violating trajectories, we say that vacuum is inconclusive. 

Of the 66 vacua, we find that \textbf{44 are safe, 10 are violating, and 12 are inconclusive.} Before discussing these results in detail, we must make some disclaimers about their interpretation:
\begin{itemize}
    \item These are the results of numerical searches. Vacua labeled as safe have not been \textit{proven} to not admit PI violating initial data, but instead safe here means that no data have been found in the search. We have gathered extensive numerical evidence that PI-violating initial data in these vacua do not exist on the loci in moduli space where we have searched, but cannot rule out their existence elsewhere in moduli space. 
    \item The vacua found in \cite{McaMor24,McaNal26WIP} themselves come with several disclaimers. These come in two broad classes: flux quantization and unknown corrections to the EFT. It was pointed out in \cite{FrePol02} that certain exotic toroidal orientifolds only admit flux vacua with purely even flux quanta. It is not known whether the same is true of CY3 orientifolds, and the vacua studied here have odd quanta. If CY3 orientifolds also require even flux quanta, these models might not be well-defined, but the results of the calculations here are nonetheless valid EFT calculations.
    \item Conversely, unknown corrections to the EFT, such as loop corrections to the Kahler potential and $\alpha'$ corrections to KPV, are known to exist, but we do not at present know how to compute them. 10\% relative shifts in the energy between vacua, or in the overall scale of $V_{up}$, are enough to change whether some vacua are PI-safe or not, and it is entirely plausible that these corrections will change the potential by more than that, or even destroy the vacua altogether. On the other hand, within the EFT constructed in \cite{McaMor24,McaNal26WIP} and used here, we are confident that all numerical errors are well below the threshold where they could qualitatively change any of the outcomes.  
    \item The KKLT construction imagines breaking supersymmetry by confining an anti-D3 brane at the end of a KS throat \cite{KKLT}. The minimum for the antibrane energy that this calculation relies on is metastable, and antibranes in KS throats have a finite lifetime \cite{KacPea01}. Thus, even if we find that a given KKLT vacuum is PI-safe, it will eventually decay via antibrane annihilation, regardless of the existence of a weak fake superpotential. Nevertheless, our results are powerful enough to rule out the existence of decaying solutions in the EFT of \cite{McaMor24,McaNal26WIP}, neglecting these nonperturbative effects.
\end{itemize}
Thus, these results are best thought of as being statements about the EFT considered here.

With those disclaimers out of the way, we can discuss our results in more detail. Of the 44 safe vacua, after extensive numerical searches we find that 32 come from compactifications where we can find no lower-energy vacua; we suspect that these are the global minima of their potentials, and hence are globally PI-safe by the DEC theorem of \cite{Fol22}, but as mentioned above cannot yet prove this. Conversely, 12 of the safe vacua live in compactifications with other, lower energy vacua to which they could in principle decay; 6 of these even share a moduli space with other vacua which we have found to be PI-violating. While these vacua can in principle decay, along the regions in moduli space where we have reconstructed a weak fake superpotential they obey a PMT, and so cannot decay on this locus.

The ten vacua where we find PI-violating initial data are summarized in Table \ref{tab:violators}. We find violations in both massive and tachyonic vacua; in particular, both tachyonic vacua found in the scan admit PI-violating initial data. In some sense, these violations are a much stronger result than the safe vacua:  unlike the safe vacua, where we can only rule out the existence of PI violations on a sublocus of moduli space, here we have explicit PI-violating initial data in hand, with no remaining uncertainty. 

\begin{figure}
    \centering
    \begin{subfigure}{0.90\textwidth}
        \centering
        \includegraphics[
            width=\linewidth,
            height=0.46\textheight,
            keepaspectratio
        ]{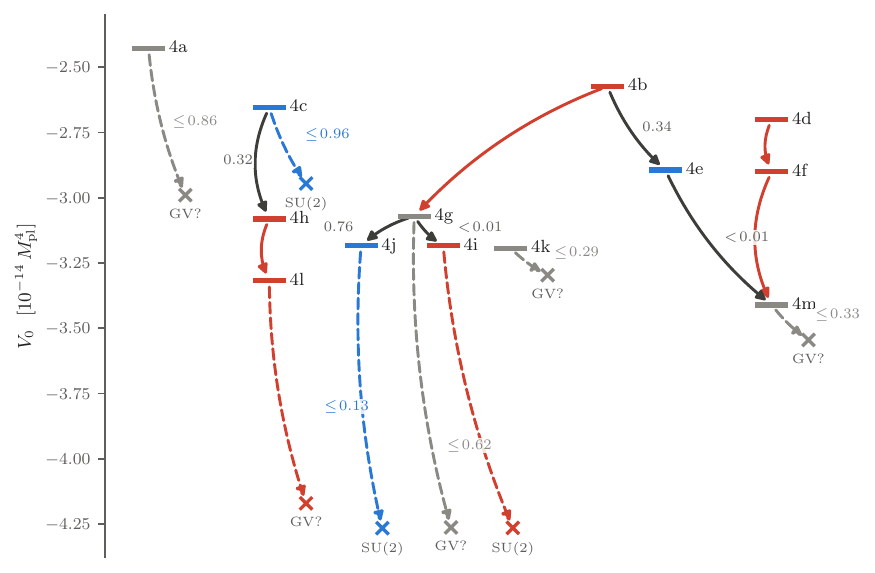}
        \caption{Compactification 4.}
        \label{fig:decays-c4}
    \end{subfigure}

    \medskip

    \begin{subfigure}[t]{0.48\textwidth}
        \centering
        \includegraphics[
            width=\linewidth,
            height=0.32\textheight,
            keepaspectratio
        ]{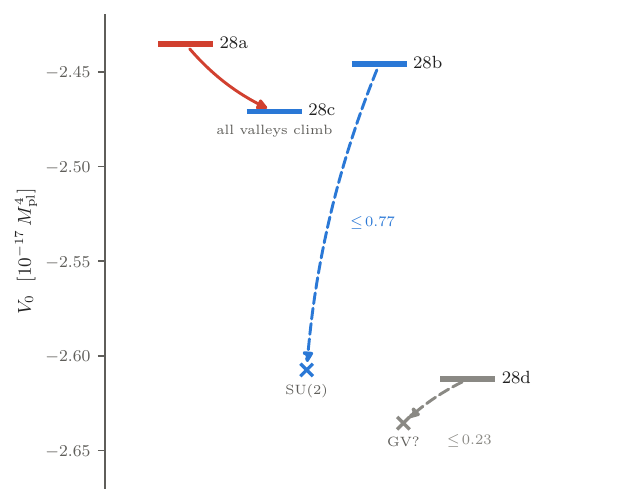}
        \caption{Compactification 28.}
        \label{fig:decays-c28}
    \end{subfigure}
    \hfill
    \begin{subfigure}[t]{0.48\textwidth}
        \centering
        \includegraphics[
            width=\linewidth,
            height=0.32\textheight,
            keepaspectratio
        ]{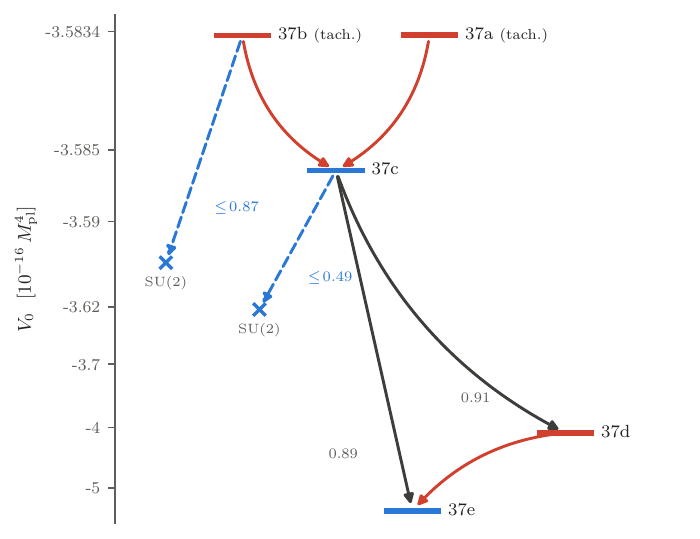}
        \caption{Compactification 37.}
        \label{fig:decays-c37}
    \end{subfigure}

    \caption{The three compactifications in which we find vacua with PI-violating initial data. For each compactification, we list all of its vacua and their PI safety status: blue vacua are safe, red vacua admit explicit PI-violating initial data, and grey vacua are inconclusive. For each vacuum, we also plot some of its possible decay routes and whether or not they have been seen to admit PI-violating initial data, using the same color scheme. For each trajectory where no violation has been found, we list the largest Penrose ratio $\calR$ that we have found.}
    \label{fig:decay-graphs}
\end{figure}

All violating vacua originate in one of three compactifications, each of which has a rich landscape of vacua; within each of these compactifications, we have schematically plotted the vacua and their allowed and disallowed decays in Figure \ref{fig:decay-graphs}. In one of these, compactification 4 in the notation of Table \ref{tab:vacua}, we have found thirteen distinct nonsupersymmetric AdS vacua, of which six are found to admit PI-violating initial data.

\begin{table}[t]
\centering
\begin{tabular}{lll}
\hline
Starting vacuum & Path destination & Penrose ratio $\calR$ \\
\hline
4b          & 4g       & 1.068360 \\
4d        & 4f       & 1.074204 \\
4f       & 4m    & 1.119323 \\
4h         & 4l   & 1.004091 \\
4i          & SU(2) wall        & 1.571822 \\
4l          & nontoric        & 1.030763 \\
28a        & 28c & 1.020226 \\
37a     & 37c        & 1.000688 \\
37b        & 37c        & 1.000680 \\
37d          & 37e        & 1.172654 \\
\hline
\end{tabular}
\caption{The ten vacua in which we found PI-violating initial data. For each vacuum, we list one of the vacua to which it can decay, and the PI ratio $\calR$ of one choice of initial data demonstrating the decay. Vacuum 4l admits PI-violating initial data along a slice of moduli space that descends past the toric Kahler cone, through several nontoric phases we can construct and into a phase which we cannot construct; nevertheless, we can find the violation explicitly, inside phases where we can compute the potential reliably.}
\label{tab:violators}
\end{table}

In principle, to numerically evolve the PI-violating initial data we have found in these models and determine whether they evolve to a big crunch or a naked singularity, we must keep track of all $2(h^{1,1}+h^{2,1}+1)$ real fields. This is computationally infeasible, but one can make partial progress by modeling it as a one-dimensional problem along the ray in moduli space hosting the violating trajectory. This one-dimensional approximation is much more tractable, and we performed the truncation and evolution for one explicit set of PI-violating initial data originating from the 4b$\to$4g transition. We found numerical evidence that the one-dimensional truncation of that initial data slice evolves to a big crunch geometry, with the AdS boundary collapsing in finite time. We find this suggestive that the full-dimensional models will evolve to big crunches as well, but we certainly have not proven this in general.

\section{A Stable Nonsupersymmetric DGKT Vacuum?}
\label{sec:dgkt}
So far we have considered toy models for, and genuine examples of, the KKLT construction, which is formulated in IIB string theory. In IIA string theory, the most explicit known flux compactifications are due to DeWolfe, Giryavets, Kachru, and Taylor (DGKT) \cite{DewGir05}; we now turn to a study of the PI in the simplest nonsupersymmetric DGKT model, the ``skew-whiffed" vacua first studied in \cite{NarTri10}.   

The standard skew-whiffed model in \cite{NarTri10} has, in addition to two degenerate true vacua, a tachyonic false vacuum that will be our main object of study in this section. Although there is a generic expectation that nonsupersymmetric AdS vacua should always admit \textit{some} decay route \cite{Har10}, for instance from flux-changing decays or CdL instantons leading to lower-energy vacua, we find a hint that these skew-whiffed vacua may well be absolutely stable because they admit no known decay channels. DGKT is formulated in IIA string theory, where the relevant branes are D-even branes, so the possible mediators for flux-changing transitions in the RR sector are D2, D4, D6, and D8 branes. Moreover, to mediate such a transition, such a brane must be \textit{superextremal}: its charge must be bigger than its tension, in appropriate units. In \cite{MarQui22} it was shown that, to first subleading order in a $g_s$ expansion about the background of \cite{DewGir05}, D2 branes are subextremal and the relevant D4 and D6 brane states are exactly extremal, and so none of these can mediate a decay. This leaves only D8-brane bound states as possible mediators, but \cite{MarQui22} further pointed out that there exist choices of flux in the skew-whiffed models-- namely those without space-filling D6 branes-- for which the tadpole bound rules out those decays. Thus, at least to the order studied in \cite{MarQui22}, there are no known allowed flux-changing decays. We will see below that the tachyonic false vacuum in this model admits a global weak fake superpotential for \textit{any} choice of fluxes, including those that prohibit D8-mediated decays, and therefore does not admit CdL decays to the true vacua, at least with standard boundary conditions. Compactifications with those flux quanta, therefore, at present admit no known decay channels. We find this to be a tantalizing hint that \textit{these models might be absolutely stable}, although we certainly do not claim to have proven that.

The basic setting for DGKT is massive IIA supergravity \cite{Rom85}. In DGKT, the relevant fluxes are those corresponding to the RR field strengths $F_0$ and $F_4$ and the NS-NS field strength $H_3$, and the relevant orientifolds are O6 planes; the full effective action, including more general flux choices than those used in DGKT, was worked out in \cite{GriLou04}. In DGKT, the tadpole bound stems from O6 planes, and only the flux quanta of $F_0$ and $H_3$ enter into the tadpole constraint; this means that $F_4$ flux is totally unbounded, and can be taken to be very large in order to yield classically well-controlled vacua with parametric scale separation and weak coupling.

DGKT considered compactifying massive IIA on a particular $T^6$ orientifold of the form $X = T^6/\bbZ_3^2$; this orientifold has $h^{2,1}(X)=0$ and $h^{1,1}(X)=12$, so including the universal hypermultiplet there are 13 total complex moduli fields. DGKT found an infinite sequence of flux choices that supersymmetrically stabilized all 13 of these moduli, and \cite{NarTri10} demonstrated that simply flipping the signs of certain flux quanta stabilizes all moduli nonsupersymmetrically. Importantly, in these nonsupersymmetric vacua, called skew-whiffed vacua in \cite{NarTri10} and the following literature, the scalar potential for the moduli fields is known totally explicitly in closed form; that makes the skew-whiffed vacua a natural testing ground for PI violations.

Of the 12 Kahler moduli of the toroidal orientifold $X$, three come from the untwisted sector and the remaining nine come from blowups. Following \cite{DewGir05} and subsequent works such as \cite{NarTri10,HerTeg07,AraVig21}, we omit the nine blowup moduli and work only with the untwisted sector Kahler moduli and the universal hypermultiplet. Thus, the model we consider has four complex scalars, \begin{subequations} the IIA axion-dilaton\footnote{As opposed to the axiodilaton in IIB.} \eq{N = \xi + i\sigma,} where $\sigma$ is defined in terms of the four-dimensional dilaton $D$ as \eq{\sigma = \sqrt{2}e^{-D},} and the three Kahler moduli \eq{t_i  = b_i + iv_i} for $i=1,2,3$. In terms of the $t_i$, we have that the CY volume is given by \eq{\calV = \kappa v_1v_2v_3,} where $\kappa = 81$ is the unique nonvanishing intersection number on the $T^6/\bbZ_3\times\bbZ_3$ orientifold.\end{subequations}

Their dynamics are governed by the superpotential \eq{W = \sum_{i=1}^3e_it_i - m_0\kappa t_1t_2t_3 - pN,\label{eq:dgktW}} where $e_i,m_0,p$ are integer flux quanta, and the Kahler potential \eq{K = -\log\left(8\kappa v_1v_2v_3\right) - 4\log{\sigma} + \log4.\label{eq:dgktK}} The O6 tadpole bound requires \eq{m_0p<0,} and the supersymmetric vacua constructed in \cite{DewGir05} require \eq{m_0e_i < 0~ \forall i.} Conversely, the ``skew-whiffed" nonsupersymmetric vacua studied in \cite{NarTri10} require the opposite sign, \eq{m_0e_i > 0~ \forall i;\label{eq:skewWhiffFluxes}} these are the vacua we will study here. 

Once the superpotential and the Kahler potential have been specified, we construct as usual the F-term potential, which in the skew-whiffed case is given explicitly as \cite{NarTri10} \eq{V = \frac{2}{\calV\sigma^4}\left[\vphantom{\displaystyle\sum_{(i,j,k) = (1,2,3) + \text{cyc}}} \frac{p^2\sigma^2}{4}   + 2m_0p\calV\sigma  + m_0^2\calV^2\left(1+\sum_{i=1}^3 \frac{b_i^2}{v_i^2}\right) + \left(\sum_{i=1}^3 e_ib_i - p\xi - m_0\kappa b_1b_2b_3\right)^2 \right.  \nonumber \\ + \left.\sum_{(i,j,k) = (1,2,3) + \text{cyclic}} v_i^2\left(e_i-m_0\kappa b_jb_k\right)^2\right].\label{eq:dgktV}} Importantly, none of our results depend on the actual choice of flux quanta as long as they obey the skew-whiffed condition in Eq. \ref{eq:skewWhiffFluxes}; they set the overall scale of the potential but do not affect any of the calculations we perform. 

Solving $\nabla V = 0$ gives three vacua. The first, which we will call $q_0$, occurs at \eq{t_i = i \frac{1}{\left|e_i\right|} \sqrt{\frac{5}{3}\left|\frac{e_1e_2e_3}{\kappa m_0}\right|}, ~ ~ ~ N = i\frac{\sqrt{2}}{|p|} \sqrt{\frac{160}{27} \left|\frac{e_1e_2e_3}{\kappa m_0}\right|};} we will write these vevs as $t_i = iv_{i,0}, N = i\sigma_0$ for later convenience. At this vacuum, we find that the cosmological constant $V_0$ is given by \eq{V_0 = -\sqrt{\frac{4}{15}} \left(\frac{27}{160}\right)^2 \frac{p^4\kappa^{3/2}\left|m_0\right|^{5/2}}{\left|e_1e_2e_3\right|^{3/2}}.} In terms of the AdS length $L^2 = 3/|V_0|$, we have that the spectrum of canonically normalized masses is given by \eq{m^2L^2 = (-2, 18, 18, 18, 40, 40, 40, 70).} We therefore have a single tachyonic mode, which we will call $\phi$ throughout. The other two vacua, which we will call $q_\pm$, are exactly degenerate, and are related by the axionic reflection symmetry $(b_i,\xi)\to (-b_i,-\xi)$; they occur at \eq{t_i =  \frac{v_{i,0}}{\sqrt{5}} \left(\pm1 + 2i\right), ~ ~ ~ N = \frac{\sigma_0}{\sqrt{5}} \left(\mp1 + 2i\right).} These vacua have cosmological constants \eq{V_\pm = \frac{25\sqrt{5}}{48}V_0} and masses \eq{m^2L_\pm^2 = (6, 18, 18, 18, 48, 48, 48, 96).} 

\begin{figure}
    \centering
    \includegraphics[width=\linewidth]{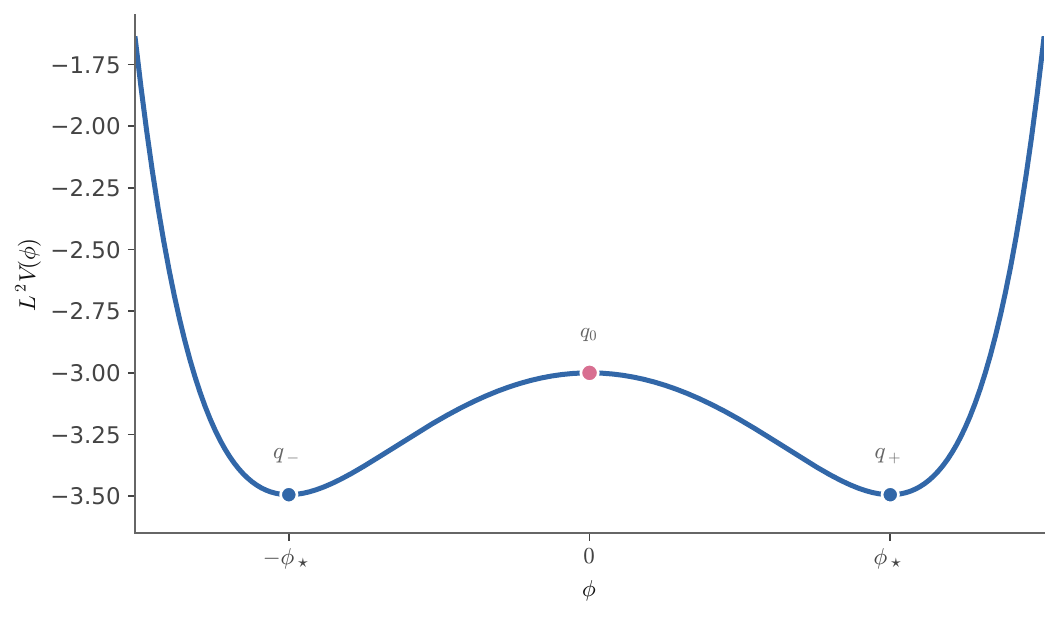}
    \caption{The potential of the nonsupersymmetric DGKT vacuum along the canonically normalized tachyonic direction $\phi$. All three vacua are visible along this slice of moduli space; $q_\pm$ furnish the global minima of the potential, and $q_0$ is a PI-safe tachyonic vacuum which surprisingly enjoys a positive mass theorem for standard boundary conditions.}
    \label{fig:dgktV}
\end{figure}

We thus have a landscape of three vacua to consider. Note that, although the potential in Eq. \ref{eq:dgktV} takes the form of a standard F-term potential, these vacua are nonsupersymmetric, in the sense that $DW \neq 0$ \cite{NarTri10}. We have plotted the potential along $\phi$ in Figure \ref{fig:dgktV}; the massive vacua $q_\pm$ lie a geodesic distance \eq{\phi_* = \sqrt{\frac{7}{2}}\log\frac{1+\sqrt{5}}{2} = 0.9003...}  away from $q_0$ along this axis. In fact, $V_\pm$ is the global minimum of the potential; the EFTs of these vacua therefore obey the DEC, and they are guaranteed to be PI-safe by the theorem of \cite{Fol22}. 

Conversely, $q_0$ is tachyonic, and therefore potentially PI violating. However, we can prove that this vacuum is also PI-safe. In particular, we construct a global weak fake superpotential, i.e. a function $P$ such that \eq{V \geq 2\left|\nabla P\right|^2 - 3P^2\label{eq:dgktWeakFakeGeneral}} everywhere on moduli space. Define \eq{R = 2\sum_{i=1}^3 \left|\frac{t_i - i v_{i,0}}{t_i + iv_{i,0}}\right|^2 + 8 \left|\frac{N-i\sigma_0}{N+i\sigma_0}\right|^2.} We then claim that \eq{P = \frac{1}{L}\left(1+\frac{R}{4}\right)\label{eq:dgktP}} furnishes a weak fake superpotential. The proof goes in two parts. First we prove that $P$ furnishes a weak fake superpotential outside the ball $R\leq 1/2$, and then we prove that it also furnishes one inside this ball. This somewhat artificial split occurs because the proof methods in the two regimes are different: outside the ball we prove it analytically, and inside we prove it numerically, using interval arithmetic. Note that this is genuinely a proof, not just numerical evidence. 

Write \eq{\calE = V + 3P^2 - 2\left|\nabla P\right|^2,\label{eq:calEdefdgkt}} so that what we need to show is $\calE \geq 0$ everywhere. As outlined above, we begin by showing this outside the ball $R\leq \frac{1}{2}$; this is essentially an algebraic exercise. The Kahler potential in Eq. \ref{eq:dgktK} determines the metric on moduli space; it turns out to be a hyperbolic metric, and one can use it to show that \eq{|\nabla R|^2 \leq 4R.} Plugging this and the definition of $P$ in terms of $R$ into $\calE$, we find that \eq{\calE \geq V + \frac{1}{L^2}\left(3+R+\frac{3R^2}{16}\right).} As stated above, we have a global lower bound, $V\geq -25\sqrt{5}/16L^2$. We therefore have that \eq{\calE \geq \frac{1}{L^2}\left(-\frac{25\sqrt{5}}{16} + 3 + R + \frac{3R^2}{16}\right).} For $R>1/2$, we then have \eq{L^2\calE \geq 3 + \frac{1}{2}+\frac{3}{64} - \frac{25\sqrt{5}}{16} \geq 0,} as desired. 

Conversely, for $R\leq1/2$, we do not know of an analytic proof. Instead, we use a rigorous proof strategy from numerical analysis called interval arithmetic; see e.g. \cite{Tuc11} for a review, and \cite{Joh17} for the precise implementation used. The idea is straightforward; we construct a bounding box for the ball inside which the analytic proof is insufficient, and then subdivide that box into smaller ones. Inside each box, we rigorously demonstrate a nonnegative lower bound for $\calE$: \eq{L^2\calE \geq \frac{3R^2}{16};} this establishes that $\calE\geq0$ everywhere inside the ball, and completes the proof. 

We have therefore proven that the tachyonic, nonsupersymmetric skew-whiffed DGKT vacuum admits a weak fake superpotential. By the calibration theorem in Sec. \ref{sec:theorem}, this vacuum is PI-safe in standard quantization: no spherically symmetric initial data with standard boundary conditions can ever violate the Penrose Inequality. Even more surprisingly, the weak fake superpotential implies that, with standard boundary conditions, \textit{this theory enjoys a positive mass theorem} \cite{Bou84,Tow84,EldJoy14}! At the level of the eight-field, two-derivative EFT, then, this vacuum is stable: even though the lower-energy vacua $q_\pm$ are present in the same moduli space, the $q_0$ vacuum does not admit CdL instantons leading to those vacua, and so is stable against decays to them, subject to standard boundary conditions. This does not rule out decay by nonperturbative string theoretic mechanisms, such as brane-flux annihilation, but it is nonetheless a strong statement at the EFT level that goes against some conjectures in the existing literature. In fact, in this model more is true: there exist choices of flux in this skew-whiffed model, namely those with no spacetime-filling D6 branes, for which there are no known brane-mediated decay routes \cite{MarQui22}. With standard boundary conditions, then, there are no known channels through which this false vacuum can decay, and it may well be absolutely stable. 

So far we have only considered standard boundary conditions. However, the tachyonic mode $\phi$ has mass $m_\phi^2L^2 = -2$, so it is inside the window where designer gravity boundary conditions can be considered. With standard boundary conditions, DGKT is always PI-safe and enjoys a PMT, but with designer boundary conditions the story is less simple. As usual with designer boundary conditions, we work with \eq{\phi \sim \frac{\alpha}{r} + \frac{f\alpha^2}{r^2} + \cdots} for some $f$; different choices of fake superpotential can establish PI safety and PMTs for different values of $f$. For instance, with the weak fake superpotential in Eq. \ref{eq:dgktP}, we find that the generalized Witten-Nester bound in Eq. \ref{eq:designerEsgeq} becomes \eq{L^2M \geq  w(\alpha),} so we have a PMT only for boundary conditions where $w(\alpha)\geq0$ for all $\alpha$; this includes both the standard $\alpha=0$ boundary conditions and the $\beta=0$ Neumann conditions for any $\alpha$, but for the AdS-preserving designer boundary conditions in Eq. \ref{eq:designerf} this only occurs for $f=0$. Other choices of weak fake superpotential yield looser bounds on $f$; the function \eq{P_{1/3} = \frac{1}{L}\left(1+\frac{R}{4}-\frac{R^{3/2}}{18}\right)} also furnishes a weak fake superpotential, this time with $s = 1/3$, and so we find that AdS-preserving designer boundary conditions with $|f|\leq 1/3$ are PI-safe. The best such bound we have been able to find is \eq{|f|\leq \frac{4}{5};} all AdS-preserving designer boundary conditions in this range are PI-safe and enjoy a PMT, as was true in standard quantization. 

For $f$ outside of this interval, there is no guarantee of PI-safety, and indeed we can construct explicit violations. Take $r_0 = 1$ in units with $L=1$, and consider the field profile \eq{\phi(r) = -\frac{3}{r+3}.} Asymptotically we have \eq{\phi(r) \sim \frac{-3}{r} + \frac{9}{r^2} + \cdots,} so this is compatible with AdS-preserving designer boundary conditions with $f = 1$. These initial data have Penrose ratio \eq{\calR = 1.13... > 1,} and so violate the SAdS-PI. As with all of the PI-violating initial data we have seen in this work, we have strong numerical evidence that these initial data evolve into a big crunch geometry, rather than a naked singularity. Thus, while standard quantization is PI-safe in DGKT, generic boundary conditions need not be.

\section{Where do the Proofs Fail?}
\label{sec:fail}
Let us now consider the unfortunate state of affairs we find ourselves in: (a) we have three different proofs of the AdS Penrose Inequality under rather reasonable assumptions such as the covariant holographic entanglement entropy prescription, the absence of trapped surfaces outside of horizons,  and the duality between high energy thermal ensembles and AdS black holes, and yet (b) we have violations of the AdS Penrose Inequality in rather pedestrian excitations of vacua in type IIB string theory. This presents a \textit{prima facie} unpalatable dilemma for both authors, who happen to be proponents of string theory.

As explained in Sec.~\ref{sec:real},  not all SUSY-broken AdS vacua support violations, and the ones that do feature violations are still unexpectedly well-behaved: violations occur only in a field excursion from one vacuum to another via false vacuum decay (or bubble of nothing). How, then, do such relatively reasonable geometries violate the assumptions of the various proofs of the AdS PI? 

Below we go through the various assumptions of each proof and identify the ones that are violated by the vacua that are not PI-safe in Sec.~\ref{sec:real}. We will not go through the proof from the existence of a fake superpotential lower bound, as that has already been discussed in Sec.~\ref{sec:theorem}: the fake superpotential lower bound exists locally for any vacuum, but for those vacua that feature violations, the fake superpotential ceases to exist as CdL bubbles come into existence.  We focus on the two remaining proofs: the first, from weak cosmic censorship, and the second, from holography. 

\paragraph{PI from WCC:} Recall that this proof, outlined in Sec.~\ref{sec:PIReview}, assumes that (1) trapped surfaces lie behind event horizons, (2) causal horizons obey an area theorem, and (3) black holes equilibrate to stationarity.  The first assumption is hard to test without a known analytic solution for all time, but fortunately the bubble of nothing violations in a compactification of eleven-dimensional SUGRA with SUSY violating boundary conditions (Sec.~\ref{sec:cosh}) provide precisely such a case study.  In that example, it is true that there are no trapped surfaces within causal contact of $\mathscr{I}$. Nevertheless,  whether trapped surfaces lie behind \textit{event} horizons or not is in fact a subtle question here that also relates to assumption (2).  Certainly it is true that for any future-infinite causal worldline $\gamma$,  trapped surfaces lie behind $\partial J^{-}[\gamma]$, the causal horizon generated by $\gamma$, and it is also true that this horizon obeys an area theorem.  However,  if we turn our attention to \textit{event} horizons, defined as $\partial J^{-}[i^{+}]$, we find a problem: $i^{+}$ does not exist in this geometry due to the big crunch -- thus the event horizon is empty. Thus trapped surfaces could lie behind all possible causal horizons, but do not in any meaningful sense lie behind an event horizon.  Finally, let us comment on assumption (3). It is not clear that black holes in a vacuum unstable to CdL necessarily equilibrate to stationarity: the equilibrium state may require a timescale over which a true vacuum or bubble of nothing will nucleate. In the former case, we expect the equilibrium to be achieved for a black hole in the true vacuum, whereas in the latter there is no equilibrium black hole at all. 

\paragraph{PI from Holography:} Let us now consider the holographic proof.  Recall that its assumptions are as follows: (1) well-posedness of the characteristic initial value problem in General Relativity, (2) the covariant holographic entanglement entropy prescription or, equivalently, its justification via the gravitational path integral in the classical limit, (3) the duality between high energy thermal ensembles and (microcanonically) stable black holes. Considering the evidence in favor of (1) in the absence of any kind of SUSY considerations, we shall immediately drop it as a possible failure point. This leaves us with (2) and (3) as possible failed assumptions in the unstable vacuum. Both (2) and (3) may be justified via different applications of the gravitational path integral (GPI)~\cite{LewMal13, DutFau19, Mar18}; in particular, since our initial data are time symmetric,  these are just two different applications of the Euclidean GPI.  It is certainly possible that the GPI fails to compute an entropy for unstable false vacua even prior to any decay; however, a less radical explanation exists: a failure of the duality between the microcanonical thermofield double and black holes in a false vacuum that are not PI-safe.  Consider a theory allowing initial data containing field profiles that asymptote to either a true vacuum or a false one vulnerable to CdL decay to the true vacuum; such a theory could plausibly be obtained by viewing states in the expansion about the false vacuum as highly excited states in the theory of the true vacuum. In such a theory, there is no a priori reason to expect black holes in the false vacuum to dominate the thermal ensemble in the GPI; more plausibly, the dominant contributions would be black hole states in the true vacuum. This raises the possibility that some version of Ryu-Takayanagi could hold  even in these unstable vacua -- at least within a WDW patch of a single time slice of the asymptotic boundary, or with a UV cutoff in the dual theory/IR cutoff in the bulk. 

\vspace{0.4cm}

It is also worth emphasizing here that \textit{most} of the non-SUSY AdS vacua we found do not violate the PI, suggesting a possible scenario in which \textit{all} of the assumptions of \textit{both} proofs are well-motivated in such vacua. This is despite the explicitly broken supersymmetry and any potential nonperturbative instabilities (e.g. KPV), and is especially plausible in skew-whiffed DGKT vacua, for which certain flux choices admit no known brane-mediated decay routes \cite{MarQui22}. 

\section{What does it mean?}
\label{sec:conclusion}
In this paper, we have investigated the Penrose Inequality in several nonsupersymmetric anti-de Sitter vacua of string theory, arising from both KKLT \cite{KKLT,McaMor24,McaNal26WIP} and DGKT \cite{DewGir05,NarTri10} compactifications. Using the basic tool of weak fake superpotentials, we found that, while certain vacua are PI-safe (sometimes provably so!), others admit PI-violating initial data. In particular, we found that the ``skew-whiffed" DGKT vacuum is PI safe, assuming standard boundary conditions and spherical symmetry, but some KKLT compactifications admit rich landscapes of nearby vacua, some of which are PI-violating. 

This poses an interesting puzzle: if string theory admits PI violations, and we believe that our universe arises from string theory, why haven't we observed any such violations in nature? Is string theory ``bad"? That would be troubling, but fortunately there is no actual contradiction: our results suggest that PI violations in string theory come in the form of vacuum decay, and indeed we have rigorously shown that (at least in AdS vacua, although see below for speculation about dS vacua like ours) no initial data sets lying entirely within the barrier of some vacuum can ever violate the PI. Thus, since our universe has yet to decompactify, denucleate, or otherwise decay, we do not expect to have observed PI violations. While we initially hoped to find that string theory never admits PI violations, we find this interpretation in terms of vacuum decay to be reassuring. 

In fact, Penrose's original definition of a naked singularity~\cite{Pen79} excluded big crunch singularities as required for vacuum decay. While the PI is not a good discriminator between big crunch/false vacuum-type singularities and the more pathological naked singularities considered in~\cite{Pen79}, which would be detectable to an infinitely long-lived observer sitting at the asymptotic boundary, it is encouraging for string theory that its modus operandi for violating the PI appears -- within the data set we have investigated -- to be purely that of vacuum decay. While we have certainly not proven that naked singularities are impossible, we have rigorously shown that any PI-violating initial data must necessarily include field excursions outside of the basin of the false vacuum. It is plausible that sufficiently fine-tuned initial data will evolve to a naked singularity in the false vacuum, but we expect that generic initial data outside the basin of a false vacuum will correspond to vacuum decay.

We close with several open questions:

\begin{itemize}
    \item \textbf{How important is spherical symmetry?} Throughout, we have assumed spherical symmetry, essentially for technical convenience. Indeed, our main theorem states that weak fake superpotentials imply PI safety, but only assuming standard boundary conditions and spherical symmetry; conversely, weak fake superpotentials imply positive mass theorems even in the absence of spherical symmetry. We have already seen that allowing for more general boundary conditions allows for PI violations even in the presence of a weak fake superpotential; does relaxing spherical symmetry do the same? In particular, the ostensibly PI-safe vacua in Sec. \ref{sec:real} are outside the mass window where designer boundary conditions apply; do non-spherically symmetric initial data allow them to violate the Penrose inequality?
    \item \textbf{What distinguishes vacua where the PI is or is not violated?} \cite{McaMor24} observed two classes of behavior in dS compactifications: some compactifications allowed a single dS vacuum, and others allowed many nearby vacua. In the AdS compactifications studied in Sec. \ref{sec:real}, we have observed even more fine-grained differences, this time between vacua in the same compactification: some vacua are PI safe, even when the DEC theorem of \cite{Fol22} does not guarantee their safety, and others admit PI-violating initial data. What distinguishes these two classes? Given the data specifying a KKLT vacuum, what is the smallest amount of information needed to determine whether or not it admits PI-violating initial data? Obviously, knowing that a vacuum is the global minimum of its compactification or that it admits a global weak fake superpotential would be enough to rule out PI violations, but can we predict the result of computing a weak fake superpotential from some more minimal dataset? 
    \item \textbf{Are these theories holographic?}  Some nonsupersymmetric AdS vacua admit a positive mass theorem, at least at the EFT level, and so are much more stable than previously thought. This also implies that they are  consistent with a quasi-local version of RT. What does this mean for nonsupersymmetric  holographic duality in AdS? Our results suggest that PI violations are closely tied to vacuum decay; is there some approximate sense in which these nonsupersymmetric vacua admit a holographic dual until they decay? What about vacua which are the global minimum of their potential, and so are guaranteed to be PI-safe?  One intriguing hint comes from e.g. the DGKT example, which is PI-safe in standard boundary conditions but not in designer gravity. The designer gravity boundary conditions are holographically dual to a multitrace deformation of the boundary theory \cite{Wit01b,KlebWit99}; can we exploit these results to learn anything about the holographic dual of DGKT and other theories in the designer mass window that become PI-violating only in the presence of these boundary conditions?  
    
    \item \textbf{Is DGKT absolutely stable?} We showed that the skew-whiffed DGKT vacua admit a global weak fake superpotential, and so are PI-safe with standard boundary conditions. There exist choices of flux quanta in this model for which there are no known brane-mediated ten-dimensional solutions that allow them to decay \cite{MarQui22}, so, with standard boundary conditions and spherical symmetry, there are at present no known decay routes by which even the false vacuum in this model can decay. This raises the intriguing possibility that this model might be \textit{absolutely stable}, which goes against conventional wisdom that all nonsupersymmetric AdS vacua should decay.  If these vacua really are stable, then their holographic dual would plausibly be a conventional conformal field theory. There exists a standard argument that nonsupersymmetric AdS vacua should not admit standard holographic duals, because the lifetime of the vacuum is essentially one AdS time \cite{HorOrg07,Har10}: assume there is a nonvanishing probability $\Gamma$ to nucleate a decay bubble per unit volume; AdS has infinite volume, so such a bubble nucleates immediately, and hits the boundary in AdS time. However, if these DGKT vacua really are absolutely stable, then $\Gamma=0$ and the argument does not apply. If that is indeed the case, what would be the appropriate boundary theory? The boundary theory, if it exists, is expected to have \textit{integer conformal dimensions} \cite{ApeMon22,ApeCon22,Qui22} for certain operators. 
    
    \item \textbf{What about de Sitter?} Because of the absence of an AD or ADM mass, there is no obvious way to formulate a Penrose Inequality in de Sitter space. However, given that we have established PMTs for some nonsupersymmetric AdS vacua, it is tempting to ask whether some of the dS vacua constructed in \cite{McaMor24,McaNal26WIP} might similarly be more stable than is naively thought. Indeed, from the perspective of KKLT, whether any given compactification admits AdS or dS vacua is rather accidental, depending only on the relative tuning of certain complex structure parameters. The effective potentials defining dS and nonsupersymmetric AdS vacua are therefore very similar, and in fact it is possible in principle for a single KKLT compactification to admit vacua with both signs of the cosmological constant.\footnote{See \cite{Har26} for a holographic discussion of such a scenario.} Can dS vacua ever experience any of the same protections -- weak fake superpotentials, PMTs, and ultimately PI safety -- as the nicest non-SUSY AdS vacua? In the limit of infinite CY volume, the potential of any KKLT vacuum, whether AdS or dS, tends to zero, so any dS vacuum admits a natural candidate decay route: a CdL instanton to ten-dimensional flat space. However, we have seen throughout that some ``obvious" decay routes are forbidden, as we showed in e.g. DGKT with standard boundary conditions. Are there any dS vacua for which this decay is forbidden? 
\end{itemize}

\section*{Acknowledgements} It is a pleasure to thank Raphael Bousso, Daniel Harlow, Shamit Kachru, Juan Maldacena, Liam McAllister, Washington Taylor, and Edward Witten for valuable discussions, and Daniel Harlow and Natalie Paquette for helpful comments on a draft. We are also grateful to the authors of \cite{McaNal26WIP} for allowing us to include as-yet-unpublished vacua in this work.
The work of NE is supported in part by the Department of Energy under Early Career Award DE-SC0021886 during earlier stages of this work and by DOE grant number DE-SC0012567 towards its end. The work of NE is further supported by  the HEP-QIS program under grant number DE-SC0025937, by the Heising-Simons Foundation under grant no. 2023-4430, by the Moore Foundation via the Black Hole Initiative, and by the MIT Physics Department. RN is a Pappalardo Postdoctoral Fellow in the MIT Department of Physics. RN was also partly funded during this work  by the Heising-Simons Foundation under grant no. 2023-4430. 

Computational aspects of this work made use of MIT's subMIT cluster \cite{Aco25}. In the course of the preparation of this paper, we used AI models including OpenAI's ChatGPT Astra 6 and Sol 5.6, and Anthropic's Claude Opus 5.5 and Fable 5.1, as well as certain older models. In particular, the theorem in Sec. \ref{sec:thm} was first proven by Astra; the authors subsequently understood and verified that proof. No AI-generated text appears in this paper, and the authors take responsibility for the accuracy of all scientific content herein.

\bibliographystyle{jhep}
\bibliography{all}

@article{ChoGer69,
  author  = {Choquet-Bruhat, Yvonne and Geroch, Robert},
  title   = {Global Aspects of the {C}auchy Problem in General Relativity},
  journal = {Communications in Mathematical Physics},
  volume  = {14},
  pages   = {329--335},
  year    = {1969},
  doi     = {10.1007/BF01645389}
}

@article{Mar09,
    author = "Mars, Marc",
    title = "{Present status of the Penrose inequality}",
    eprint = "0906.5566",
    archivePrefix = "arXiv",
    primaryClass = "gr-qc",
    doi = "10.1088/0264-9381/26/19/193001",
    journal = "Class. Quant. Grav.",
    volume = "26",
    pages = "193001",
    year = "2009"
}

@article{HuiIlm01,
  author  = {Huisken, Gerhard and Ilmanen, Tom},
  title   = {The Inverse Mean Curvature Flow and the
             {Riemannian Penrose Inequality}},
  journal = {Journal of Differential Geometry},
  volume  = {59},
  number  = {3},
  pages   = {353--437},
  year    = {2001},
  doi     = {10.4310/jdg/1090349447}
}

@misc{Bra99,
      title="{Proof of the Riemannian Penrose Conjecture Using the Positive Mass Theorem}", 
      author={Hubert L. Bray},
      year={1999},
      eprint={math/9911173},
      archivePrefix={arXiv},
      primaryClass={math.DG},
      url={https://arxiv.org/abs/math/9911173}, 
}

@article{HubLiu04,
  author = {Hubeny, Veronika E. and Liu, Xiao and Rangamani, Mukund and Shenker, Stephen},
  title = {Comments on Cosmic Censorship in {AdS/CFT}},
  journal = {Journal of High Energy Physics},
  volume = {2004},
  number = {12},
  pages = {067},
  year = {2004},
  doi = {10.1088/1126-6708/2004/12/067},
  eprint = {hep-th/0403198},
  archivePrefix = {arXiv}
}

@article{Chr99,
  author  = {Christodoulou, Demetrios},
  title   = {On the Global Initial Value Problem and the Issue of Singularities},
  journal = {Classical and Quantum Gravity},
  volume  = {16},
  number  = {12A},
  pages   = {A23--A35},
  year    = {1999},
  doi     = {10.1088/0264-9381/16/12A/302}
}

@book{Chr91,
  author    = {Chru{\'s}ciel, Piotr T.},
  title     = {On Uniqueness in the Large of Solutions of {Einstein}'s Equations (``Strong Cosmic Censorship'')},
  series    = {Proceedings of the Centre for Mathematics and its Applications, Australian National University},
  volume    = {27},
  publisher = {Centre for Mathematics and its Applications, Australian National University},
  address   = {Canberra},
  year      = {1991},
  isbn      = {0-7315-0443-7},
  url       = {https://maths.anu.edu.au/files/CMAProcVol27-Complete.pdf}
}

@article{EngFol24a,
    author = "Engelhardt, Netta and Folkestad, {\AA}smund and Levine, Adam and Verheijden, Evita and Yang, Lisa",
    title = "{Cryptographic Censorship}",
    eprint = "2402.03425",
    archivePrefix = "arXiv",
    primaryClass = "hep-th",
    doi = "10.1007/JHEP01(2025)122",
    journal = "JHEP",
    volume = "01",
    pages = "122",
    year = "2025"
}

@article{FriSch93,
	archiveprefix = {arXiv},
	author = {Friedman, John L. and Schleich, Kristin and Witt, Donald M.},
	doi = {10.1103/PhysRevLett.71.1486},
	eprint = {gr-qc/9305017},
	journal = {Phys. Rev. Lett.},
	note = {[Erratum: Phys.Rev.Lett. 75, 1872 (1995)]},
	pages = {1486--1489},
	reportnumber = {PRINT-93-0448 (SANTA-BARBARA,ITP), NSF-ITP-93-80},
	title = {{Topological censorship}},
	volume = {71},
	year = {1993}}

@article{Bou84,
	author = {Boucher, W.},
	journal = {Nucl. Phys.},
	pages = {282},
	slaccitation = {%%CITATION = NUPHA,B242,282;%%},
	title = {POSITIVE ENERGY WITHOUT SUPERSYMMETRY},
	volume = {B242},
	year = {1984}}

@article{BreFre82,
	author = {Breitenlohner, Peter and Freedman, Daniel Z.},
	journal = {Phys. Lett.},
	pages = {197},
	slaccitation = {%%CITATION = PHLTA,B115,197;%%},
	title = {POSITIVE ENERGY IN ANTI-DE {S}ITTER BACKGROUNDS AND GAUGED EXTENDED SUPERGRAVITY},
	volume = {B115},
	year = {1982}}

@article{ColDel80,
	author = "{S. Coleman and F. De Luccia}",
	journal = {Phys. Rev. D},
	pages = {3305-3315},
	title = {Gravitational effects on and of vacuum decay},
	volume = {21},
	year = {1980}}

@article{DewGir05,
	author = {DeWolfe, Oliver and Giryavets, Alexander and Kachru, Shamit and Taylor, Washington},
	eprint = {hep-th/0505160},
	journal = {JHEP},
	pages = {066},
	slaccitation = {%%CITATION = HEP-TH 0505160;%%},
	title = {Type {IIA} moduli stabilization},
	volume = {07},
	year = {2005}}

@article{EngWal18,
	archiveprefix = {arXiv},
	author = {Engelhardt, Netta and Wall, Aron C.},
	doi = {10.1007/JHEP05(2019)160},
	eprint = {1806.01281},
	journal = {JHEP},
	pages = {160},
	primaryclass = {hep-th},
	title = {{Coarse Graining Holographic Black Holes}},
	volume = {05},
	year = {2019}}

@article{GalSch99,
	archiveprefix = {arXiv},
	author = {Galloway, G. J. and Schleich, K. and Witt, D. and Woolgar, E.},
	doi = {10.1016/S0370-2693(01)00335-5},
	eprint = {hep-th/9912119},
	journal = {Phys. Lett.},
	pages = {255-262},
	primaryclass = {hep-th},
	slaccitation = {%%CITATION = HEP-TH/9912119;%%},
	title = {{The AdS / CFT correspondence conjecture and topological censorship}},
	volume = {B505},
	year = {2001}}

@incollection{GerHor79,
	author = {Geroch, Robert P. and Horowitz, G. T.},
	booktitle = {General Relativity: An Einstein Centenary Survey},
	pages = {212-293},
	publisher = {Cambridge University Press},
	slaccitation = {%%CITATION = INSPIRE-151495;%%},
	title = {{Global structure of spacetimes}},
	year = {1979}}

@article{GibHul83,
	author = {Gibbons, G. W. and Hull, C. M. and Warner, N. P.},
	journal = {Nucl. Phys.},
	pages = {173},
	slaccitation = {%%CITATION = NUPHA,B218,173;%%},
	title = {THE STABILITY OF GAUGED SUPERGRAVITY},
	volume = {B218},
	year = {1983}}

@article{GidKac01,
	author = {Giddings, Steven B. and Kachru, Shamit and Polchinski, Joseph},
	eprint = {hep-th/0105097},
	journal = {Phys. Rev. D},
	pages = {106006},
	slaccitation = {%%CITATION = HEP-TH 0105097;%%},
	title = {Hierarchies from fluxes in string compactifications},
	volume = {66},
	year = {2002}}

@article{HerHor04,
	archiveprefix = {arXiv},
	author = {Hertog, Thomas and Horowitz, Gary T.},
	doi = {10.1088/1126-6708/2004/07/073},
	eprint = {hep-th/0406134},
	journal = {JHEP},
	pages = {073},
	primaryclass = {hep-th},
	slaccitation = {%%CITATION = HEP-TH/0406134;%%},
	title = {{Towards a big crunch dual}},
	volume = {07},
	year = {2004}}

@article{HerHor05,
	archiveprefix = {arXiv},
	author = {Hertog, Thomas and Horowitz, Gary T.},
	doi = {10.1088/1126-6708/2005/04/005},
	eprint = {hep-th/0503071},
	journal = {JHEP},
	pages = {005},
	primaryclass = {hep-th},
	slaccitation = {%%CITATION = HEP-TH/0503071;%%},
	title = {{Holographic description of AdS cosmologies}},
	volume = {04},
	year = {2005}}

@article{HubRan07,
	archiveprefix = {arXiv},
	author = {Hubeny, Veronika E. and Rangamani, Mukund and Takayanagi, Tadashi},
	doi = {10.1088/1126-6708/2007/07/062},
	eprint = {0705.0016},
	journal = {JHEP},
	pages = {062},
	primaryclass = {hep-th},
	reportnumber = {DCPT-07-13, KUNS-2069},
	slaccitation = {%%CITATION = ARXIV:0705.0016;%%},
	title = {{A Covariant holographic entanglement entropy proposal}},
	volume = {0707},
	year = {2007}}

@article{KKLT,
	author = {Kachru, Shamit and Kallosh, Renata and Linde, Andrei and Trivedi, Sandip P.},
	eprint = {hep-th/0301240},
	journal = {Phys. Rev. D},
	pages = {046005},
	slaccitation = {%%CITATION = HEP-TH 0301240;%%},
	title = {De {S}itter vacua in string theory},
	volume = {68},
	year = {2003}}

@article{LewMal13,
	archiveprefix = {arXiv},
	author = {Lewkowycz, Aitor and Maldacena, Juan},
	doi = {10.1007/JHEP08(2013)090},
	eprint = {1304.4926},
	journal = {JHEP},
	pages = {090},
	primaryclass = {hep-th},
	slaccitation = {%%CITATION = ARXIV:1304.4926;%%},
	title = {{Generalized gravitational entropy}},
	volume = {1308},
	year = {2013}}

@article{Pen65,
	author = {Penrose, Roger},
	doi = {10.1103/PhysRevLett.14.57},
	issue = {3},
	journal = {Phys. Rev. Lett.},
	month = {Jan},
	numpages = {0},
	pages = {57--59},
	publisher = {American Physical Society},
	title = {Gravitational Collapse and Space-Time Singularities},
	url = {https://link.aps.org/doi/10.1103/PhysRevLett.14.57},
	volume = {14},
	year = {1965}}

@article{Pen69,
	author = {Penrose, R.},
	journal = {Riv. Nuovo Cim.},
	note = {[Gen. Rel. Grav.34,1141(2002)]},
	pages = {252-276},
	slaccitation = {%%CITATION = RNCIB,1,252;%%},
	title = {{Gravitational collapse: The role of general relativity}},
	volume = {1},
	year = {1969}}

@article{RyuTak06,
	archiveprefix = {arXiv},
	author = {Ryu, Shinsei and Takayanagi, Tadashi},
	doi = {10.1103/PhysRevLett.96.181602},
	eprint = {hep-th/0603001},
	journal = {Phys.Rev.Lett.},
	pages = {181602},
	primaryclass = {hep-th},
	reportnumber = {NSF-KITP-06-11},
	slaccitation = {%%CITATION = HEP-TH/0603001;%%},
	title = {{Holographic derivation of entanglement entropy from AdS/CFT}},
	volume = {96},
	year = {2006}}

@article{DutFau19,
	archiveprefix = {arXiv},
	author = {Dutta, Souvik and Faulkner, Thomas},
	eprint = {1905.00577},
	month = {5},
	primaryclass = {hep-th},
	title = {{A canonical purification for the entanglement wedge cross-section}},
	year = {2019}}

@article{Mar18,
	archiveprefix = {arXiv},
	author = {Marolf, Donald},
	doi = {10.1007/JHEP09(2018)114},
	eprint = {1808.00394},
	journal = {JHEP},
	pages = {114},
	primaryclass = {hep-th},
	title = {{Microcanonical Path Integrals and the Holography of small Black Hole Interiors}},
	volume = {09},
	year = {2018}}

@article{Wit81,
	author = {Witten, Edward},
	doi = {10.1007/BF01208277},
	journal = {Commun. Math. Phys.},
	pages = {381},
	reportnumber = {Print-81-0057 (PRINCETON)},
	slaccitation = {%%CITATION = CMPHA,80,381;%%},
	title = {{A Simple Proof of the Positive Energy Theorem}},
	volume = {80},
	year = {1981}}

@article{EngWal17b,
	archiveprefix = {arXiv},
	author = {Engelhardt, Netta and Wall, Aron C.},
	doi = {10.1103/PhysRevLett.121.211301},
	eprint = {1706.02038},
	journal = {Phys. Rev. Lett.},
	number = {21},
	pages = {211301},
	primaryclass = {hep-th},
	slaccitation = {%%CITATION = ARXIV:1706.02038;%%},
	title = {{Decoding the Apparent Horizon: Coarse-Grained Holographic Entropy}},
	volume = {121},
	year = {2018}}

@inproceedings{Pen79,
	author = {Penrose, R.},
	booktitle = {{General Relativity: An Einstein Centenary Survey}},
	pages = {581-638},
	slaccitation = {%%CITATION = INSPIRE-1215196;%%},
	title = {{Singularities and time-asymmetry}},
	year = {1979}}

@inproceedings{Pen73,
	author = {{Penrose}, R.},
	booktitle = {Sixth Texas Symposium on Relativistic Astrophysics},
	doi = {10.1111/j.1749-6632.1973.tb41447.x},
	editor = {{Hegyi}, D.~J.},
	pages = {125},
	series = {Annals of the New York Academy of Sciences},
	title = {{Naked Singularities}},
	volume = 224,
	year = 1973}

@article{EngHor19,
	archiveprefix = {arXiv},
	author = {Engelhardt, Netta and Horowitz, Gary T.},
	doi = {10.1103/PhysRevD.99.126009},
	eprint = {1903.00555},
	journal = {Phys. Rev. D},
	number = {12},
	pages = {126009},
	primaryclass = {hep-th},
	title = {{Holographic argument for the Penrose inequality in AdS spacetimes}},
	volume = {99},
	year = {2019}}

@article{ItkOz11,
	archiveprefix = {arXiv},
	author = {Itkin, Igor and Oz, Yaron},
	doi = {10.1016/j.physletb.2012.01.007},
	eprint = {1106.2683},
	journal = {Phys. Lett. B},
	pages = {307--308},
	primaryclass = {hep-th},
	title = {{Penrose Inequality for Asymptotically AdS Spaces}},
	volume = {708},
	year = {2012}}

@article{Fol22,
	archiveprefix = {arXiv},
	author = {Folkestad, {\AA}smund},
	doi = {10.1103/PhysRevLett.130.121501},
	eprint = {2209.00013},
	journal = {Phys. Rev. Lett.},
	number = {12},
	pages = {121501},
	primaryclass = {hep-th},
	reportnumber = {MIT-CTP/5458},
	title = {{Penrose Inequality as a Constraint on the Low Energy Limit of Quantum Gravity}},
	volume = {130},
	year = {2023}}

@article{McaMor24,
    author = "McAllister, Liam and Moritz, Jakob and Nally, Richard and Schachner, Andreas",
    title = "{Candidate de Sitter vacua}",
    eprint = "2406.13751",
    archivePrefix = "arXiv",
    primaryClass = "hep-th",
    reportNumber = "CERN-TH-2024-090",
    doi = "10.1103/PhysRevD.111.086015",
    journal = "Phys. Rev. D",
    volume = "111",
    number = "8",
    pages = "086015",
    year = "2025"
}

@article{Fol25,
    author = "Folkestad, {\r{A}}smund",
    title = "{Holographic time crystals vs. Penrose}",
    eprint = "2502.01723",
    archivePrefix = "arXiv",
    primaryClass = "hep-th",
    doi = "10.1007/JHEP03(2026)203",
    journal = "JHEP",
    volume = "03",
    pages = "203",
    year = "2026"
}

@article{NarTri10,
    author = "Narayan, Prithvi and Trivedi, Sandip P.",
    title = "{On The Stability Of Non-Supersymmetric AdS Vacua}",
    eprint = "1002.4498",
    archivePrefix = "arXiv",
    primaryClass = "hep-th",
    reportNumber = "TIFR-TH-10-05",
    doi = "10.1007/JHEP07(2010)089",
    journal = "JHEP",
    volume = "07",
    pages = "089",
    year = "2010"
}

@article{KreSka00,
    author = "Kreuzer, Maximilian and Skarke, Harald",
    title = "{Complete classification of reflexive polyhedra in four-dimensions}",
    eprint = "hep-th/0002240",
    archivePrefix = "arXiv",
    reportNumber = "HUB-EP-00-13, TUW-00-07",
    doi = "10.4310/ATMP.2000.v4.n6.a2",
    journal = "Adv. Theor. Math. Phys.",
    volume = "4",
    pages = "1209--1230",
    year = "2000"
}

@article{Bat93,
    author = "Batyrev, Victor V.",
    title = "{Dual polyhedra and mirror symmetry for Calabi-Yau hypersurfaces in toric varieties}",
    eprint = "alg-geom/9310003",
    archivePrefix = "arXiv",
    journal = "J. Alg. Geom.",
    volume = "3",
    pages = "493--545",
    year = "1994"
}

@article{DemKim21,
    author = "Demirtas, Mehmet and Kim, Manki and McAllister, Liam and Moritz, Jakob and Rios-Tascon, Andres",
    title = "{Small cosmological constants in string theory}",
    eprint = "2107.09064",
    archivePrefix = "arXiv",
    primaryClass = "hep-th",
    doi = "10.1007/JHEP12(2021)136",
    journal = "JHEP",
    volume = "12",
    pages = "136",
    year = "2021"
}

@article{McaNal26WIP,
	author = {Compagnin, Federico and MacFadden, Nate and McAllister, Liam and Moritz, Jakob and Nally, Richard and Schachner, Andreas},
	eprint = {in preparation},
	title = "{A Landscape of Candidate de Sitter Vacua}",
	year = {2026}}

@misc{KleNalWIP,
  author = {Kleban, Matthew and Nally, Richard and Norton, Cameron},
  note = {In preparation}
}

@article{Tow84,
    author = "Townsend, P. K.",
    title = "{Positive Energy and the Scalar Potential in Higher Dimensional (Super)gravity Theories}",
    reportNumber = "PRINT-84-0914 (CAMBRIDGE)",
    doi = "10.1016/0370-2693(84)91610-1",
    journal = "Phys. Lett. B",
    volume = "148",
    pages = "55--59",
    year = "1984"
}

@article{KleStr00,
    author = "Klebanov, Igor R. and Strassler, Matthew J.",
    title = "{Supergravity and a confining gauge theory: Duality cascades and chi SB resolution of naked singularities}",
    eprint = "hep-th/0007191",
    archivePrefix = "arXiv",
    reportNumber = "IASSNS-HEP-00-56, PUPT-1944",
    doi = "10.1088/1126-6708/2000/08/052",
    journal = "JHEP",
    volume = "08",
    pages = "052",
    year = "2000"
}

@article{KacPea01,
    author = "Kachru, Shamit and Pearson, John and Verlinde, Herman L.",
    title = "{Brane / flux annihilation and the string dual of a nonsupersymmetric field theory}",
    eprint = "hep-th/0112197",
    archivePrefix = "arXiv",
    reportNumber = "SLAC-PUB-9083, SU-ITP-01-50, PUPT-2019",
    doi = "10.1088/1126-6708/2002/06/021",
    journal = "JHEP",
    volume = "06",
    pages = "021",
    year = "2002"
}

@article{MarMca11,
    author = "Marsh, David and McAllister, Liam and Wrase, Timm",
    title = "{The Wasteland of Random Supergravities}",
    eprint = "1112.3034",
    archivePrefix = "arXiv",
    primaryClass = "hep-th",
    doi = "10.1007/JHEP03(2012)102",
    journal = "JHEP",
    volume = "03",
    pages = "102",
    year = "2012"
}

@book{Hub92,
    author = "Hubsch, Tristan",
    title = "{Calabi-Yau manifolds: A Bestiary for physicists}",
    isbn = "978-981-02-1927-7",
    publisher = "World Scientific",
    address = "Singapore",
    year = "1994"
}

@book{HorKat03,
    author = "Hori, K. and Katz, S. and Klemm, A. and Pandharipande, R. and Thomas, R. and Vafa, C. and Vakil, R. and Zaslow, E.",
    title = "{Mirror symmetry}",
    publisher = "AMS",
    address = "Providence, USA",
    series = "Clay mathematics monographs",
    volume = "1",
    year = "2003"
}

@article{Wal66,
    author = "Wall, C. T. C.",
    title = "{Classification problems in differential topology. V}",
    doi = "10.1007/BF01389738",
    journal = "Invent. Math.",
    volume = "1",
    number = "4",
    pages = "355--374",
    year = "1966"
}

@article{ChaCon23,
    author = "Chandra, Aditi and Constantin, Andrei and Fraser-Taliente, Cristofero S. and Harvey, Thomas R. and Lukas, Andre",
    title = "{Enumerating Calabi-Yau Manifolds: Placing Bounds on the Number of Diffeomorphism Classes in the Kreuzer-Skarke List}",
    eprint = "2310.05909",
    archivePrefix = "arXiv",
    primaryClass = "hep-th",
    doi = "10.1002/prop.202300264",
    journal = "Fortsch. Phys.",
    volume = "72",
    number = "5",
    pages = "2300264",
    year = "2024"
}

@article{GenMac23,
    author = "Gendler, Naomi and MacFadden, Nate and McAllister, Liam and Moritz, Jakob and Nally, Richard and Schachner, Andreas and Stillman, Mike",
    title = "{Counting Calabi-Yau Threefolds}",
    eprint = "2310.06820",
    archivePrefix = "arXiv",
    primaryClass = "hep-th",
    reportNumber = "CERN-TH-2023-189",
    month = "10",
    year = "2023"
}

@article{KacKal03b,
    author = "Kachru, Shamit and Kallosh, Renata and Linde, Andrei D. and Maldacena, Juan Martin and McAllister, Liam P. and Trivedi, Sandip P.",
    title = "{Towards inflation in string theory}",
    eprint = "hep-th/0308055",
    archivePrefix = "arXiv",
    reportNumber = "SLAC-PUB-9669, SU-ITP-03-18, TIFR-TH-03-06",
    doi = "10.1088/1475-7516/2003/10/013",
    journal = "JCAP",
    volume = "10",
    pages = "013",
    year = "2003"
}

@article{DemRio22,
    author = "Demirtas, Mehmet and Rios-Tascon, Andres and McAllister, Liam",
    title = "{CYTools: A Software Package for Analyzing Calabi-Yau Manifolds}",
    eprint = "2211.03823",
    archivePrefix = "arXiv",
    primaryClass = "hep-th",
    month = "11",
    year = "2022"
}

@article{DemMca20,
    author = "Demirtas, Mehmet and McAllister, Liam and Rios-Tascon, Andres",
    title = "{Bounding the Kreuzer-Skarke Landscape}",
    eprint = "2008.01730",
    archivePrefix = "arXiv",
    primaryClass = "hep-th",
    doi = "10.1002/prop.202000086",
    journal = "Fortsch. Phys.",
    volume = "68",
    pages = "2000086",
    year = "2020"
}

@article{MacOre26,
    author = "MacFadden, Nate and Orevkov, Stepan Yu. and Stepniczka, Michael",
    title = "{Further bounding the Kreuzer-Skarke landscape}",
    eprint = "2602.16909",
    archivePrefix = "arXiv",
    primaryClass = "hep-th",
    doi = "10.1007/JHEP10(2026)062",
    journal = "JHEP",
    volume = "10",
    pages = "062",
    year = "2026"
}

@article{BerHub16,
    author = "Berglund, Per and Hubsch, Tristan",
    title = "{A Generalized Construction of Calabi-Yau Models and Mirror Symmetry}",
    eprint = "1611.10300",
    archivePrefix = "arXiv",
    primaryClass = "hep-th",
    doi = "10.21468/SciPostPhys.4.2.009",
    journal = "SciPost Phys.",
    volume = "4",
    number = "2",
    pages = "009",
    year = "2018"
}

@article{MacShe25,
    author = "MacFadden, Nate and Sheridan, Elijah",
    title = "{Calabi-Yau threefolds from vex triangulations}",
    eprint = "2512.14817",
    archivePrefix = "arXiv",
    primaryClass = "hep-th",
    doi = "10.1007/JHEP09(2026)202",
    journal = "JHEP",
    volume = "09",
    pages = "202",
    year = "2026"
}

@article{DemKim23,
    author = "Demirtas, Mehmet and Kim, Manki and McAllister, Liam and Moritz, Jakob and Rios-Tascon, Andres",
    title = "{Computational Mirror Symmetry}",
    eprint = "2303.00757",
    archivePrefix = "arXiv",
    primaryClass = "hep-th",
    reportNumber = "MIT-CTP/5528",
    doi = "10.1007/JHEP01(2024)184",
    journal = "JHEP",
    volume = "01",
    pages = "184",
    year = "2024"
}

@article{GukVaf99,
    author = "Gukov, Sergei and Vafa, Cumrun and Witten, Edward",
    title = "{CFT's from Calabi-Yau four folds}",
    eprint = "hep-th/9906070",
    archivePrefix = "arXiv",
    reportNumber = "HUTP-99-A034, IASSNS-HEP-99-52, PUPT-1864",
    doi = "10.1016/S0550-3213(00)00373-4",
    journal = "Nucl. Phys. B",
    volume = "584",
    pages = "69--108",
    year = "2000",
    note = "[Erratum: Nucl.Phys.B 608, 477--478 (2001)]"
}

@article{DemKim19,
    author = "Demirtas, Mehmet and Kim, Manki and McAllister, Liam and Moritz, Jakob",
    title = "{Vacua with Small Flux Superpotential}",
    eprint = "1912.10047",
    archivePrefix = "arXiv",
    primaryClass = "hep-th",
    doi = "10.1103/PhysRevLett.124.211603",
    journal = "Phys. Rev. Lett.",
    volume = "124",
    number = "21",
    pages = "211603",
    year = "2020"
}

@article{DemKim20,
    author = "Demirtas, Mehmet and Kim, Manki and McAllister, Liam and Moritz, Jakob",
    title = "{Conifold Vacua with Small Flux Superpotential}",
    eprint = "2009.03312",
    archivePrefix = "arXiv",
    primaryClass = "hep-th",
    doi = "10.1002/prop.202000085",
    journal = "Fortsch. Phys.",
    volume = "68",
    pages = "2000085",
    year = "2020"
}

@article{Wit96a,
    author = "Witten, Edward",
    title = "{Nonperturbative superpotentials in string theory}",
    eprint = "hep-th/9604030",
    archivePrefix = "arXiv",
    reportNumber = "IASSNS-HEP-96-29",
    doi = "10.1016/0550-3213(96)00283-0",
    journal = "Nucl. Phys. B",
    volume = "474",
    pages = "343--360",
    year = "1996"
}

@article{Wit96b,
    author = "Witten, Edward",
    title = "{Five-brane effective action in $M$-theory.}",
    eprint = "hep-th/9610234",
    archivePrefix = "arXiv",
    reportNumber = "IASSNS-HEP-96-101",
    doi = "10.1016/S0393-0440(97)80160-X",
    journal = "J. Geom. Phys.",
    volume = "22",
    pages = "103--133",
    year = "1997"
}

@article{Wit96c,
    author = "Witten, Edward",
    title = "{Phase transitions in M theory and F theory}",
    eprint = "hep-th/9603150",
    archivePrefix = "arXiv",
    reportNumber = "IASSNS-HEP-96-26",
    doi = "10.1016/0550-3213(96)00212-X",
    journal = "Nucl. Phys. B",
    volume = "471",
    pages = "195--216",
    year = "1996"
}

@article{GriLou04,
    author = "Grimm, Thomas W. and Louis, Jan",
    title = "{The Effective action of type IIA Calabi-Yau orientifolds}",
    eprint = "hep-th/0412277",
    archivePrefix = "arXiv",
    doi = "10.1016/j.nuclphysb.2005.04.007",
    journal = "Nucl. Phys. B",
    volume = "718",
    pages = "153--202",
    year = "2005"
}

@book{Tuc11,
  author    = {Warwick Tucker},
  title     = {Validated Numerics: A Short Introduction to Rigorous Computations},
  publisher = {Princeton University Press},
  year      = {2011},
  isbn      = {9780691147819}
}

@article{Joh17,
  author  = {Fredrik Johansson},
  title   = {{Arb}: Efficient Arbitrary-Precision Midpoint-Radius
             Interval Arithmetic},
  journal = {IEEE Transactions on Computers},
  volume  = {66},
  number  = {8},
  pages   = {1281--1292},
  year    = {2017},
  doi     = {10.1109/TC.2017.2690633}
}

@article{EldJoy14,
    author = "Elder, Benjamin and Joyce, Austin and Khoury, Justin and Tolley, Andrew J.",
    title = "{Positive energy theorem for $P(X,\phi)$ theories}",
    eprint = "1405.7696",
    archivePrefix = "arXiv",
    primaryClass = "hep-th",
    doi = "10.1103/PhysRevD.91.064002",
    journal = "Phys. Rev. D",
    volume = "91",
    number = "6",
    pages = "064002",
    year = "2015"
}

@article{Rom85,
    author = "Romans, L. J.",
    editor = "Salam, A. and Sezgin, E.",
    title = "{Massive N=2a Supergravity in Ten-Dimensions}",
    reportNumber = "NSF-ITP-85-148",
    doi = "10.1016/0370-2693(86)90375-8",
    journal = "Phys. Lett. B",
    volume = "169",
    pages = "374",
    year = "1986"
}

@article{HerTeg07,
    author = "Hertzberg, Mark P. and Tegmark, Max and Kachru, Shamit and Shelton, Jessie and Ozcan, Onur",
    title = "{Searching for Inflation in Simple String Theory Models: An Astrophysical Perspective}",
    eprint = "0709.0002",
    archivePrefix = "arXiv",
    primaryClass = "astro-ph",
    reportNumber = "MIT-CTP-3859, SU-ITP-07-13, SLAC-PUB-12778",
    doi = "10.1103/PhysRevD.76.103521",
    journal = "Phys. Rev. D",
    volume = "76",
    pages = "103521",
    year = "2007"
}

@article{AraVig21,
    author = "Aragam, Vikas and Paban, Sonia and Rosati, Robert",
    title = "{The Multi-Field, Rapid-Turn Inflationary Solution}",
    eprint = "2010.15933",
    archivePrefix = "arXiv",
    primaryClass = "hep-th",
    reportNumber = "UTTG-10-2020",
    doi = "10.1007/JHEP03(2021)009",
    journal = "JHEP",
    volume = "03",
    pages = "009",
    year = "2021"
}

@article{CveDuf99,
    author = "Cvetic, Mirjam and Duff, M. J. and Hoxha, P. and Liu, James T. and Lu, Hong and Lu, J. X. and Martinez-Acosta, R. and Pope, C. N. and Sati, H. and Tran, Tuan A.",
    title = "{Embedding AdS black holes in ten-dimensions and eleven-dimensions}",
    eprint = "hep-th/9903214",
    archivePrefix = "arXiv",
    reportNumber = "UPR-0840-T, CTP-TAMU-11-99, RU-99-4-B",
    doi = "10.1016/S0550-3213(99)00419-8",
    journal = "Nucl. Phys. B",
    volume = "558",
    pages = "96--126",
    year = "1999"
}

@article{AmsMar06,
    author = "Amsel, Aaron J. and Marolf, Donald",
    title = "{Energy Bounds in Designer Gravity}",
    eprint = "hep-th/0605101",
    archivePrefix = "arXiv",
    doi = "10.1103/PhysRevD.74.064006",
    journal = "Phys. Rev. D",
    volume = "74",
    pages = "064006",
    year = "2006",
    note = "[Erratum: Phys.Rev.D 75, 029901 (2007)]"
}

@article{AmsHer07,
    author = "Amsel, Aaron J. and Hertog, Thomas and Hollands, Stefan and Marolf, Donald",
    title = "{A Tale of two superpotentials: Stability and instability in designer gravity}",
    eprint = "hep-th/0701038",
    archivePrefix = "arXiv",
    doi = "10.1103/PhysRevD.77.049903",
    journal = "Phys. Rev. D",
    volume = "75",
    pages = "084008",
    year = "2007",
    note = "[Erratum: Phys.Rev.D 77, 049903 (2008)]"
}

@article{Nes81,
title = {A new gravitational energy expression with a simple positivity proof},
journal = {Physics Letters A},
volume = {83},
number = {6},
pages = {241-242},
year = {1981},
issn = {0375-9601},
doi = {https://doi.org/10.1016/0375-9601(81)90972-5},
url = {https://www.sciencedirect.com/science/article/pii/0375960181909725},
author = {James M. Nester}
}

@article{FreNun04,
    author = "Freedman, D. Z. and Nunez, Carlos and Schnabl, M. and Skenderis, K.",
    title = "{Fake supergravity and domain wall stability}",
    eprint = "hep-th/0312055",
    archivePrefix = "arXiv",
    reportNumber = "MIT-CTP-3450, ITFA-2003-55",
    doi = "10.1103/PhysRevD.69.104027",
    journal = "Phys. Rev. D",
    volume = "69",
    pages = "104027",
    year = "2004"
}

@article{HerHor04b,
    author = "Hertog, Thomas and Horowitz, Gary T.",
    title = "{Designer gravity and field theory effective potentials}",
    eprint = "hep-th/0412169",
    archivePrefix = "arXiv",
    doi = "10.1103/PhysRevLett.94.221301",
    journal = "Phys. Rev. Lett.",
    volume = "94",
    pages = "221301",
    year = "2005"
}

@article{HenMar06,
    author = "Henneaux, Marc and Martinez, Cristian and Troncoso, Ricardo and Zanelli, Jorge",
    title = "{Asymptotic behavior and Hamiltonian analysis of anti-de Sitter gravity coupled to scalar fields}",
    eprint = "hep-th/0603185",
    archivePrefix = "arXiv",
    reportNumber = "CECS-PHY-06-04, NSF-KITP-06-20, ULB-TH-06-05",
    doi = "10.1016/j.aop.2006.05.002",
    journal = "Annals Phys.",
    volume = "322",
    pages = "824--848",
    year = "2007"
}

@article{Wit01b,
    author = "Witten, Edward",
    title = "{Multitrace operators, boundary conditions, and AdS / CFT correspondence}",
    eprint = "hep-th/0112258",
    archivePrefix = "arXiv",
    month = "12",
    year = "2001"
}

@article{KlebWit99,
    author = "Klebanov, Igor R. and Witten, Edward",
    title = "{AdS / CFT correspondence and symmetry breaking}",
    eprint = "hep-th/9905104",
    archivePrefix = "arXiv",
    reportNumber = "PUPT-1863, IASSNS-HEP-99-49",
    doi = "10.1016/S0550-3213(99)00387-9",
    journal = "Nucl. Phys. B",
    volume = "556",
    pages = "89--114",
    year = "1999"
}

@article{FrePol02,
    author = "Frey, Andrew R. and Polchinski, Joseph",
    title = "{N=3 warped compactifications}",
    eprint = "hep-th/0201029",
    archivePrefix = "arXiv",
    reportNumber = "NSF-ITP-01-77",
    doi = "10.1103/PhysRevD.65.126009",
    journal = "Phys. Rev. D",
    volume = "65",
    pages = "126009",
    year = "2002"
}

@article{Har26,
    author = "Harlow, Daniel",
    title = "{Observers, $\alpha$-parameters, and the Hartle-Hawking state}",
    eprint = "2602.03835",
    archivePrefix = "arXiv",
    primaryClass = "hep-th",
    reportNumber = "MIT-CTP/5900",
    month = "2",
    year = "2026"
}

@article{MarQui22,
    author = "Marchesano, Fernando and Quirant, Joan and Zatti, Matteo",
    title = "{New instabilities for non-supersymmetric AdS$_{4}$ orientifold vacua}",
    eprint = "2207.14285",
    archivePrefix = "arXiv",
    primaryClass = "hep-th",
    reportNumber = "IFT-UAM/CSIC-22-77",
    doi = "10.1007/JHEP10(2022)026",
    journal = "JHEP",
    volume = "10",
    pages = "026",
    year = "2022"
}

@article{Aco25,
    author = "Acosta, P. and others",
    title = "{SubMIT: A Physics Analysis Facility at MIT}",
    eprint = "2506.01958",
    archivePrefix = "arXiv",
    primaryClass = "cs.DC",
    reportNumber = "MIT-CTP/5848",
    doi = "10.1007/s41781-025-00156-1",
    journal = "EPJ RI",
    volume = "10",
    number = "1",
    pages = "2",
    year = "2026"
}

@article{CandHor85,
    author = "Candelas, P. and Horowitz, Gary T. and Strominger, Andrew and Witten, Edward",
    title = "{Vacuum configurations for superstrings}",
    reportNumber = "NSF-ITP-84-170",
    doi = "10.1016/0550-3213(85)90602-9",
    journal = "Nucl. Phys. B",
    volume = "258",
    pages = "46--74",
    year = "1985"
}

@article{FauHor10,
    author = "Faulkner, Thomas and Horowitz, Gary T. and Roberts, Matthew M.",
    title = "{New stability results for Einstein scalar gravity}",
    eprint = "1006.2387",
    archivePrefix = "arXiv",
    primaryClass = "hep-th",
    doi = "10.1088/0264-9381/27/20/205007",
    journal = "Class. Quant. Grav.",
    volume = "27",
    pages = "205007",
    year = "2010"
}

@article{XiaYan22,
    author = "Xiao, Zi-Qing and Yang, Run-Qiu",
    title = "{Penrose inequality in holography}",
    eprint = "2204.12239",
    archivePrefix = "arXiv",
    primaryClass = "hep-th",
    doi = "10.1103/PhysRevD.107.026013",
    journal = "Phys. Rev. D",
    volume = "107",
    number = "2",
    pages = "026013",
    year = "2023"
}

@article{HorOrg07,
    author = "Horowitz, Gary T. and Orgera, Jacopo and Polchinski, Joe",
    title = "{Nonperturbative Instability of $AdS_5 \times S^5/\mathbb{Z}_k$}",
    eprint = "0709.4262",
    archivePrefix = "arXiv",
    primaryClass = "hep-th",
    doi = "10.1103/PhysRevD.77.024004",
    journal = "Phys. Rev. D",
    volume = "77",
    pages = "024004",
    year = "2008"
}

@article{Har10,
    author = "Harlow, Daniel",
    title = "{Metastability in Anti de Sitter Space}",
    eprint = "1003.5909",
    archivePrefix = "arXiv",
    primaryClass = "hep-th",
    reportNumber = "SU-ITP-10-11",
    month = "3",
    year = "2010"
}

@article{ApeMon22,
    author = "Apers, Fien and Montero, Miguel and Van Riet, Thomas and Wrase, Timm",
    title = "{Comments on classical AdS flux vacua with scale separation}",
    eprint = "2202.00682",
    archivePrefix = "arXiv",
    primaryClass = "hep-th",
    doi = "10.1007/JHEP05(2022)167",
    journal = "JHEP",
    volume = "05",
    pages = "167",
    year = "2022"
}

@article{ApeCon22,
    author = "Apers, Fien and Conlon, Joseph P. and Ning, Sirui and Revello, Filippo",
    title = "{Integer conformal dimensions for type IIa flux vacua}",
    eprint = "2202.09330",
    archivePrefix = "arXiv",
    primaryClass = "hep-th",
    doi = "10.1103/PhysRevD.105.106029",
    journal = "Phys. Rev. D",
    volume = "105",
    number = "10",
    pages = "106029",
    year = "2022"
}

@article{Qui22,
    author = "Quirant, Joan",
    title = "{Noninteger conformal dimensions for type IIA flux vacua}",
    eprint = "2204.00014",
    archivePrefix = "arXiv",
    primaryClass = "hep-th",
    reportNumber = "IFT-UAM/CSIC-22-35",
    doi = "10.1103/PhysRevD.106.066017",
    journal = "Phys. Rev. D",
    volume = "106",
    number = "6",
    pages = "066017",
    year = "2022"
}
\end{document}